\documentclass[3p,twocolumn]{elsarticle}

\usepackage{mathrsfs}
\usepackage{amsmath,amsthm}  
\usepackage{amssymb}
\usepackage{color}

\usepackage{url}  

\renewcommand{\leq}{\leqslant}
\renewcommand{\geq}{\geqslant}

\def\eqdef{\stackrel{\mbox{\tiny def}}{=}}     
\def\eqlaw{\stackrel{\mbox{\tiny (law)}}{=}}     
\newcommand{\ket}[1]{|\kern.3ex#1\kern.3ex\rangle}
\newcommand{\bra}[1]{\langle\kern.3ex #1 \kern.3ex|}
\newcommand{\scalar}[2]{\langle\kern.3ex{#1}\kern.3ex|\kern.3ex{#2}\kern.3ex\rangle}

\newcommand{\mean}[1]{\left\langle #1\right\rangle}
\newcommand{\smean}[1]{\langle #1\rangle}
\newcommand{\EXP}[1]{\mathrm{e}^{#1}}         

\newcommand{\re}{\mathop{\mathrm{Re}}\nolimits}      
\newcommand{\im}{\mathop{\mathrm{Im}}\nolimits}      
\newcommand{\tr}[1]{\mathop{\mathrm{Tr}}\nolimits\left\{ #1 \right\}}  
\newcommand{\str}[1]{\mathop{\mathrm{Tr}}\nolimits\big\{ #1 \big\}}  

\newcommand{\heaviside}{\theta_\mathrm{H}}  

\def\I{{\rm i}}
\newcommand{\deriv}[2]{\frac{\mathrm{d}#1}{\mathrm{d}#2}}
\newcommand{\derivp}[2]{\frac{\partial #1}{\partial #2}}
\newcommand{\derivf}[2]{\frac{\delta #1}{\delta #2}}
\def\D{{\rm d}}                  

\newcommand\identity{\mathbf{1}}

\def\Wt{\tau_\mathrm{W}} 
\def\Ht{\tau_\mathrm{H}}
\def\Nc{N}
\def\Sm{\mathcal{S}}
\def\WSm{\mathcal{Q}}    

\def\rdc{R_\mathrm{dc}}
\def\rr{r_q} 
\def\DoS{\nu}
\def\EThouless{\varepsilon_\mathrm{Th}}
\def\PotEnerg{\mathscr{U}}

\def\mass{m}

\begin{document}

\title{Wigner time delay and related concepts \\ Application to transport in coherent conductors \tnoteref{t1}} 
\tnotetext[t1]{Contribution to a special issue ``\textit{Frontiers in quantum electronic transport -- in memory of Markus B\"uttiker}''}


\author[LPTMS]{Christophe Texier}
\ead{christophe.texier@u-psud.fr}

\address[LPTMS]{LPTMS, CNRS, Univ. Paris Sud, Universit\'e Paris Saclay, B\^at. 100, F-91405 Orsay, France}

\begin{abstract}
  The concepts of Wigner time delay and Wigner-Smith matrix allow to characterize temporal aspects of a quantum scattering process.
  The article reviews the statistical properties of the Wigner time delay for disordered systems~;
  the case of disorder in 1D with a chiral symmetry is discussed and the relation with exponential functionals of the Brownian motion underlined.
  Another approach for the analysis of time delay statistics is the random matrix approach, from which we review few results. 
  As a pratical illustration, we briefly outline a theory of non-linear transport and AC transport developed by B\"uttiker and coworkers, where the concept of Wigner-Smith time delay matrix is a central piece allowing to describe screening properties in out-of-equilibrium coherent conductors.

\vspace{0.25cm}
  
  \hfill
  \textsf{Published online the 8 october 2015 in: Physica E \textbf{82}, 16--33 (2016).}
  
  
  \begin{center}
  $\rhd$ \textsc{ This arXiv version was updated and extended after the publication in Physica. } $\lhd$
  \end{center}  
\end{abstract}

\begin{keyword}
\PACS 73.23.-b \sep 73.20.Fz \sep 72.15.Rn
\end{keyword}



\maketitle


\setcounter{tocdepth}{2}
\tableofcontents


\section{Introduction}

The purpose of this article is to review several results linked to a concept which has been very influential in the work of Markus B\"uttiker, namely the concept of Wigner time delay, and several of its extensions -- traversal time, Wigner-Smith matrix, injectance, etc. 
In my opinion, this topic allows one to have a good flavor of Markus B\"uttiker's style as a physicist~:
a combination between formal developments with mathematical elegance and motivations from pratical questions of fundamental condensed matter physics.
Besides, the choice of this theme is also related to a more personal anecdote as it was the subject of my first scientific exchange with Markus B\"uttiker, 
which I rediscovered during the preparation of this article, finding in a box a copy of an old email, dated the 8th July 1997, addressed to me and my PhD advisor Alain Comtet. Here are few sentences  extracted from this message~:
``{\it 
I noticed yesterday your paper on the cond-mat network \cite{ComTex97}. I like to react to two things: (...)
In your introduction you jump from the work of Fyodorov and Sommers immediatly to the work of Brouwer \textit{et al}. Both of these are of course fine works. But the work of Brouwer \textit{et al.} as they make clear in very strong terms takes its starting point from my work by Gopar and Mello \cite{GopMelBut96}, and that alone should have been a good enough reason to not simply leave it out.
(...)
Therefore, it is my hope that you can give this work the place it deserves.
Sincerely,
Markus B\"uttiker}''.
Of course, he was absolutely right and we soon after amended our paper according to his remarks.
Despite this somewhat awkward beginning, Markus B\"uttiker hired me as a postdoctoral assistant at the University of Geneva two years later and I enjoyed very much the pleasant and stimulating atmosphere of the Physics Department.
I especially appreciated the freedom which Markus gave to his postdocs and learnt a lot from our regular exchanges.


The notion of time delay was introduced in the fifties by Eisenbud and Wigner in the context of the quantum theory of scattering. It allows to capture temporal aspects of the scattering process~\cite{Eis48,Wig55,TsaOsb75,BolOsb76,Lyu77}.
The most fundamental aspects of time delays have been reviewed by Carvalho and Nussenzveig~\cite{CarNus02}.
Some other articles have reviewed more specific aspects~:
Beenakker has considered the case of wave guides in the localised regime \cite{Bee01}.
In Ref.~\cite{Kot05}, Kottos has reviewed the random matrix approach for ergodic systems (chaotic cavities) and the case of (non ergodic) disordered systems --diffusive or strongly localised. We have emphasized in \cite{ComDesTex05} the connection with the theory of exponential functional of the Brownian motion.
The purpose of the present article is to review few results on time delays from different contexts and to emphasize the diversity of the physical situations where this concept has found some applications.

The outline is the following~:
in Section \ref{sec:TimeDelays} we discuss several definitions of times characterizing the scattering process.
Section \ref{sec:DisorderedSystems} reviews the statistical analysis of time delays for disordered systems. In Section \ref{sec:RandomMatrices}, we discuss few aspects of the random matrix approach.
Section \ref{sec:InjectanceEtc} introduces several generalized concepts (partial DoS, injectance, etc.) which will find pratical applications for the analysis of non-linear transport (Section~\ref{sec:Nonlinear}) and AC transport (Section~\ref{sec:ACtransport}).


\section{Wigner time delay and other characteristic times}
\label{sec:TimeDelays}

\subsection{Scattering on the half line}
\label{subsec:ScattHalfLine}

\begin{figure}[!ht]
\centering
\includegraphics[scale=0.75]{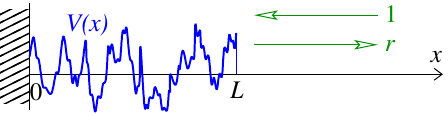}
\caption{\it A plane wave is sent on a potential living on the interval $[0,L]$.}
\label{fig:ScattHalfLine}
\end{figure}

The most simple situation allowing to introduce the concept of time delay is the scattering problem on the half line (this also corresponds to project a rotational invariant problem in an orbital momentum channel, in higher dimensions).
We consider the Schr\"odinger equation 
\begin{equation}
  \label{eq:Schrodinger}
  -\psi_\varepsilon''(x)+V(x)\psi_\varepsilon(x)=\varepsilon\,\psi_\varepsilon(x)
\end{equation}
for $x\in\mathbb{R}^+$ (with $\hbar^2/(2m)=1$), describing the scattering of a particle by a potential defined on a finite interval $[0,L]$ (Fig.~\ref{fig:ScattHalfLine}). At $x=0$ we choose to impose a Dirichlet boundary condition, $\psi_\varepsilon(0)=0$.
In the ``asymptotic'' region ($x>L$), the stationary scattering state of energy $\varepsilon=k^2$ is the superposition of an incident wave $\EXP{-\I k(x-L)}$ and a reflected wave $r\,\EXP{+\I k(x-L)}$~:
\begin{equation}
  \label{eq:ScatteringState}
  \psi_\varepsilon(x) \underset{x>L}{=} 
  \frac{1}{\sqrt{2\pi\hbar v}}\left(\EXP{-\I k(x-L)}+r\,\EXP{\I k(x-L)}\right)
  \:.
\end{equation}
The normalisation constant, involving the group velocity $v=(1/\hbar)\D\varepsilon/\D k=2k$, corresponds to associate a measure $\D\varepsilon$ to the eigenstate.~\footnote{ 
  This choice of normalisation ensures that orthonormalisation reads 
  \begin{equation}
    \scalar{\psi_\varepsilon}{\psi_{\varepsilon'}}=\delta(\varepsilon-\varepsilon')
  \end{equation}
  and closure relation 
  $\int\D\varepsilon\,\ket{\psi_\varepsilon}\bra{\psi_\varepsilon}=1$
  \cite{Tex15book}.  
  }
The reflection probability amplitude has unit modulus as a consequence of current conservation~: $r=\EXP{\I\delta_r}$.  
The Wigner time delay is defined as the derivative of the reflection phase
\begin{equation}
  \label{eq:DefinitionWignerTimeDelay}
  \Wt(\varepsilon) 
  = -\I \hbar \, r^*\derivp{r}{\varepsilon}
  = \hbar\derivp{\delta_r(\varepsilon)}{\varepsilon} 
  \:.
\end{equation}
It measures the time spent by a wave packet of energy $\varepsilon$ in the domain $[0,L]$.
This can be easily understood by considering the time evolution of a wave packet 
\begin{equation}
\Phi(x;t)=\int_0^\infty\D\varepsilon\,\hat{\Phi}(\varepsilon)\,\psi_\varepsilon(x)\,\EXP{-\I\varepsilon t}
\:,
\end{equation}
 where $\hat{\Phi}(\varepsilon)$ is a narrow function centered around an energy $\varepsilon_0$ (from now on, we will set $\hbar=1$ for simplicity).
Normalisation of the wave function is 
$\int\D x\,\big|\Phi(x;t)\big|^2=\int_0^\infty\D\varepsilon\,\big|\hat{\Phi}(\varepsilon)\big|^2=1$.
We can split the wave packet into an incident part and a reflected part~:
$\Phi(x;t)=\Phi_\mathrm{inc}(x;t)+\Phi_\mathrm{ref}(x;t)$.
In the free region $x>L$, neglecting the effect of dispersion,
$\Phi_\mathrm{inc}(x;t)\simeq(1/\sqrt{4\pi})\int_0^\infty\D\varepsilon\,\varepsilon^{-1/4}\,\hat{\Phi}(\varepsilon)\,\EXP{-\I\sqrt{\varepsilon}(x-L)-\I\varepsilon t}$ can be rewritten under the form 
\begin{equation}
\Phi_\mathrm{inc}(x;t)\propto f(-x+L-v_0t)\:, 
\end{equation}
where $v_0$ is the group velocity at energy $\varepsilon_0$ and $f$ the function $f(x)\approx\int\D K\,\hat{\Phi}(\varepsilon_0+v_0K)\,\EXP{\I Kx}$. 
In the reflected wave packet, the expansion of the phase shift produces a shift in time and we obtain 
\begin{equation}
\Phi_\mathrm{ref}(x;t)\propto f\big(x-L-v_0t+v_0\,\Wt(\varepsilon_0)\big)
\end{equation}
(cf. also chapter 10 of the book \cite{Tex15book}).
The interpretation of $\Wt$ as the delay in time relies on following the motion of a wave packet with sufficiently narrow dispersion in energy. 
As trivial illustrations we can consider the case of an impenetrable region when $\delta_r=\pi$, leading to $\Wt=0$, and a free region ($V(x)=0$), leading to $\delta_r=\pi+2kL$ and $\Wt=2L/v$.

\begin{figure}[!ht]
\centering
\includegraphics[width=0.4\textwidth]{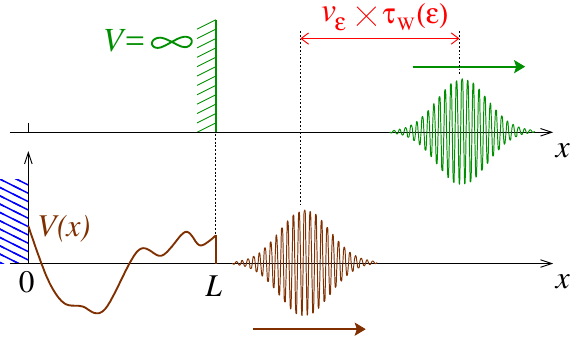}
\caption{\it The Wigner time delay $\Wt(\varepsilon)$ measures the delay of the reflected wave packet with group velocity $v_\varepsilon$ (bottom), with respect to the wave packet reflected on the boundary of the scattering region (top). (Dispersion is neglected here).}
\label{fig:wavepacket}
\end{figure}

\subsection{The scattering problem in one dimension~: several characteristic times}
\label{subsec:CharacteristicTimes}

\begin{figure}[!ht]
\centering
\includegraphics[scale=0.75]{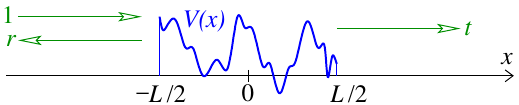}
\caption{\it The stationary scattering state $\psi_{\varepsilon,L}(x)$ describes the scattering of an incident plane wave from the left, which is encoded by a reflection amplitude $r$ and a transmission amplitude~$t$.}
\label{fig:ScattFullLine}
\end{figure}

The scattering problem on the infinite line illustrates that, although the situation is still simple, many other characteristic times can be already introduced.
We consider a plane wave sent from $-\infty$ (or $+\infty$) on a potential defined on the interval $[-L/2,L/2]$ (Fig.~\ref{fig:ScattFullLine}). 
The scattering properties can be encoded in two stationary scattering states $\psi_{\varepsilon,L}(x)$ (Fig.~\ref{fig:ScattFullLine}) and $\psi_{\varepsilon,R}(x)$, controlled by two pairs of reflection/transmission amplitudes $r,\,t$ and $r',\,t'$ characterizing transmission from the left and from the right, respectively.
A general scattering state is a linear combination
\begin{equation}
  \label{eq:GeneralScatteringState}
  \Psi_\varepsilon(x) 
  = A_L\, \psi_{\varepsilon,L}(x) +  A_R\, \psi_{\varepsilon,R}(x)
  \:,
\end{equation}
where $A=(A_L,A_R)$ are the amplitudes of the two incoming plane waves.
They are related to the amplitudes $B=(B_L,B_R)$ of the outgoing plane waves by the $2\times2$ scattering matrix $B=\Sm\,A$~:
\begin{equation}
  \label{eq:2by2Smatrix}
  \Sm = 
  \begin{pmatrix}
     r & t' \\ t & r' 
  \end{pmatrix}
  \:.
\end{equation}

As a consequence of current conservation, the scattering matrix is a  unitary matrix $\Sm\in U(2)$, parametrised by four independent real parameters, what is made clear by expressing $\Sm$ in the polar representation~:
\begin{equation}
  \label{eq:PolarDecomposition}
  \Sm = 
  \EXP{\I\Phi_f/2}
  \begin{pmatrix}
    \sqrt{1-\mathcal{T}}\, \EXP{\I\alpha}  & \I\sqrt{\mathcal{T}}\, \EXP{-\I\chi} \\
    \I\sqrt{\mathcal{T}}\, \EXP{\I\chi}    & \sqrt{1-\mathcal{T}}\, \EXP{-\I\alpha}
  \end{pmatrix}
  \:.
\end{equation}
$\mathcal{T}=|t|^2=|t'|^2\in[0,1]$ denotes the transmission probability. The three independent phases are a global phase $\Phi_f=\det\Sm$, called the ``\textit{Friedel phase}'', the phase $\alpha$ controlling the left/right asymmetry and a magnetic phase $\chi$ (which can be removed by a gauge transformation in the 1D case).
As the scattering process is characterised by a matrix, there is much more freedom to define characteristic times, as we now discuss.

\subsubsection{Friedel phase and Wigner time delay}

The most simple generalization of the definition \eqref{eq:DefinitionWignerTimeDelay} is 
\begin{align}
  \Wt(\varepsilon) &= -\frac{\I}{\Nc} \, 
  \tr{ \Sm^\dagger \derivp{\Sm}{\varepsilon} }
  = -\frac{\I}{\Nc} \, \derivp{\ln\det\Sm}{\varepsilon}
  \\
  &=\frac{1}{\Nc} \, \derivp{\Phi_f}{\varepsilon}
  \:,
\end{align}
where $\Nc$ is the number of scattering channels (here $\Nc=2$).
Although this is an important quantity, as we will explain in the \S~\ref{subsec:KreinFriedel}, $\Wt$ does not have a simple interpretation as a wave packet's delay in time.

\subsubsection{Transmission and reflection group delays}

A plane wave sent on the scattering region from the left is splitted into two parts to which may be associated two phases~: a reflection phase $\delta_r=\Phi_f/2+\alpha$ and a transmission phase $\delta_t=\Phi_f/2+\chi+\pi/2$ (we also introduce $\delta_{r'}=\Phi_f/2-\alpha$). 
Extending the argument exposed in \S~\ref{subsec:ScattHalfLine} would lead to introduce three characteristic times~: one \textit{transmission group delay time} $\check{\tau}_t=\partial\delta_t/\partial\varepsilon$, and two \textit{reflection group delay times} $\check{\tau}_r=\partial\delta_r/\partial\varepsilon$ and $\check{\tau}_{r'}=\partial\delta_{r'}/\partial\varepsilon$
 (note that $\alpha=0$ or $\pi$, and thus $\check{\tau}_r=\check{\tau}_{r'}$ for a symmetric potential).

\subsubsection{Partial time delays~: diagonalisation and derivation}

The two scattering states $\psi_{\varepsilon,\alpha}(x)$, with $\alpha\in\{L,\,R\}$, describing a particle incoming from the left/right (Fig.~\ref{fig:ScattFullLine}), may be recombined in order to form other basis of solutions.
An important case corresponds to the two partial scattering states, which behave ``asymptotically'' as~\footnote{
  For a symmetric potential $V(x)=V(-x)$, the two partial waves are the symmetric and antisymmetric solutions (i.e. $C_1=D_1$ and $C_2=-D_2$).
} 
\begin{equation}
  \label{eq:PartialWave}
  \phi_{\varepsilon,a}(x)
  \underset{|x|>L/2}{=}\big[C_a\heaviside(-x)+D_a\heaviside(x)\big]\cos(k|x|+\eta_a) 
\end{equation}
with $a\in\{1,\,2\}$. $\heaviside(x)$ is the Heaviside function.
Decomposing $\phi_{\varepsilon,a}(x)$ over the left/right scattering states, as in Eq.~\eqref{eq:GeneralScatteringState}, leads to~\cite{Tex02,Tex15book} $\det(\Sm-\EXP{2\I\eta_a}\,\mathbf{1}_2)=0$, i.e. the two partial waves are eigenvectors of the scattering matrix with eigenvalues $\{\EXP{2\I\eta_a}\}$. We can write~:
\begin{equation}
  \label{eq:DiagonalisationS}
  \Sm = \mathcal{U} 
  \begin{pmatrix}
    \EXP{2\I\eta_1} & 0 \\ 0 & \EXP{2\I\eta_2}
  \end{pmatrix}
  \mathcal{U}^\dagger
  \:,
\end{equation}
where $\mathcal{U}$ is a unitary matrix and $\{\eta_a\}$ the scattering phase shifts. 
In particular, using the polar decomposition, one finds the expressions of the two phase shifts
\begin{align}
  &\EXP{2\I\eta_{1,2}} 
  \\\nonumber
  &= \EXP{\I\Phi_f/2}
  \left( 
     \sqrt{1-\mathcal{T}}\,\cos\alpha
     \pm\I\sqrt{1-(1-\mathcal{T})\cos^2\alpha}
  \right)
  \:,
\end{align}
leading to $2(\eta_{1}+\eta_{2})=\Phi_f$, as it should.
Two other characteristic times could be introduced, which are the two \textit{partial time delays}
\begin{equation}
  \label{eq:DefPartial}
  \tilde{\tau}_a = 2 \, \derivp{\eta_a}{\varepsilon}
  \:.
\end{equation}
As the partial waves \eqref{eq:PartialWave} may be combined in order to form wave packets, the partial time delays can be interpreted as delays in time for such wave packets.

\subsubsection{Proper time delays~: derivation and diagonalisation}

The determination of the partial time delays involves a diagonalisation of the scattering matrix and a derivation of its eigenvalues.
The converse of these two operations leads to introduce the \textit{proper time delays} $\{\tau_a\}$, which are eigenvalues of the Wigner-Smith time delay matrix
\begin{equation}
  \label{eq:DefinitionWS}
  \WSm = -\I \, \Sm^\dagger\derivp{\Sm}{\varepsilon}
  = \mathcal{V} 
  \begin{pmatrix}
    \tau_1 & 0 \\ 0 & \tau_2
  \end{pmatrix}
  \mathcal{V}^\dagger
  \:,
\end{equation}
where $\mathcal{V}$ is a unitary matrix, which differs from $\mathcal{U}$ involved in \eqref{eq:DiagonalisationS} in general (see \ref{Appendix:PartialVsProper}).
A related matrix, which will play a role, is
\begin{equation}
  \label{eq:DefinitionWStilde}
  \widetilde{\WSm} = -\I \, \derivp{\Sm}{\varepsilon}\Sm^\dagger = \Sm\,\WSm\,\Sm^\dagger
  \:.
\end{equation}
Hermiticity, $\WSm = \WSm^\dagger$ and $\widetilde\WSm = \widetilde\WSm^\dagger$, follows from unitarity of $\Sm$.
Obviously, $\WSm$ and $\widetilde{\WSm}$ are characterised by the same spectrum of eigenvalues $\{\tau_a\}$, as they are related by a unitary transformation.
Proper time delays and partial time delays have an important difference~:
being derivatives of the $\Sm$-matrix eigenvalues, the partial time delays are intrinsic properties of the scattering process, whereas the proper time delays depend on the particular choice of basis in which the $\Sm$-matrix is expressed (see \ref{Appendix:PartialVsProper}).
Although they do not coincide in general, they satisfy the sum rule
\begin{equation}
  -\I\derivp{}{\varepsilon}\ln\det\Sm
  =\sum_{a=1}^\Nc \tilde{\tau}_a = \sum_{a=1}^\Nc \tau_a = \Nc\,\Wt
\end{equation}
where $\Nc$ is the number of scattering channels (here $\Nc=2$).

\subsubsection{Other characteristic times}

The time delay interpretation of the various characteristic times deduced from scattering phases may lead to paradoxal conclusions in the presence of tunneling barriers, like superluminal propagation.
For this reason ``clock approaches'' have been proposed in order to provide more satisfactory answers to the question ``how much time needs a wave packet to travel in a given region~?''.
We refer to the review articles \cite{HauSto89,LeaAer89,But90,LanMar94,But02a,CarNus02}~\footnote{
  The introduction of Ref.~\cite{BenJay02}, which introduces a sojourn time from wave attenuation, provides a good overview.
} and B\"uttiker's contributions \cite{ButLan82,But83,ButLan85}.

\subsection{Krein-Friedel relation and Virial expansion}
\label{subsec:KreinFriedel}

An important aspect behind the notion of time delay is its link with the spectral properties of open systems with continuous spectra, which can be understood as follows.
Let us come back to the simple situation of scattering on the semi-infinite line for simplicity (Fig.~\ref{fig:ScattHalfLine}).
The scattering states \eqref{eq:ScatteringState} satisfying the Dirichlet boundary condition $\psi_\varepsilon(0)=0$ may be used as a basis in order to consider the spectral problem on the \textit{finite} interval $[0,L]$, which involves a second boundary condition at $x=L$. We choose a Neumann boundary condition $\psi_\varepsilon'(L)=0$ for instance. 
The eigenenergies $\{\varepsilon_n\}$ are solutions of the quantisation equation $\delta_r(\varepsilon_n)=2n\pi$ for $n$ integer.  
When two successive energies are sufficiently close, we can expand the relation 
$\delta_r(\varepsilon_{n+1})-\delta_r(\varepsilon_n)=2\pi$ as 
$\delta_r'(\varepsilon_n)\,\delta \varepsilon_n\simeq2\pi$, where $\delta \varepsilon_n=\varepsilon_{n+1}-\varepsilon_n$ is the level spacing. We deduce the expression of the density of states $\DoS(\varepsilon)\simeq1/\delta\varepsilon_n\simeq\Wt(\varepsilon)/(2\pi\hbar)$.
A more precise connection between spectral and scattering properties was obtained by Friedel~\cite{Fri58} and Smith~\cite{Smi60}, who established the relation
\begin{equation}
  \label{eq:FriedelSmith}
  \int_0^L\D x\, \left|\psi_\varepsilon(x)\right|^2 
  = \frac{1}{2\pi}\left( \Wt(\varepsilon) +\frac{\sin\delta_r(\varepsilon)}{2\varepsilon} \right)
\end{equation}
(see also Ref.~\cite{TexCom99}). 
The left hand side of the equation is simply related to the local density of states (DoS)~\footnote{
  The choice of the normalisation for the scattering state \eqref{eq:ScatteringState} is important.
} 
\begin{equation}
  \label{eq:LDoS}
  \DoS(x;\varepsilon)=\bra{x}\delta(\varepsilon-H)\ket{x}=|\psi_\varepsilon(x)|^2
  \:.
\end{equation}
Thus, Eq.~\eqref{eq:FriedelSmith} shows that the Wigner time delay provides a measure of the DoS of the interval $[0,L]$, i.e. of the integral of the local DoS over the support of the potential. 
It may be written in the more general form~\cite{TexBut03}
\begin{equation}
  \label{eq:TexierButtiker2003}
  \DoS(\varepsilon) =\int_0^L\D x\, \DoS(x;\varepsilon)
  = \frac{1}{2\I\pi}
  \tr{ \Sm^\dagger\derivp{\Sm}{\varepsilon} + \frac{\Sm-\Sm^\dagger}{4\varepsilon} }
  \:,
\end{equation}
so that it also applies to more general situations such as the one of Fig.~\ref{fig:ScattFullLine}, for a potential vanishing outside $[0,L]$, or even complex structures like metric graphs~\cite{Tex02,TexBut03,TexDeg03}.
Such equalities are known as Krein-Friedel relations \cite{Fri52,Kre53,Fri58,Bus62} (or Birman-Krein formula \cite{BirKre62}).~\footnote{
  The Krein-Friedel relation is sometimes written in terms of the spectral shift function 
  $$\Delta(\varepsilon)=\int_{-\infty}^{+\infty}\D x\, \big[\DoS(x;\varepsilon)-\DoS_0(x;\varepsilon)\big]\:,$$ which measures the variation of the DoS due to the introduction of the scattering potential, where $\DoS_0(x;\varepsilon)$ is the free local DoS.
  The precise relationship between the two formulations has been discussed in Ref.~\cite{Tex02} for the 1D situation. 
  When the scattering matrix is defined like in \eqref{eq:ScatteringState}, we have
  $$\Delta(\varepsilon)=(2\I\pi)^{-1}\partial_\varepsilon\big[\ln\det\Sm-2\I\sqrt{\varepsilon}L\big]\:.$$
  Note that the second term can be absorbed in a redefinition of the phases of $\Sm$ (see Appendix of Ref.~\cite{Tex02} for details).
}
They have a long history and first appeared in the virial expansion of the equation of state of real gases 
\begin{equation}
 p=nT\,\big[1+B_2(T)\,n+B_3(T)\,n^2+\cdots\big]
 \:,
 \end{equation} 
 where $p$ is the pressure, $n$ the density and $T$ the temperature.
As shown by Beth and Uhlenbeck~\cite{UhlBet36,BetUhl37}, the second virial coefficient can be written as 
\begin{equation}
B_2(T)=-2^{-5/2}\Lambda_T^3(\pm1+16Z_\mathrm{int}) 
\end{equation}
for bosons ($+$) and fermions ($-$),
where $\Lambda_T=\sqrt{2\pi\hbar^2/(mT)}$ is the thermal length and $m$ the mass of the particles.
The first contribution to $B_2$ encodes the quantum correlations arising from the symmetrisation postulate, whereas the second term describes the correlations due to the interaction, $Z_\mathrm{int}$ being the partition function of the two body problem (in the relative coordinates)
\begin{equation}
  \label{eq:Zint}
  Z_\mathrm{int} = \sum_{\ell}(2\ell+1)
  \int_0^\infty\frac{\D\varepsilon}{2\pi\hbar}\tau_\ell(\varepsilon)\,\EXP{-\varepsilon/T}
  \:,
\end{equation}
where summation runs over orbital momentum.
$\tau_\ell=2\partial\eta_\ell/\partial\varepsilon$ is the time delay related to the scattering phase shift~\footnote{
  Note that \eqref{eq:Zint} assumes that the scattering phase shifts are measured with respect to the free eigenstates, $\psi_\ell(r)\sim(1/r)\big[\EXP{-\I kr}+\EXP{\I kr+2\I\eta_\ell}\big]$, as it is done usually in scattering theory, instead of~\eqref{eq:ScatteringState}.
}
$\eta_\ell$ characterizing the scattering in the channel of orbital momentum $\ell$ (cf. \S~77 of \cite{LanLif66e})~; in the presence of bound states, the partition function receives additional contributions.
This approach was later made more systematic by Dashen, Ma and Bernstein~\cite{DasMaBer69}.
Note that a similar analysis of real gases based on time delay can be developed within a purely classical frame, as beautifully explained by Ma~\cite{Ma85}.

\subsection{Few remarks}

Characterisation of the spectral properties of open systems has important applications as it allows to express several physical observables in terms of scattering properties at the heart of the Landauer-B\"uttiker description of quantum transport.
For example, \eqref{eq:TexierButtiker2003} allows to measure the charge inside a conductor in terms of its scattering matrix, an idea which turns out to be central in many developments of B\"uttiker involving screening, which are reviewed below.
Several remarks~:
\begin{enumerate}[(i)]

\item 
  The Krein-Friedel relation was studied in metric graphs in \cite{Tex02,TexBut03}.
  In such systems with non trivial topology, the system might exhibit so-called ``\textit{bound states in the continuum}'' (BIC), i.e. a discrete spectrum \textit{superimposed} onto the continuum spectrum (this occurs when symmetries allow some bound states to remain uncoupled to the continuum)~\cite{NeuWig29,StiHer75}.
  In this case, the Krein-Friedel relation only characterises the continuous part of the spectrum \cite{TexBut03,TexDeg03,Tex10hdr} (see \cite{Tex02,TexBut03} for illustrations).

\item
  A related problem concerns the role of the transmission phase. As it is clear from the polar representation \eqref{eq:PolarDecomposition}, the Friedel phase $\Phi_f$, which enters the relation \eqref{eq:TexierButtiker2003} as $\det\Sm=\EXP{\I\Phi_f}$, is related to the transmission phase $\delta_t=(\Phi_f+\pi)/2$ in 1D, as the magnetic field can always be removed by a gauge transformation on the infinite line. Thus the DoS can be as well related to $\delta_t$ in 1D~\cite{AviBan85}.
  If the system has a complex topology, the transmission may vanish which causes transmission phase jumps. As a consequence of these phase jumps, the Friedel phase and the transmission phase differ~; the density of states is then only related to the Friedel phase, as discussed by Taniguchi and B\"uttiker~\cite{TanBut99}.
  This question has been examined numerically in quantum dots by L\'evy Yeyati and B\"uttiker~\cite{LevBut00}.

\item
  In  Ref.~\cite{SouSuz02}, the Krein-Friedel relation was considered in the quasi-one-dimensional situation (wave guide) which is relevant in order to describe the electric contacts of mesoscopic structures.

\item
  A formulation of the time dependent transport was developed in \cite{Pas92} (using the Keldysh formalism) and was shown to involve the time delay matrix expressed in terms of Green's functions.
  
\item
  In Subsection~\ref{subsec:ScattHalfLine}, the Wigner time delay has been introduced by considering the delay of the position of a wave packet with narrow dispersion in energy. This corresponds to follow the \textit{mean} position of the particle, cf. Fig.~\ref{fig:wavepacket} (where averaging would be taken with respect to measurements in a given quantum state). The remark also holds for all times introduced in Section~\ref{sec:TimeDelays}.
  Elaborating on Refs.~\cite{Lyu83,BluSmi88}, Smilansky has introduced a distribution of time delay for a particle injected in channel $i$ and detected in channel $f$~\cite{Smi17}
  \begin{equation}
    \label{eq:Smilansky2017}
    P_{fi}(\tau) = \frac{1}{2\pi}
    \left|
      \int_0^\infty\D\varepsilon\,\hat{\Phi}(\varepsilon)\,\Sm_{fi}(\varepsilon)\,\EXP{-\I\varepsilon\tau}
    \right|^2
    \:,
  \end{equation}
  where the wave packet envelope $\hat{\Phi}(\varepsilon)$ is a real and narrow normalised function.
  Smilansky has also discussed the classical limit of the distribution.
  In the simple case of scattering on the half line, we can check that the mean time is related to the Wigner time delay $\Wt(\varepsilon)=-\I\Sm(\varepsilon)^*\Sm'(\varepsilon)$~:
  \begin{align}
    \int\D\tau\,P(\tau)\,\tau 
    &= 
    -\I\int\D\varepsilon\,
    \big[\hat{\Phi}(\varepsilon)\Sm(\varepsilon)\big]^*
    \big[\hat{\Phi}(\varepsilon)\Sm(\varepsilon)\big]'
    \nonumber\\
    &= \int\D\varepsilon\, \hat{\Phi}(\varepsilon)^2\,\Wt(\varepsilon)
  \end{align}
  where we have used unitarity $|\Sm|^2=1$ and normalisation $\int_0^\infty\D\varepsilon\,\hat{\Phi}(\varepsilon)^2=1$. 
  For several channels, a similar relation with the Wigner-Smith time delay matrix \eqref{eq:DefinitionWS} only holds with an additional summation 
  \begin{equation}
    \sum_f \int\D\tau\,P_{fi}(\tau)\,\tau 
    = \int\D\varepsilon\, \hat{\Phi}(\varepsilon)^2\,\WSm_{ii}
    \:.
  \end{equation}
  Similarly, a summation over the ingoing channel involves the matrix \eqref{eq:DefinitionWStilde}
  \begin{equation}
    \sum_i \int\D\tau\,P_{fi}(\tau)\,\tau 
    = \int\D\varepsilon\, \hat{\Phi}(\varepsilon)^2\,\widetilde{\WSm}_{ff}
    \:.
  \end{equation}
  
  We stress that the distribution \eqref{eq:Smilansky2017} describes the fluctuations of time due to the stochasticity of the scattering process in \textit{a given sample}. It should \textit{not} be confused with the time delay distributions studied below in Sections~\ref{sec:DisorderedSystems} and \ref{sec:RandomMatrices}, which characterize the \textit{sample to sample} (mesoscopic) fluctuations of the Wigner time delay $\Wt(\varepsilon)$ or the related quantities (Wigner-Smith matrix, proper times, partial times, etc).
\end{enumerate}


\section{Disordered systems (localised regime)}
\label{sec:DisorderedSystems}

The review \cite{ComDesTex05} has underlined the close relationship between Wigner time delay for 1D disordered systems and exponential functionals of the Brownian motion, which have been widely studied in the mathematical literature \cite{Yor00} (see also \cite{ComMonYor98} for a brief review). 
Let us first recall few properties which will be useful for the following.

\subsection{Exponential functionals of the Brownian motion}
\label{subsec:ExpFunct}

We introduce the random variable
\begin{equation}
  \label{eq:ExponFunct}
  Z_X^{(\mu)}
  =\int_0^X\D x\, \EXP{-2(\mu\, x + B(x))}
  \:,
\end{equation}
where $B(x)$ is a normalised Brownian motion starting from $B(0)=0$ (Wiener process).
The distribution of \eqref{eq:ExponFunct} was found in \cite{MonCom94,ComMon96} (see also \cite{ComMonYor98})~:
\begin{align}
  \label{eq:DistribExpFunct}
  &\psi_X^{(\mu)}(Z) = 2\frac{\EXP{-1/(2Z)}}{(2Z)^{1+\mu}} 
  \sum_{0\leq n<\mu/2}
  \frac{(-1)^{n}(\mu-2n)}{\Gamma(1+\mu-n)}
  \nonumber\\
  &\hspace{1.5cm}\times\EXP{-2Xn(\mu-n)}\,(2Z)^n\, L_n^{\mu-2n}\left(\frac{1}{2Z}\right)
  \nonumber\\
  &\hspace{-0.25cm}
  +\frac{\EXP{-1/(4Z)}}{2\pi^2(2Z)^{(1+\mu)/2}}
  \int_0^\infty\D s\,s\,\sinh(\pi s)\,
  \left|\Gamma\left(\frac{\I s-\mu}{2}\right)\right|^2\,
  \nonumber\\
  &\hspace{0.75cm}\times W_{(1+\mu)/2,\I s/2}\left(\frac{1}{2Z}\right)
  \EXP{-(X/2)(\mu^2+s^2)}
  \:,
\end{align}
where $L_n^\alpha(x)$ is a Laguerre polynomial and $W_{\mu,\nu}(z)$ the Whittaker function~\cite{gragra}.
Eq.~\eqref{eq:DistribExpFunct} shows that the random variable admits the limit law 
\begin{equation}
  \psi_\infty^{(\mu)}(Z) = \frac{1}{2^\mu\Gamma(\mu)\,Z^{1+\mu}}\,\EXP{-1/(2Z)}
  \quad\mbox{for }\mu>0
  \:.
\end{equation}
For finite $X$, the functional is characterised by exponential moments
\begin{equation}
  \smean{\big(Z_X^{(\mu)}\big)^n} 
  \underset{X\to\infty}{\simeq} 
  2^{-n} \frac{\Gamma(n-\mu)}{\Gamma(2n-\mu)}\,\EXP{2n(n-\mu)X}
  \:.
\end{equation}
The precise expression of the moments can be found in Ref.~\cite{OshMogMor93} for~$\mu=0$ and Ref.~\cite{MonCom94} in the general case.~\footnote{
  Note however that Eq.~4.4 of Ref.~\cite{MonCom94} for $\mu=0$ contains a misprint~:
  $C_{2n}^k$ should be replaced by $C_{2n}^{n-k}$.
}

\subsection{Disorder on the half line - Single parameter scaling and universality}
\label{subsec:DOHL}

The interest for time delays in disordered systems has started with the work of Sulem and coworkers~\cite{FriFroSchSul73} on stochastic resonances (see also \cite{PenKirCas86}). The idea that localised states could produce sharp resonances was later developed by Azbel~\cite{Azb83}.
The first statistical analysis of the Wigner time delay in a one-dimensional situation was provided by Jayannavar, Vijayagovindan and Kumar \cite{JayVijKum89}. 
These authors have studied the time delay distribution $P_L(\tau)$ for the Schr\"odinger equation \eqref{eq:Schrodinger} on the half line (Fig.~\ref{fig:ScattHalfLine}) when $V(x)$ is a Gaussian white noise.
They have identified the existence of a limit law with power law tail $P_\infty(\tau)\sim\tau^{-2}$ for an infinitely long disordered region, $L\to\infty$, although their expression of  $P_\infty(\tau)$ was partly incorrect (with a non vanishing distribution as $\tau\to0$). 
Similar conclusions were obtained in Ref.~\cite{JosJay98} when the potential $V(x)$ is the \textit{integral} of a white noise, i.e. a Brownian motion.
As it was recognised later~\cite{RamKum01}, the incorrect non vanishing $P_\infty(0)$ of Refs.~\cite{JayVijKum89,JosJay98} was due to some inappropriate averaging over the fast phase variable in the weak disorder regime.
Short after, Heinrichs identified that the exponential moments, $\mean{\tau^n}\sim\exp[2n(n-1)L/\xi]$ where $\xi$ is the localisation length, are characteristic of a log-normal tail $P_L(\tau)$ when the length $L$ is finite~\cite{Hei90}.
Inspired by the work of Faris and Tsay \cite{FarTsa94}, we established in Ref.~\cite{ComTex97} the connection between the Wigner time delay and exponential functionals of the Brownian motion.
We analysed two different disordered models exhibiting qualitatively different spectral and localisation properties, namely the Schr\"odinger Hamiltonian $H=-\partial_x^2+V(x)$ and the supersymmetric Hamiltonian $H=-\partial_x^2+\mass(x)^2+\mass'(x)$ [i.e. the square of the Dirac Hamiltonian~\eqref{eq:Dirac}], where the potential $V(x)$ and the mass $\mass(x)$ are Gaussian white noises.
This analysis suggested the universality of the statistical properties, which was later demonstrated in Ref.~\cite{TexCom99}.
This can be understood from the integral representation \eqref{eq:FriedelSmith}~:
in the \textit{weak disorder} regime we may write $\Wt\simeq2\pi\int_0^L\D x\,|\psi_\varepsilon(x)|^2$.
Using that the logarithm of the envelope of the wave function is a Brownian motion with drift, we have related the time delay to the functional \eqref{eq:ExponFunct}~\cite{TexCom99,ComDesTex05}~\footnote{
  An equality in law $\eqlaw$ relates two quantities with the same statistical properties.
  For example, the well-known scaling properties of the Brownian motion may be conveniently written
  $B(\lambda x)\eqlaw\lambda^{1/2}B(x)$, where $B(x)$ is a Brownian motion. 
}
\begin{equation}
  \label{eq:EqualityInLaw1}
  \Wt \eqlaw 2\tau_\xi\,Z_{L/\xi}^{(1)}
  \:,
\end{equation}
where $\xi$ is the localisation length~\footnote{
  For the model \eqref{eq:Schrodinger} with $\mean{V(x)V(x')}=\sigma\,\delta(x-x')$, the localisation length is $\xi\simeq8\varepsilon/\sigma$ for weak disorder $\sigma^{2/3}\ll\varepsilon$.
} 
and $\tau_\xi=\xi/v=\xi/(2k)$ is the time needed for a ballistic motion on the localisation length $\xi$.
The value of the drift $\mu=1$ originates from the equality 
\begin{equation}
  \label{eq:SPS}
  \mean{ \ln|\psi_\varepsilon(x)| } 
  \simeq \mathrm{Var}\left(\ln|\psi_\varepsilon(x)|\right)
  \quad\mbox{ for } x\gg\xi  
  \:,
\end{equation}
valid in the weak disorder regime \cite{AntPasSly81}. 
Eq.~\eqref{eq:SPS} is known as ``\textit{single parameter scaling}'' \cite{AndThoAbrFis80,CohRotSha88}, as the full distribution of $\ln|\psi_\varepsilon(x)|$ (or of the conductance) is characterised by a unique length scale, the localisation length defined by
$1/\xi=\lim_{x\to\infty}(1/x)\ln|\psi_\varepsilon(x)|$.
This question has been rediscussed more recently for 1D disordered systems \cite{DeyLisAlt00,DeyLisAlt01,SchTit03,RamTex14}.
The universality of the Wigner time delay statistical properties are thus understood as a direct consequence of the universality of the localisation properties in 1D in the weak disorder regime.~\footnote{
  The time delay distribution was also studied in the non universal (strong disorder) regime in \cite{TexCom99} for a particular model.
} 
In other terms, Eq.~\eqref{eq:EqualityInLaw1} means that the distribution of the Wigner time delay is, in the universal (weak disorder) regime, 
\begin{equation}
  \label{eq:ComtetTexier1997}
  P_L(\tau) = \frac{1}{2\tau_\xi} \, \psi_{L/\xi}^{(1)}\left(\frac{\tau}{2\tau_\xi}\right)
  \:.
\end{equation}
If the length $L$ of the disordered region goes to infinity we get the limit law~\cite{ComTex97,Tex99,TexCom99} 
\begin{equation}
\label{eq:ComtetTexier1997LimitLaw}
  P_\infty(\tau)=\frac{\tau_\xi}{\tau^2}\,\EXP{-\tau_\xi/\tau}
  \:.
\end{equation}
The limits $L\to\infty$ and $\tau\to\infty$ do not commute~: for finite $L$ the tail of the distribution is log-normal~\cite{Hei90,TexCom99},
$P_L(\tau)\sim\exp\big[-(\xi/8L)\ln^2(\tau/\tau_\xi)\big]$ for $\tau\to\infty$.
The leading term of the moments was also given in Ref.~\cite{ComTex97}~:
\begin{align}
  \mean{\tau}&=2vL \\
  \label{eq:MomentsTDUniversal}
  \mean{\tau^n}&\simeq\frac{(n-2)!}{(2n-2)!}\tau_\xi^n\EXP{2n(n-1)L/\xi}
  \hspace{0.5cm}
  \mbox{for }
  n>1
  \:,
\end{align}
confirming the main exponential behaviour found earlier by Heinrichs~\cite{Hei90} and providing the pre-exponential factor (a more precise expression of the moments was given in Ref.~\cite{TexCom99}).
Note that the log-normal distribution is characterised by moments of the form $\exp[c\,n^2]$~; the fact that the moments \eqref{eq:MomentsTDUniversal} increase with a lower rate with $n$, as $\sim\exp[2n(n-1)L/\xi]$, reflects the existence of a limit law with tail $\tau^{-2}$ (see \S~\ref{subsec:ExpFunct}).
The importance of Azbel resonances was emphasized as it allowed to recover the moments by some heuristic argument \cite{TexCom99}.
A lattice model was studied in Ref.~\cite{OssKotGei00}, leading to the same conclusions.

\subsection{Disorder in 1D with a chiral symmetry}
\label{subsec:DWCS}

The localisation properties of disordered systems are mainly controlled by their dimensionality \cite{AbrAndLicRam79} and symmetries.
A first classification of symmetries was given by Wigner and Dyson in the context of random matrix theory, depending on the presence of time reversal symmetry and spin rotational symmetry, leading to the orthogonal ($\beta=1$), unitary ($\beta=2$) and symplectic classes ($\beta=4$)~\cite{Meh04}.
These three symmetry classes were later completed by others, identified according to the presence or not of two other types of discrete symmetries which might occur in condensed matter physics~: the chiral (or sublattice) symmetry and the  particle-hole symmetry. 
These has led to the classification in terms of ten symmetry classes \cite{Zir96,AltZir97,EveMir08}.
A one-dimensional disordered model with a chiral symmetry which has attracted a lot of attention is the Dirac equation with a random mass (see \cite{BouComGeoLeD90,ComTex98,TexHag10} for reviews)~\footnote{
  The case of higher dimensions is also very important~: besides that it appears in several contexts of condensed matter physics (graphene, etc), it is of fundamental interest as it was shown that all symmetry classes can be represented within the 2D Dirac equation with randomness~\cite{BerLec02}.
}
\begin{equation}
  \label{eq:DiracEquation}
  \mathcal{H}_D\, \Psi(x) = \varepsilon\,\Psi(x)
\end{equation}
with
\begin{equation}
  \label{eq:Dirac}
  \mathcal{H}_D =\I\sigma_2\partial_x+\sigma_1\,\mass(x)
  \:,
\end{equation}
where $\sigma_i$ are the Pauli matrices.
A chiral symmetry is the anticommutation of the Hamiltonian with a unitary operator, here 
$\sigma_3\mathcal{H}_D\sigma_3=-\mathcal{H}_D$.
The Wigner time delay distribution for this model was obtained in Ref.~\cite{ComTex97} in the ``\textit{universal}'' (weak disorder) regime, leading to the same conclusions as for the Schr\"odinger equation, Eqs.~(\ref{eq:EqualityInLaw1},\ref{eq:ComtetTexier1997}), provided that the localisation length is modified, $\tau_\xi=\xi\simeq2/g$ for $\varepsilon\gg g$ (in the Dirac equation, the velocity is equal to unity).
The energy $\varepsilon$ can be viewed as a parameter which tunes the chiral symmetry breaking~: 
at the symmetry point $\varepsilon=0$, where chiral symmetry holds, the Wigner time delay presents different properties as we now discuss.

The scattering problem on the half line for the Dirac equation is settled as follows~:
we consider again the situation of Fig.~\ref{fig:ScattHalfLine} where the mass is non zero on the interval $[0,L]$.
In the free region we can decompose the spinor of energy $\varepsilon>0$ as the combination of an incoming plane wave and a reflected plane wave~:
\begin{equation}
  \label{eq:ScattStateDirac}
  \Psi_\varepsilon(x) \underset{x>L}{=} 
  \begin{pmatrix}
    1 \\ \I
  \end{pmatrix}
  \EXP{-\I\varepsilon(x-L)}
  + 
  \begin{pmatrix}
    -1 \\ \I
  \end{pmatrix}
  \EXP{\I\varepsilon(x-L)+\I\delta_r(\varepsilon)}
  \:.
\end{equation}
Following \cite{ComTex97,Tex99} we can parametrize the spinor on $[0,L]$ as 
$\Psi(x)=\rho(x)\,\big(\sin\theta(x),-\cos\theta(x)\big)$, 
which leads to the equation for the phase 
\begin{equation}
  \label{eq:DifferentialEquationForThePhaseTheta}
  \derivp{\theta(x)}{x} = \varepsilon + \mass(x)\, \sin\left[2\theta(x)\right]
  \:.
\end{equation}
The matching with \eqref{eq:ScattStateDirac} shows that the phase shift is given by 
$\delta_r=2\theta(L)$. 
We choose the initial condition $\theta(0)=0$ or $\pi/2$, which corresponds to an infinite mass $\mass(x)=+\infty$ or $\mass(x)=-\infty$ for $x<0$, respectively (these boundary conditions confine the particle on $\mathbb{R}_+$~; they are the only ones which do not break the chiral symmetry \cite{TexHag10}).
Following \cite{ComTex97} we introduce the variable 
$\mathcal{Z}(x)=2\,\partial\theta(x)/\partial\varepsilon$ 
which obeys 
\begin{equation}
  \label{eq:SDEforZ}
  \derivp{\mathcal{Z}(x)}{x} = 2 + 2\mass(x)\, \mathcal{Z}(x)\,\cos\left[2\theta(x)\right]
\end{equation}
with initial condition $Z(0)=0$ and provides the value of the Wigner time delay $\Wt=\mathcal{Z}(L)$.

The analysis of the symmetry point $\varepsilon=0$ is easy~: the phase remains locked at $0$ or $\pi/2$ (corresponding to $\mass(x)=\pm\infty$ for $x<0$), therefore one can integrate \eqref{eq:SDEforZ}. On gets
\begin{equation}
  \label{eq:WTD-Dirac-zero}
  \Wt = 2\int_0^L\D x\, \EXP{\pm2\int_x^L\D x'\,\mass(x')}
  \quad
  \mbox{at }
  \varepsilon=0
  \:.
\end{equation}

For the sake of concreteness, we now consider the situation where the mass is a Gaussian white noise 
\begin{align}
\label{eq:RandomMass}
\begin{cases}
\mean{\mass(x)\mass(x')}_c=g\,\delta(x-x')
\\
 \mean{\mass(x)}=\mu\,g
 \end{cases}
\end{align}
where $\mean{XY}_c=\mean{XY}-\mean{X}\mean{Y}$. 
The representation \eqref{eq:WTD-Dirac-zero} makes clear the identity
\begin{equation}
  \Wt  \eqlaw \frac{2}{g}\, Z_{gL}^{(\mp\mu)}
  \quad
  \mbox{at }
  \varepsilon=0
  \:,
\end{equation}
which is distributed according to~\footnote{
  It is interesting to compare \eqref{eq:DistribWTDDirac} for $\varepsilon=0$ with the high energy ($\varepsilon\gg g$) distribution \eqref{eq:ComtetTexier1997} obtained in the absence of a chiral symmetry, which takes the explicit form
  \begin{equation}
    P_L(\tau) = ({g}/{4}) \, \psi_{gL/2}^{(1)}\left({g\tau}/{4}\right)
  \quad
  \mbox{at }
  \varepsilon\gg g
   \:.    
  \end{equation}
}
\begin{align}
  \label{eq:DistribWTDDirac}
  P_L(\tau) =\frac{g}{2} \, \psi_{gL}^{(\mp\mu)}\left(\frac{g\tau}{2}\right)
  \quad
  \mbox{at }
  \varepsilon=0
  \:,
\end{align}
where the dimensionless function was defined above, Eq.~\eqref{eq:DistribExpFunct}.

\subsubsection{The critical case $\mean{\mass(x)}=0$}

The distribution for $\mu=0$ was obtained in \cite{SteCheFabGog99} (although the distribution has been already determined in another context \cite{MonCom94}, what was used in \cite{Tex99}).
It is characterised by a log-normal tail $P_L(\tau)\sim\exp\big[-1/(8gL)\ln^2(\tau)\big]$ and exponential moments 
\begin{align}
  \mean{\tau^n} \simeq \frac{(n-1)!}{(2n-1)!}g^{-n}\EXP{2n^2gL}
    \hspace{0.5cm}
  \mbox{for }
  n\geq1
  \:.
\end{align}
Although the distribution has no limit law, the large $L$ limit presents the power law behaviour \cite{MonCom94}
\begin{equation}
  \label{eq:TailDirac}
  P_L(\tau) \underset{L\to\infty}{\simeq}
  \frac{1}{\sqrt{2\pi gL}\,\tau}\EXP{-1/(g\tau)}
\end{equation}
cut off by the log-normal tail for $\tau\to\infty$.
The absence of a limit law may be associated with the delocalisation of the model for $\varepsilon=0$, as all time scales extracted from the distribution increase with $L$ (see \cite{BouComGeoLeD90,ComTex98,TexHag10} for reviews).

\subsubsection{The case $\mean{\mass(x)}=\mu g\neq0$~: time delay as a probe for zero mode}

We now analyse the case $\mean{\mass(x)}=\mu g\neq0$ which was not considered in the literature.
Although the spectral properties of the model are invariant under the change of the sign of the mass, this is not the case for the scattering properties~: neither for the phase distribution~\cite{GraTex16} nor for the Wigner time delay distribution, as we demonstrate here. 
The distribution $P_L(\tau)$ has an interesting property~: the existence of a limit law is correlated with the choice of the boundary condition at $x=0$ and the sign of the average mass $\mean{\mass(x)}=\mu g$.
If we choose the boundary condition corresponding to $\mass(x)=\pm\infty$ for $x<0$, we obtain the limit law only for $\mp\mu>0$
\begin{equation}
  \label{eq:DWTDsusy}
  P_\infty(\tau) = \frac{g}{\Gamma(|\mu|)\,(g\tau)^{1+|\mu|}}\,\EXP{-1/(g\tau)} 
  \:.
\end{equation}
Conversely, for $\pm\mu>0$, there is no limit law as $L\to\infty$.
We can correlate the existence of the limit law with the presence of a chiral zero mode located at the boundary $x=0$, what occurs when the sign of the mass changes at the boundary~: 
\begin{center}
\begin{tabular}{l|l|l|l|}
$\mass(x)$    & $\mean{\mass(x)}$ & zero & limit\\
for $x<0$ & for $x>0$     & mode & law $P_\infty(\tau)$ \\
\hline
$-\infty$ & $>0$ & yes & yes \\
$-\infty$ & $<0$ & no  & no \\
$+\infty$ & $>0$ & no  & no \\
$+\infty$ & $<0$ & yes & yes \\
\hline
\end{tabular}
\end{center}
The existence of zero modes in the multichannel Dirac equation with random mass was recently studied in Ref.~\cite{GraTex16}, where their topological nature was underlined.

\subsection{Related questions and remarks}

\begin{enumerate}[(i)]

\item
  An alternative derivation of the stationary distribution \eqref{eq:ComtetTexier1997LimitLaw} was proposed in Ref.~\cite{RamKum00}, by using the relation between the time delay and the reflection coefficient in the presence of a constant imaginary component in the potential (wave amplification), i.e. making use of analytic properties of the scattering matrix. This point is furher discussed in the next paragraph.
   
\item 
  A discrete Anderson model with Cauchy disorder (Lloyd model~\cite{Llo69}) was studied in Ref.~\cite{DeyLisAlt00}, for which the relation between the two cumulants of $\ln|\psi_\varepsilon(x)|$ presents an extra factor $2$ (i.e. drift $\mu=1/2$), compared with the usual form for the single parameter scaling, Eq.~\eqref{eq:SPS}.~\footnote{
    Localisation for potentials with power law distribution was studied in \cite{BieTex08,DeyLisAlt01,Llo69,TitSch03}.
  }
  Some time ago, I performed some (unpublished) numerical investigations, which had shown that the distribution of the Wigner time delay is however still characterised by a power law tail $\tau^{-2}$ (and not $\tau^{-3/2}$). It would thus be interesting to clarify this observation in connection with the discussion of the section.

\item
  Correlations in energy $\mean{\Wt(\varepsilon)\Wt(\varepsilon')}$ were analysed by Titov and Fyodorov~\cite{TitFyo00}.
  
\item 
  Time delay and resonance width are closely related (see the review \cite{Kot05}).
  The resonance width distribution was studied in \cite{KunSha08,GurSha12}.

\item
  We have focused the discussion on the Wigner time delay characterizing the reflection problem on the half line.
  The time delay describing the transmitted wave through a disordered medium was analysed in \cite{BolLamFalPriEps99}.

\end{enumerate}


\subsection{Time delay, absorption/amplication and analyticity (Dirac case)}

We discuss here an interesting connection between time delay and the question of absorption/amplification.
This relation, consequence of the analytic properties of the scattering matrix, is completly general and has been used in various contexts~: 1D random Schr\"odinger Hamiltonians \cite{PraKum94,RamKum00}, 
multichannel disordered models \cite{BeeBro01} (standard class) or chiral/BdG clasees \cite{TitBroFurMud01} and also in Ref.~\cite{Fyo03} (\ref{app:ButtikerTexier} also follows from this idea).
The idea is simple~:
given the $\Nc\times\Nc$ scattering matrix (without absorption) $\Sm(\varepsilon)$, 
the case with a uniform absorption/amplification (in space) is described by the reflection matrix
can be accounted for through the subsitution
\begin{equation}
  \label{eq:SubstitutionEnergy}
  \varepsilon \longrightarrow \tilde\varepsilon=\varepsilon + {\I\,\Gamma}/{2}
  \:.
\end{equation}
Thus, the evolution is not anymore unitary and the reflection matrix
\begin{align}
  \label{eq:SdaggerSabsorption}
   \left[\Sm(\tilde\varepsilon)\right]^\dagger\Sm(\tilde\varepsilon)
   &= \mathbf{1}_\Nc + \I\,\Gamma \,\Sm^\dagger\partial_\varepsilon\Sm
   +\mathcal{O}(\Gamma^2)
   \nonumber\\
   &= \mathbf{1}_\Nc - \Gamma \,\mathcal{Q}+\mathcal{O}(\Gamma^2)
\end{align}
encodes useful information. 
In the one channel case, the weak absorption/amplification limit of the reflection coefficient  
\begin{equation}
  \label{eq:MappingRTau}
  R = |\Sm(\tilde\varepsilon)|^2 = 1 - \Gamma \, \tau_\mathrm{W} +\mathcal{O}(\Gamma^2)
\end{equation}
is related to the Wigner time delay at lowest order in $\Gamma$ [the same holds in general if one traces Eq.~\eqref{eq:SdaggerSabsorption}].
This expression shows that the absorption probability (for $\Gamma>0$) is proportional to the time spent in the scattering region, at lowest order in $\Gamma$.

We consider below the scattering problem for the one-dimensional Dirac equation with a random mass, which is more rich than the Schr\"odinger case discussed previously in Refs.~\cite{PraKum94,RamKum00}.

\subsubsection{The scattering problem}
\label{subsec:TheScattPb}

We start from the Dirac equation (\ref{eq:DiracEquation},\ref{eq:Dirac}).
We can restrict to $\varepsilon\geq0$ without loss of generality.
In order to introduce a reflection coefficient in a more general setting, we consider the scattering problem on the full line~: the scattering state incoming from the right is 
\begin{align}
  &\Psi^{(R)}_\varepsilon(x) =
  \\\nonumber
  & \begin{cases}
          t  
          \begin{pmatrix}
            1 \\ \I
          \end{pmatrix}
          \EXP{-\I\varepsilon x} 
            &\mbox{for }x<0 
          \\ 
          \alpha(x) \,  
          \begin{pmatrix}
            1 \\ \I
          \end{pmatrix}
          \EXP{-\I\varepsilon x}  
          - \beta(x)\, 
          \begin{pmatrix}
            -1 \\ \I
          \end{pmatrix}
          \EXP{\I\varepsilon x} 
          & 
          \mbox{for }0<x<L 
          \\ 
          \begin{pmatrix}
            1 \\ \I
          \end{pmatrix}
          \EXP{-\I\varepsilon (x-L)}  
          + r'\, 
          \begin{pmatrix}
            -1 \\ \I
          \end{pmatrix}
          \EXP{\I\varepsilon (x-L)} 
          & \mbox{for }x>L 
      \end{cases}
\end{align}
injecting this form into \eqref{eq:DiracEquation}, we get two differential equations for the two functions
\begin{align}
  \label{eq:EqDiffAlpha}
  \deriv{\alpha(x)}{x} &= \mass(x)\, \beta(x)\,\EXP{+2\I\varepsilon x}  
  \\
  \label{eq:EqDiffBeta}
  \deriv{\beta(x)}{x} &= \mass(x)\, \alpha(x)\,\EXP{-2\I\varepsilon x}  
\end{align}
Matching at boundaries gives $\alpha(0)=t$ and $\beta(0)=0$ while $\alpha(L)=\EXP{+\I\varepsilon L}$ and $\beta(L)=-r'\EXP{-\I\varepsilon L}$.
The reflection coefficient $r'$ is most naturally obtained by introducing the variable 
$
  \Upsilon(x) \eqdef {\beta(x)}/{\alpha(x)}
$, 
which obeys 
\begin{equation}
  \label{eq:EqDiffUpsilon}
  \deriv{\Upsilon(x)}{x} 
  = \mass(x)\left(\EXP{-2\I\varepsilon x}  
  - \Upsilon(x)^2\EXP{+2\I\varepsilon x}\right)
\end{equation}
for initial condition  
$
  \Upsilon(0) = 0 
$.
The reflection coefficient is then given by
$
  \Upsilon(L) = - r'\, \EXP{-2\I\varepsilon L}
$.
Knowing $\Upsilon(x)$, the transmission coefficient could be in principle deduced by dividing \eqref{eq:EqDiffAlpha} by $\alpha$ and integrating~:
one gets
$t=\exp\big\{\I\varepsilon L - \int_0^L\D x\,\mass(x)\,\Upsilon(x)\,\EXP{2\I\varepsilon x}\big\}$.

At this point, it is convenient to parametrize the complex variable $\Upsilon(x)=-\rho(x)\,\EXP{2\I\theta(x)-2\I\varepsilon x}$ in terms of two real variables $\rho$ and $\theta$. The reflection amplitude is $r'=+\rho(L)\,\EXP{2\I\theta(L)}$. Thus
\begin{align}
  \label{eq:DEtheta0}
  \deriv{\theta}{x} &=  \varepsilon  + \mass(x)\, \frac{\rho^2+1}{2\rho}\, \sin2\theta 
  \\
  \label{eq:DErho0}
  \deriv{\rho}{x}   &= 2\mass(x)\, (\rho^2-1)\, \cos2\theta
\end{align}
for the initial conditions~\footnote{
  The initial condition for the phase follows from the perturbative expression of the reflection coefficient~:
  from the Lippmann-Schwinger formula we obtain 
  \begin{equation}
    \psi(x) \simeq \psi^{(0)}(x) + \int\D x'\, G_0^\mathrm{R}(x,x';k^2)V(x')\psi^{(0)}(x')
  \end{equation}
  hence 
  \begin{equation}
    r' = -\frac{\I}{2k}\int_0^L\D x'\,V(x')\,\EXP{-2\I kx'} + \mathcal{O}(V^3)
  \end{equation}
  thus $r'/|r'|\to-\I$ in the limit $L\to0$.
}    
$\theta(0) = \frac{\pi}{4}$ and $\rho(0)=0$.

If instead we impose the boundary conditions $\theta(0) = 0$ and $\rho(0)=1$, corresponding to confinement on the positive axis (i.e. scattering on the half line), we see that the amplitude remains locked at $\rho(x)=1\ \forall\,x$  and we recover Eq.~\eqref{eq:DifferentialEquationForThePhaseTheta} for the phase.

\subsubsection{Absorption/amplification}

We now add the effect of absorption or amplification by introducing a complex term in the Hamiltonian \eqref{eq:Dirac}
\begin{equation}
  \mathcal{H} = \I\sigma_2\, \partial_x +\sigma_1\, \mass(x) + A_0
  \hspace{0.25cm}\mbox{with }
  A_0 = -\I\Gamma/2
\end{equation}
$\Gamma>0$ for absorption and $\Gamma<0$ for ampli\-fi\-ca\-tion.
This is equivalent to perform the substitution \eqref{eq:SubstitutionEnergy} in the Dirac equation \eqref{eq:DiracEquation}.
Performing the same steps as in \S~\ref{subsec:TheScattPb}, we obtain that Eqs.~(\ref{eq:DEtheta0},\ref{eq:DErho0}) are replaced by 
\begin{align}
  \label{eq:DEtheta}
  \deriv{\theta}{x} &= \re(\tilde\varepsilon) + \mass(x)\, \frac{\rho^2+1}{2\rho}\, \sin2\theta 
  \\
  \label{eq:DErho}
  \deriv{\rho}{x} &=  -2\im(\tilde\varepsilon)\,\rho + \mass(x)\, (\rho^2-1)\, \cos2\theta
\end{align}

\subsubsection{Dirac point ($\varepsilon=0$)}

We first consider the Dirac point ($\varepsilon=\re\tilde\varepsilon=0$), where chiral symmetry holds.
As we have seen in \S~\ref{subsec:DWCS}, the phase is locked at $\theta=\pi/2$ (this ensures the existence of a zero mode and a limit law for the time delay when $\mu>0$). 
Then 
\begin{equation}
  \deriv{\rho}{x} =  -\Gamma\,\rho - \mass(x)\, (\rho^2-1)
\end{equation}
We consider $R=\rho^2$, which obeys 
\begin{equation}
  \label{eq:EqDiffR}
  \deriv{R}{x} = -2\Gamma\,R + 2\,\mass(x)\,\sqrt{R}\,(1-R)
\end{equation}
(at $x=L$, $R(L)$ is the reflection probability).
We see on this equation that the diffusion term in Eq.~\eqref{eq:EqDiffR} vanishes at $R=1$. 
Only the term $-2\Gamma\,R$ of the drift allows the process to cross the point $R=1$.
For absorption, the drift is negative, therefore the distribution has support $[0,1]$.
For amplification it has support $[1,\infty[$.
Assuming that $\mass(x)$ is a Gaussian white noise, Eq.~\eqref{eq:RandomMass}, the (forward) generator describing the diffusion \eqref{eq:EqDiffR} is
\begin{align}
  \mathscr{G}_R^\dagger 
  &= 2 \derivp{}{R} \left( \Gamma \,R - \mu\,g\,\sqrt{R}\,(1-R) \right)
  \nonumber\\
  &+ 2g \derivp{}{R} \sqrt{R}\,(1-R)\derivp{}{R} \sqrt{R}\,(1-R)
\end{align}
i.e. the distribution of the process obeys the Fokker-Planck equation 
$\partial_xW_x(R)=\mathscr{G}_R^\dagger W_x(R)$.
For $\Gamma>0$ (absorption), we find the stationary distribution solution of $\mathscr{G}_R^\dagger W_\infty(R)=0$~:
\begin{align}
  W_\infty(R) = \frac{C}{\sqrt{R}\,(1-\sqrt{R})^{1+\mu}\,(1+\sqrt{R})^{1-\mu}}
  \EXP{-\frac{\Gamma/g}{1-R}} 
\end{align}
for $R\in[0,1]$, where $C$ is a normalisation constant.
For $\Gamma<0$ (amplification), we get
\begin{align}
  W_\infty(R) = \frac{C}{\sqrt{R}\,(\sqrt{R}-1)^{1+\mu}\,(\sqrt{R}+1)^{1-\mu}}
  \EXP{-\frac{|\Gamma|/g}{R-1}} 
\end{align}
with support $[1,\infty[$.

Using \eqref{eq:MappingRTau} for $\Gamma\to0$ allows to map the stationary distribution $W_\infty(R)$ onto \eqref{eq:DWTDsusy}. Moreover, the correspondence can be established at the level of the two generators
\begin{equation}
   \mathscr{G}_R^\dagger 
   \underset{\Gamma\to0}{\longrightarrow}
   \mathscr{G}_\tau^\dagger = \derivp{}{\tau}(2\mu\,g\,\tau-2) + 2g \derivp{}{\tau}\tau \derivp{}{\tau}\tau
\end{equation}
where $\mathscr{G}_\tau^\dagger$ describes the diffusion \eqref{eq:SDEforZ} for $\theta(x)=\pi/2$.

\subsubsection{Universal (high energy) regime}

\paragraph{Unitary evolution ---}
Let us first briefly discuss the case without absorption/amplification ($\Gamma=0$). 
In the universal (high energy) regime $\varepsilon\gg g$, we can average over the phase (fast variable) Eqs.~(\ref{eq:DEtheta},\ref{eq:DErho}), assuming a uniform distribution. 
The distribution of $\rho(x)=\EXP{-w(x)}$ (reflection coefficient) is described by the diffusion equation controlling the radial diffusion on a surface of constant negative curvature~\cite{GerVas59,AntPasSly81} (see also \cite{Tex99})~:
\begin{equation}
  \partial_x Q(w;x) \simeq 
  2\gamma\,\partial_w^2
  \left[\sinh^2w\,Q(w;x)\right]
\end{equation}
where $\gamma\simeq g/2$ is the high energy Lyapunov exponent.
The fact that one gets the same result for the Dirac (here) and Schr\"odinger \cite{AntPasSly81} equations has its origin in the universality of the high energy regime.
The distribution has no limit law for $x\to\infty$ as the reflection coefficient decays exponentially with the system size $L$, hence the process $w(x)$ grows linearly.

\paragraph{Finite absorption/amplification rate ---}
For $\Gamma\neq0$, from Eq.~\eqref{eq:DErho}, we see that the (forward) generator receives another contribution
$\mathscr{G}_w^\dagger=g\partial_w^2\sinh^2w\longrightarrow\mathscr{G}_w^\dagger=-\Gamma\partial_w+g\partial_w^2\sinh^2w$.
At his point it is more convenient to deal with the reflection probability and introduce the variable $R=\rho^2=\EXP{-2w}$. 
The new generator is given by $\mathscr{G}_R^\dagger=(\partial w/\partial R)\mathscr{G}_w^\dagger(\partial R/\partial w)$. We get finally
\begin{align}
  \mathscr{G}_R^\dagger = &2\derivp{}{R}\left( \Gamma\,R - \frac{g}{4}(1-R^2) \right)
  \nonumber\\
  &+g \derivp{}{R}\sqrt{R}(1-R)\derivp{}{R}\sqrt{R}(1-R)
  \:.
\end{align}
For finite absorption/amplification rate $\Gamma$, the reflection coefficient has a limit law for $L\to\infty$.
For $\Gamma>0$ (absorption), it is easy to show that the stationary distribution is
\begin{equation}
  \label{eq:RamakrishnaKumar2000}
  W_\infty(R)
  =\frac{\Gamma\tau_\xi}{(1-R)^2} \EXP{-\frac{\Gamma\tau_\xi}{1-R}}
\end{equation}
with support $R\in[0,1]$.
We have introduced the parameter $\tau_\xi=\xi/v=2/g$ (with $v=1$ for the Dirac equation) in order to stress the universal character of this result.
This is in exact correspondence with the distribution found in \cite{RamKum00} for the disordered Schr\"odinger equation, as it should since we are dealing here with the universal regime.
As pointed out by Ramakrishna and Kumar, in the limit $\Gamma\to0$, Eq.~\eqref{eq:RamakrishnaKumar2000} can be mapped onto the limit law of the time delay distribution \eqref{eq:ComtetTexier1997LimitLaw} by using \eqref{eq:MappingRTau}.
Here, we stress that, as at the Dirac point, this mapping can be established at the level of the generators and is thus more general than the one stated in \cite{RamKum00}~:
using \eqref{eq:MappingRTau}, it is straighforward to check that 
\begin{equation}
   \mathscr{G}_R^\dagger 
   \underset{\Gamma\to0}{\longrightarrow}
   \mathscr{G}_\tau^\dagger = \derivp{}{\tau}(g\,\tau-2) + g \derivp{}{\tau}\tau \derivp{}{\tau}\tau
\end{equation}
which is indeed the generator controlling the distribution of the Wigner time delay in the high energy limit  (given in \cite{ComTex97,Tex99}).
This is the generator of the process \eqref{eq:ExponFunct} for $\mu=1$, according to \S~\ref{subsec:DOHL}.

%
%
%


\subsection{Beyond 1D}

\subsubsection{Multichannel disordered wires}
  The extension to the quasi-1D situation (a wave guide with $N$ channels) has been considered by several authors (e.g. see \cite{MucJalPic97,TigSebStoGen99}).
  Using the relation between the time delays and the reflection probabilities in an absorbing or amplifying medium (i.e. the analyticity of the $\Sm$-matrix), Beenakker and Brouwer obtained the distribution of the proper time delays describing the scattering on a semi-infinite disordered wave guide~\cite{Bee01,BeeBro01}~:
\begin{equation}
   \label{eq:BeenakkerBrouwer2001}
   \mathcal{P}(\gamma_1,\cdots,\gamma_N)\propto\prod_{i<j}|\gamma_i-\gamma_j|^\beta
  \prod_k\EXP{-\beta\gamma_k/2}
  \:,
\end{equation}
where $\gamma_k=\tau_s/\tau_k$ is the inverse of the proper time and $\tau_s$ a characteristic time related to the disorder strength.
The result relies on the isotropy assumption (among the channels), i.e. describes the weakly disordered (quasi-1D diffusive) regime~; in this case the localisation length  increases with the number of channels as~\cite{Bee97} 
$\xi^\mathrm{(quasi\:1D)}=\xi^\mathrm{(1D)}\,\big[1+\beta(N-1)/2\big]$.
The distribution \eqref{eq:BeenakkerBrouwer2001} is a particular instance of the Laguerre ensemble of random matrix theory (RMT).
For one channel $N=1$, one recovers the limit law~\eqref{eq:ComtetTexier1997LimitLaw} by setting $\beta\tau_s/2=\tau_\xi$.
In the limit of large $N$, the marginal distribution of the rates $\gamma_a$'s is the Mar\v{c}enko-Pastur law $\rho_\mathrm{MP}(x)=(2\pi)^{-1}\sqrt{(4-x)/x}$, after rescaling $x=\gamma/N$. Accordingly, the marginal distribution of the proper times is 
\begin{equation}
  \label{eq:MarginalProperTimesDisorderedWires}
  w_\Nc^{(\beta)}(\tau) \underset{N\to\infty}{\simeq} N\,\rho(N\tau)
  \mbox{ with }
  \rho(y) = \frac{\sqrt{4y-1}}{2\pi\,y^2}
  \:. 
\end{equation}

Starting from \eqref{eq:BeenakkerBrouwer2001}, the distribution of the Wigner time delay 
$\Wt=(1/N)\sum_i\gamma_i^{-1}$ was obtained in Ref.~\cite{GraTex16b,Gra18}~:
\begin{align}
  \label{eq:GrabschTexier2016}
   &\mathscr{P}^{(\beta)}_\Nc(\tau)  
   \underset{\Nc\to\infty}{\simeq }
   \frac{C_\beta}{\tau^2}
   \\\nonumber
   &\times
   \exp\left\{
     -\frac{27\beta}{64\,\tau^2} +\left(1-\frac{\beta}{2}\right)\frac{9(2-\sqrt{3})}{4\,\tau}
   \right\}
\end{align}
where $\beta$ is the Dyson index for orthogonal ($\beta=1$) or unitary ($\beta=2$) symmetry classes. $C_\beta$ is a normalisation constant.
The distribution presents the same power law tail as in the strictly one-dimensional case $\mathscr{P}^{(\beta)}_\Nc(\tau)\sim1/\tau^2$. 
This can be understood from the fact that the physics at large time (i.e. large scale) is expected to be dominated by a single channel (the less localised one).
  
  Denoting by $\DoS_L(\varepsilon_F)$ the DoS of the multichannel disordered wire of length $L$,  \eqref{eq:GrabschTexier2016} can be interpreted as the limit law for $\Wt\simeq2\pi\,\DoS_L(\varepsilon_F)/\Nc$ in the $L\to\infty$ limit.
  The fact that all moments of $\Wt$ are infinite in this case is thus simply related to the divergence of the DoS when $L\to\infty$, as in the strictly 1D situation.

  Finally, we quote \cite{Oss18}, where the marginal distribution of the proper times was derived for the multichannel case for finite~$L$ (it converges towards \eqref{eq:MarginalProperTimesDisorderedWires} for $L\to\infty$).

\subsubsection{Higher dimensions}

\begin{enumerate}[(i)]

\item
  Some measurements of time delays were performed with electromagnetic waves in Ref.~\cite{SebLegGen99,GenSebStoTig99,ChaGen01}.

\item    
  The case of higher dimensions was considered by Ossipov and Fyodorov \cite{OssFyo05} who analysed a two dimensional situation in the weakly disordered regime. 
  The localised regime was studied by numerical simulations in Ref.~\cite{XuWan11}.

\item 
  A study of time delay at a critical point like the metal-insulator transition was performed in Ref.~\cite{KotWei02,Fyo03,OssFyo05} (see the review \cite{Kot05}).

\end{enumerate}


\section{Random matrix approach for quantum dots}
\label{sec:RandomMatrices}

In systems with ergodic properties, like a chaotic cavity with narrow contacts (Figs.~\ref{fig:qd} and \ref{fig:QRCcircuit}), or a weakly disordered cavity in the ergodic regime, it is natural to make a maximum entropy assumption \cite{MelPerSel85,FrieMel85,Bee97,MelBar99,MelKum04} leading to postulate that the scattering matrix is uniformly distributed over the unitary group (for perfect couplings at the contacts). 
Although this approach has led to many successes in the description of several properties of coherent conductors (conductance, shot noise,...) \cite{Bee97,MelKum04}, it does not provide any information about the \textit{energy dependence} of the scattering matrix, which is probed by the Wigner-Smith matrix~\eqref{eq:DefinitionWS}.
Two approaches were proposed to develop a random matrix description for the energy dependence of the scattering matrix~:
\begin{enumerate}[(i)]

\item
the ``Hamiltonian Approach'' (HA) of chaotic scattering pushed forward by Fyodorov, Savin, Sommers and coworkers, 

\item
the ``Alternative Stochastic Approach'' (ASA), pioneered by the work of Brouwer and B\"uttiker \cite{BroBut97} and mostly developed by Brouwer, Beenakker and coworkers.

\end{enumerate}

One of the first result within RMT was the calculation of the two point correlation function of the Wigner time delay~\cite{LehSavSokSom95} given below, Eq.~\eqref{eq:LehmanSavinSokolovSommers1995}.
As we have seen in the previous section, one can introduce several time scales, characterised by their distributions~: 
\begin{itemize}
\item  
  The marginal law for partial time delays 
  \begin{equation}
  \widetilde{w}_\Nc^{(\beta)}(\tau) \eqdef (1/\Nc)\sum_a\mean{\delta(\tau-\tilde{\tau}_a)}
  \:,
  \end{equation}
  which was the first to be considered by Fyodorov and Sommers \cite{FyoSom96,FyoSom97} in the unitary class ($\beta=2$), Eq.~\eqref{eq:MarginalLawTau}.~\footnote{
  Note that a similar route was followed in Ref.~\cite{SebZycZak96}, however the result of the reference is partly incorrect as explained in \cite{FyoSom97}.
 }
\item
  We also review the known results for the marginal law for the proper times 
  \begin{equation}
  w_\Nc^{(\beta)}(\tau) \eqdef (1/\Nc)\sum_i\mean{\delta(\tau-\tau_i)}
  \:.
  \end{equation}
\item
  The distribution of the Wigner time delay 
\begin{equation}
\mathscr{P}_\Nc^{(\beta)}(\tau) \eqdef \smean{\delta\big(\tau-(1/\Nc)\sum_i\tau_i\big)}
 \end{equation}
is also of great interest.   
\end{itemize}
Obviously, these distributions do not encode the full statistical information, which also require to characterise correlations (between partial/proper times, or energy correlations).

The known results corresponding to the standard (Wigner-Dyson) symmetry classes are reviewed in the next three sections. The final section reports few results obtained in other symmetry classes.

\subsection{Distribution and moments for perfect contacts}
\label{subsec:TDPC}

In this subsection, I review the results for ideal contacts, i.e. when the transmission probability between a channel of the wave guide and the QD is equal to unity.

\subsubsection{Proper time delays $\{\tau_i\}$}

Gopar, Mello and B\"uttiker obtained the Wigner time delay distribution in the one channel case for the three symmetry classes~\cite{GopMelBut96}~:
  \begin{equation}
    \label{eq:GoparMelloButtiker1996}
    \mathscr{P}_1^{(\beta)}(\tau) 
    = \frac{(\beta/2)^{\beta/2}}{\Gamma(\beta/2)}\tau^{-2-{\beta}/{2}}\,
    \EXP{-{\beta}/{(2\tau)}}
  \end{equation}
(obviously $\mathscr{P}_1^{(\beta)}(\tau) = w_1^{(\beta)}(\tau) = \widetilde{w}_1^{(\beta)}(\tau)$).
The proper times are measured in unit of the Heisenberg time $\Ht=2\pi\hbar/\Delta$, where $\Delta$ is the mean level spacing of the cavity (of more correctly the ``mean resonance spacing'' as we deal with a scattering problem).

An important advance was the work of Beenakker, Brouwer and Frahm who were able to obtain the joint distribution of the proper time delays within ASA in Ref.~\cite{BroFraBee97,BroFraBee99}.
They established the relationship with the Laguerre ensemble of random matrices~: the inverse of the symmetrised Wigner-Smith matrix~\footnote{
  The definition of $\WSm_s$ is intermediate between \eqref{eq:DefinitionWS} and \eqref{eq:DefinitionWStilde}. The three matrices have the same spectrum of eigenvalues $\{\tau_a\}$.
} 
\begin{equation}
  \WSm_s \eqdef -\I\Sm^{-1/2}\,\derivp{\Sm}{\varepsilon}\,\Sm^{-1/2}
\end{equation}
is a Wishart matrix~; i.e. $\WSm_s^{-1}=X^\dagger X$ where $X$ is a $\Nc\times(2\Nc-1+2/\beta)$ matrix with identical and independent Gaussian elements (this interpretation only holds for $\beta=1,\,2$). The joint distribution of the inverse of the proper times, $\gamma_i=\Ht/\tau_i$ being given by~\cite{BroFraBee97,BroFraBee99}
\begin{equation}
    \label{eq:BrouwerFrahmBeenakker1997}
    \mathcal{P}(\gamma_1,\cdots,\gamma_{\Nc})
    \propto
    \prod_{i<j}|\gamma_i-\gamma_j|^\beta
    \prod_k \gamma_k^{\beta N/2}\EXP{-\beta\gamma_k/2}
    \:.
\end{equation}
For $N=1$, the Gamma-law $\mathcal{P}(\gamma_1)\propto\gamma_1^{\beta/2}\EXP{-\beta\gamma_1/2}$ corresponds to  Gopar, Mello and B\"uttiker's result, Eq.~\eqref{eq:GoparMelloButtiker1996}.

In the large $\Nc$ limit, the marginal law for proper time delays takes the form
\begin{align}
   w_\Nc^{(\beta)}(\tau)
   \underset{N\to\infty}{\simeq}
   \Nc\,\rho(\Nc\,\tau)
\end{align}
where  \cite{BroFraBee99,TexMaj13}~\footnote{
 The distribution is related to the Mar\v{c}enko-Pastur law by $\rho(y)=(1/y^2)\,\rho_\mathrm{MP}(1/y)$, 
 where $$\rho_\mathrm{MP}(x)=\big[1/(2\pi x)\big]\sqrt{(x_+-x)(x-x_-)}$$
 is the distribution of the rescaled eigenvalues $\gamma=\Nc\,x$ for $\Nc\times\Nc$ matrices in the Laguerre ensemble,~Eq.~\eqref{eq:BrouwerFrahmBeenakker1997}.
}
\begin{equation}
  \rho(y) = \frac{1}{2\pi y^2}\sqrt{(x_+-y)(y-x_-)}
\end{equation}
with $x_\pm=\big(\sqrt{2}\pm1\big)^2$. 
The marginal law 
is not strictly zero out of the interval $[x_-/N,x_+/N]$ but presents finite $N$ corrections with asymptotics~\footnote{
  In the general case, the Laguerre ensemble describes $N\times N$ random Hermitian matrices $\Gamma$ with positive eigenvalues distributed according to
  $P(\Gamma)\propto(\det\Gamma)^{\alpha\beta N/2}\exp\big[-(\beta/2)\tr{\Gamma}\big]$ (Eq.~\eqref{eq:BrouwerFrahmBeenakker1997} corresponds to $\alpha=1$ and \eqref{eq:BeenakkerBrouwer2001} to $\alpha=0$).
  The Mar\v{c}enko-Pastur law $p_\Nc(\gamma)\simeq(1/\Nc)\,\rho_\mathrm{MP}(\gamma/\Nc)$ for the density of eigenvalues has support $[\Nc\,x_-,\Nc\,x_+]$ where $x_\pm=\big(\sqrt{1+\alpha}\pm1\big)^2$. 
  Finite $\Nc$ corrections to the Mar\v{c}enko-Pastur law for $\gamma\in[0,\Nc\,x_-]\cup[\Nc\,x_+,\infty[$ were obtained by Forrester in Ref.~\cite{For12}. Asymptotic expansions of Forrester's result are
  $p_\Nc(\gamma)\sim\gamma^{\alpha\beta N/2}$ for $\gamma\to0$
  and   
  $p_\Nc(\gamma)\sim\gamma^{(\alpha+2)\beta N/2}\EXP{-(\beta/2)\gamma}$ for $\gamma\to\infty$.
  }
\begin{align}
  w_\Nc^{(\beta)}(\tau)
  & \underset{\tau\to0}{\sim}
  \tau^{-2-3\beta N/2}\,\EXP{-\beta/(2\tau)}
  \\
  \label{eq:TailMarginalProperTimes}
  & \underset{\tau\to\infty}{\sim}
   \tau^{-2-\beta N/2}
   \:.
\end{align}

\begin{figure}[!ht]
\centering
\includegraphics[scale=0.5]{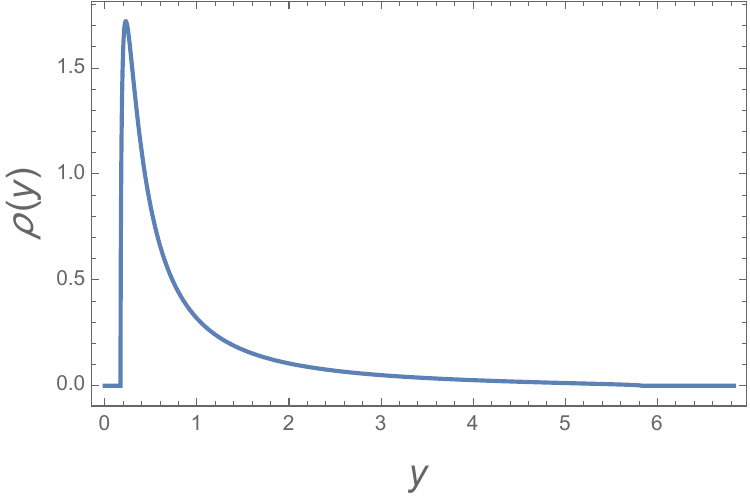}
\caption{\it Marginal law for the proper time delay.}
\end{figure}

We can deduce the moments~:
\begin{equation}
  \mean{\tau_i}=\frac{1}{\Nc}
  \hspace{0.5cm}\mbox{and}\hspace{0.5cm}
  \mathrm{Var}(\tau_i)\simeq\frac{1}{\Nc^2}
  \hspace{0.5cm}
  \forall\,\beta
  \:.
\end{equation}
The approach of Ref.~\cite{KuiSavSie14} provides the precise expression of the variance~\cite{Savin2016}~:
\begin{equation}
  \label{eq:VarianceProperTimes}
  \mathrm{Var}(\tau_i) = \frac{\Nc\big[\beta(\Nc-1)+2\big]+2}{\Nc^2(\Nc+1)(\beta\Nc-2)}
  \:.
\end{equation}
Because the moments of the sum of partial times are known [see below, Eq.~\eqref{eq:MezzadriSimm2}], using the symmetry between channels, it is possible to deduce the covariance.
Writing $\mathrm{Var}(\Nc\,\Wt)=\Nc\,\mathrm{Var}(\tau_i)+\Nc(\Nc-1)\,\mathrm{Cov}(\tau_i,\tau_j)$ and using \eqref{eq:VarianceProperTimes} provides the covariance 
\begin{equation}
\mathrm{Cov}(\tau_i,\tau_j) = -\frac{1}{\Nc^2(\Nc+1)}
\:,
\end{equation}
showing that the proper times are surprisingly independent of the symmetry index, and \textit{anti-correlated}. 
The anti-correlations are not a suprise as the proper times are the eigenvalues of an invariant random matrix, see Eq.~\eqref{eq:BrouwerFrahmBeenakker1997}.
It is interesting to rewrite the result as 
\begin{equation}
   \frac{ \mathrm{Cov}(\tau_i,\tau_j) }{\sqrt{\mathrm{Var}(\tau_i)\,\mathrm{Var}(\tau_j)}}  
   = - \frac{\beta\Nc-2}{\Nc\big[\beta(\Nc-1)+2\big]+2}
    \simeq -\frac{1}{N}
    \:,
\end{equation}
which show that they present the maximal anti-correlation (for identical variables).

Remark~:
\begin{itemize}
\item
  The joint moments of the proper times were studied in \cite{MarMarGar14}.  
\end{itemize}

\subsubsection{Partial time delays $\{\tilde\tau_a\}$}

The marginal law for partial times was obtained for $\beta=2$ in \cite{FyoSom96,FyoSom97} and for $\beta=1$ in \cite{FyoSavSom97}. 
The distribution $\forall\:\beta$ is given in Ref.~\cite{SavFyoSom01}~:
\begin{equation}
  \label{eq:MarginalLawTau}
  \widetilde{w}_N^{(\beta)}(\tau)
  = \frac{(\beta/2)^{\beta N/2}}{N\,\Gamma(\beta N/2)} \,\tau^{-2-\beta N/2}\,\EXP{-\beta/(2\tau)}
  \:.
\end{equation}
The distribution matches with \eqref{eq:GoparMelloButtiker1996} for $N=1$, as it should.
We deduce the moments 
$\mean{\tilde{\tau}_a^n}=(\beta/2)^n\Gamma(1-n+\beta\Nc/2)/\Gamma(1+\beta\Nc/2)$, i.e. 
\begin{equation}
  \label{eq:VariancePartialTimes}
  \mean{\tilde{\tau}_a}=\frac{1}{\Nc}
  \hspace{0.5cm}\mbox{and}\hspace{0.5cm}
  \mathrm{Var}(\tilde{\tau}_a)=\frac{2}{\Nc^2(\beta\Nc-2)}
\end{equation}
Thus the fluctuations of the partial time delays are reduced compared to those of the proper time delays~:
$\mathrm{Var}(\tilde{\tau}_a)\simeq2\,\mathrm{Var}(\tau_i)/(\beta\Nc)$.

Using the same remark as for proper times, we can also deduce the covariances of thr partial times~:
\begin{equation}
\mathrm{Cov}(\tilde\tau_a,\tilde\tau_b)
=
\frac{2}{N^2(N+1)(N\beta-2)}    
\:.
\end{equation}
Hence, contrary to the proper times which presents anti-correlations, the partial times present positive correlations.
We can rewrite the result as 
\begin{equation}
    \frac{ \mathrm{Cov}(\tilde\tau_a,\tilde\tau_b) }
         {\sqrt{\mathrm{Var}(\tilde\tau_a)\,\mathrm{Var}(\tilde\tau_b)}}  
    =\frac{1}{N+1} 
    \simeq +\frac{1}{N}
    \:.
\end{equation}
Note that the joint probability distribution for two partial time delays was obtained in Ref.~\cite{SavFyoSom01} (Eq.~23).

\subsubsection{Wigner time delay $\Wt$}

Despite the joint distribution for the proper times was explicitly known~\cite{BroFraBee97}, the statistical properties of their sum, the Wigner time delay $\Wt=(1/\Nc)\sum_i\tau_i=(1/\Nc)\sum_a\tilde\tau_a$, remained unknown for a while.
The distribution for $\Nc=2$ channels was deduced in Ref.~\cite{SavFyoSom01}~: 
\begin{align}
    \label{eq:Savin2001}
    &\mathscr{P}_2^{(\beta)}(\tau) =
    \frac{ \beta^{3\beta+2} \Gamma(3(\beta+1)/2) }{ \Gamma(\beta+1)\Gamma(3\beta+2) }
    \\\nonumber
    &\hspace{0.5cm}\times
    \tau^{-3(\beta+1)}\,U\left( \frac{\beta+1}2 , 2(\beta+1) ; \beta/\tau \right) \,
    \EXP{-\beta/\tau}
    \:,
\end{align}
where $U(a,b;z)$ is a Kummer function (confluent hypergeometric function)~\cite{AbrSte64}.
A method providing a systematic determination of the cumulants was proposed by Mezzadri and Simm in Ref.~\cite{MezSim13}. The authors gave explicitly the first four cumulants
\begin{align}
  \label{eq:MeanWTD}
  \mean{\Wt} &= \frac{\tau_\mathrm{H}}{N}
  \\
  \label{eq:MezzadriSimm2}
  \mean{\Wt^2}_c &= \frac{4\,\tau_\mathrm{H}^2}{N^2(N+1)(N\beta-2)}
  \\
  \label{eq:MezzadriSimm3}
  \mean{\Wt^3}_c &= \frac{96\,\tau_\mathrm{H}^3}{N^3(N+2)(N+1)(N\beta-2)(N\beta-4)}
    \\
  \label{eq:MezzadriSimm4}
  \mean{\Wt^4}_c &= 
  \left\{
  \begin{array}{l}
     \frac{96(53N^2-68N-156)\tau_\mathrm{H}^4}{N^4(N+3)(N+2)(N+1)^2(N-2)^2(N-4)(N-6)} 
     \\[0.1cm]
     \hspace{4.5cm} \mbox{ for } \beta =1
     \\[0.15cm]
     \frac{12(53N^2-77)\tau_\mathrm{H}^4}{N^4(N+3)(N+2)(N+1)^2(N-1)^2(N-2)(N-3)}
     \\[0.1cm]
     \hspace{4.5cm}  \mbox{ for } \beta =2
     \\[0.15cm]
     \frac{12(53N^2+34N-39)\tau_\mathrm{H}^4}{N^4(N+3)(N+2)(N+1)^2(2N-1)^2(N-1)(2N-3)}
     \\[0.1cm]
     \hspace{4.5cm}  \mbox{ for } \beta =4
  \end{array}
  \right.
\end{align}
It is worth emphasizing the different scalings of the variances with $\Nc$~:
\begin{align}
  \mathrm{Var}(\Nc\,\Wt)
  &=\mathrm{Var}\big(\sum_i\tau_i\big)
  \sim\mathrm{Var}(\tau_i)
  \\
  &=\mathrm{Var}\big(\sum_a\tilde\tau_a\big)
  \simeq 2\Nc\,\mathrm{Var}(\tilde\tau_a)
  \:.
\end{align}

\begin{figure}[!ht]
\centering
\includegraphics[width=0.4\textwidth]{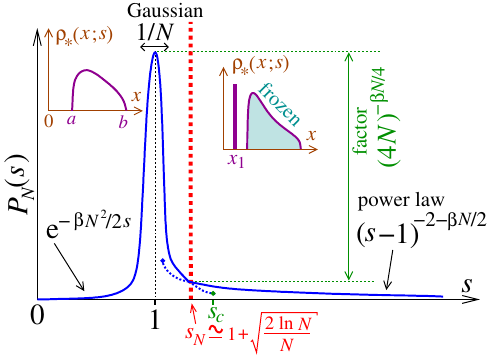}
\caption{\it Wigner time delay distribution  in the limit $N\gg1$~:
sketch of the rescaled distribution 
$P_N(s)=\tau_d\,\mathscr{P}_N^{(\beta)}(\tau=s\,\tau_d)$, where $\tau_d=h/(N\Delta)$ is the dwell time.
  The small curves are sketchs of the optimal distributions of eigenvalues of $\WSm^{-1}$ with the constraint that $\tr{\WSm}$ is fixed.
  The transition between the sharp Gaussian peak and the power law tail is associated with a phase transition in the density of eigenvalues. Figure from \cite{TexMaj13}.}
  \label{fig:TexierMajumdar2014Erratum}
\end{figure}

A systematic expression of the moments $\smean{\Wt^n}$ of the Wigner time delay for $\beta=2$ was given more recently by Novaes~\cite{Nov15a}.
We see that the variance diverges for $N\beta\leq2$, and the third cumulant for $\Nc\beta\leq4$. 
Writing the denominator of the fourth cumulant as 
$N^4(N+3)(N+2)(N+1)^2(N-2/\beta)^2(N-4/\beta)(N-6/\beta)$, we see that $\smean{\Wt^4}_c$ diverges for $\Nc\beta\leq6$. 
These observations suggest the power law tail $\mathscr{P}_\Nc^{(\beta)}(\tau)\sim\tau^{-2-\beta\Nc/2}$, conjectured in Ref.~\cite{FyoSom97} for $\beta=2$ on the basis of some heuristic argument involving resonances [i.e. identifying the tail of $\mathscr{P}_\Nc^{(\beta)}(\tau)$ with the one of $\widetilde{w}_\Nc^{(\beta)}(\tau)$, Eq.~\eqref{eq:MarginalLawTau}].
Using a Egdeworth expansion, Mezzadri and Simm concluded that the distribution weakly converges towards a Gaussian form as $\Nc$ grows.
The full distribution of the Wigner time delay for $\Nc\gg1$ was however shown to present a richer structure in Ref.~\cite{TexMaj13} where the large deviations were studied in detail, leading to the behaviours
\begin{align}
  \label{eq:DistributionWTDTexierMajumdar2013}
  \mathscr{P}_\Nc^{(\beta)}(\tau) 
  \underset{\Nc\to\infty}{\sim}
  \begin{cases}
  \tau^{-3\beta \Nc^2/4}\,\EXP{-\beta \Nc/(2\tau)} 
  \quad\mbox{for }\tau\to0
  \\[0.125cm]
  \exp\left\{-\frac{\beta\Nc^2}{8}(\Nc\tau-1)^2\right\}
  \\
  \hspace{2.75cm}\mbox{for }\tau\sim1/\Nc
  \\[0.125cm]
  (N\tau-1)^{-2-\beta\Nc/2}
  \\
  \hspace{1cm}
  \mbox{for }\Nc\tau-1\gg\sqrt{\frac2\Nc\ln\Nc}
  \end{cases}
\end{align}
The distribution is sketched in Fig.~\ref{fig:TexierMajumdar2014Erratum}.
In particular, the transition between the sharp Gaussian peak and the power law tail was shown to be related to a phase transition in the underlying Coulomb gas~\cite{TexMaj13}.

\begin{figure}[!ht]
\centering
\includegraphics[width=0.45\textwidth]{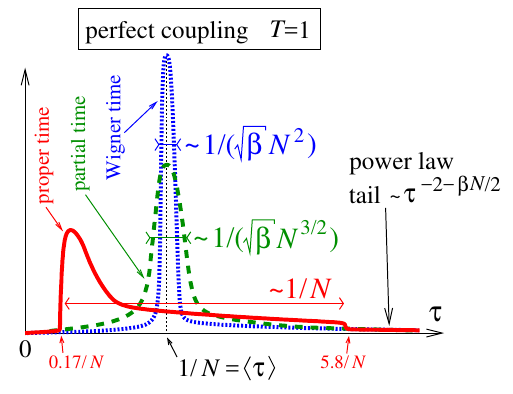}
\caption{\it Sketch of the distributions for perfect contacts. 
  Figure from Ref.~\cite{GraSavTex18}.}
\label{fig:SketchDistribPerfect}
\end{figure}

In Fig.~\ref{fig:SketchDistribPerfect}, we compare the different distributions, for proper, partial and Wigner time delays.

Few remarks~:
\begin{enumerate}[(i)]
\item
  A systematic analysis of the ``moments'' $\smean{\tr{\WSm^n}}$ within RMT was carried out by several authors~: \cite{MezSim11,MezSim12,Cun15,Nov15a,CunMezSimViv16b}.

\item
  The determination of the variance \eqref{eq:VarianceRq} and the covariance \eqref{eq:CovarianceCqRq} in Ref.~\cite{GraTex15,Gra18} (and also \cite{Viv14privcom}) has involved the correlation between the two linear statistics $\tr{\WSm}$ and $\tr{\WSm^2}$.
  A systematic analysis of the covariance $\mathrm{Cov}\big(\tr{\WSm^n},\tr{\WSm^m}\big)$ was carried out by Cunden recently \cite{Cun15}, based on the recent work~\cite{CunViv14} where a general formula for the covariance of two linear statistics was derived with the Coulomb gas method.
  More recently, the authors were also able to further analyse the correlations
  $\smean{\tr{\WSm^{n_1}}\tr{\WSm^{n_2}}\cdots}_c$ in Ref.~\cite{CunMezSimViv16}.

\item
  The distribution of the truncated sum $\sum_{a=1}^K\tau_a$ of proper times, with $K<\Nc$, was studied in Ref.~\cite{GraMajTex17b}. 
  
  The case where the sum is further restricted to the largest (or smallest) times can be analysed along the lines of Ref.~\cite{GraMajTex17,Gra18} (such an analysis allows to quantify the contribution of the fraction of eigenchannels contributing the most to the DoS).
  
\item
  The distribution of the trace of a diagonal subblock of the matrix $\WSm$ was studied in an appendix of Ref.~\cite{GraMajTex17b} in the unitary case, based on the result of \cite{SavFyoSom01}.
  The trace of the diagonal subblock $\sum_{a=1}^K\WSm_{aa}$ is more physical than $\sum_{a=1}^K\tau_a$, as it corresponds with the ``injectance'' discussed in Section~\ref{sec:InjectanceEtc}. 
  The different scaling of the two quantities with $\Nc$ was emphasized~:
  although they are characterized by the same average 
  $
  \big\langle\sum_{a=1}^K\WSm_{aa}\big\rangle
  =\big\langle\sum_{a=1}^K\tau_a\big\rangle
  =\kappa$,
  where $\kappa=K/\Nc$, the variances scale differently with the channel number~\cite{GraMajTex17b} (here for $\beta=2$)~:
  \begin{equation}
    \mathrm{Var}\left(\sum_{a=1}^K\WSm_{aa}\right)
    = \frac{\kappa\,(1+\kappa)}{\Nc^2-1}
  \end{equation}
  \begin{equation}
    \mathrm{Var}\left(\sum_{a=1}^K\tau_a\right)
    = \frac{\kappa\,(1-\kappa)}{\Nc+1} + \frac{2\kappa}{\Nc^2-1}
    \:.
  \end{equation}
  For $K=N$, the two variances coincide with \eqref{eq:MezzadriSimm2}, as it should. 
\end{enumerate}

\subsection{Distributions and moments for non-ideal contacts}
\label{subsec:TDNIC}

Less is known about the case of non-ideal contacts, when the QD is connected to the waveguides through tunnel barriers (Fig.~\ref{fig:ModelQDWeaklyCoupled}).
The distribution \eqref{eq:BrouwerFrahmBeenakker1997} of Brouwer, Frahm and Beenakker (BFB) was recently extended in Ref.~\cite{GraSavTex18}. 
The brief review  of this Subsection~\ref{subsec:TDNIC} mostly borrows from Ref.~\cite{GraSavTex18}.

\def\Sbar{\overline{\Sm}}  
\def\Wmat{\mathcal{W}}     
\def\Kmat{\mathcal{K}}     
\def\coupl{\mathcal{T}}
\def\invQ{\Gamma}      
\def\fss{\kappa} 
\def\rt{t}  
\newcommand{\mpart}[2]{\tilde{w}_{#1}^{(#2)}} 
\newcommand{\mprop}[2]{w_{#1}^{(#2)}} 
\newcommand{\dwt}[2]{\mathscr{P}_{#1}^{(#2)}} 
\def\tup{\tau_\mathrm{upper}} 
\def\tlow{\tau_\mathrm{lower}} 

\subsubsection{Wigner-Smith time delay matrix}

\begin{figure}[!ht]
\centering
\includegraphics[width=0.45\textwidth]{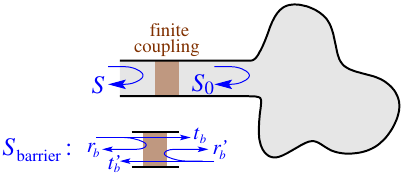}
\caption{\it Description of the model for non-ideal coupling. 
  Figure from Ref.~\cite{GraSavTex18}.}
\label{fig:ModelQDWeaklyCoupled}
\end{figure}

The derivation is based on the idea that the scattering matrix $\Sm$ for non-ideal contacts can be related to the scattering matrix $\Sm_0$ for perfect contacts\begin{equation}
  \label{eq:MappingIdealNonideal}
  \Sm = r_b + t_b' \, \left( \Sm_0^\dagger - r_b' \right)^{-1} \, t_b
  \:,
\end{equation}
where $r_b$, $r_b'$, $t_b$ and $t_b'$ are reflection and transmission matrices describing the tunneling barrier,  here supposed fixed (non random).
This is illustrated in Fig.~\ref{fig:ModelQDWeaklyCoupled} for one contact (obviously the approach is not restricted to a one contact situation).

Assuming that the matrix $\Sm_0$ is uniformly distributed over the unitary group, we deduce that the coupling between the QD and the external can be encoded in the averaged (``optical'') scattering matrix $\mean{\Sm}=r_b$.
In the following, we consider the case where all couplings are equal for simplicity, $\mean{\Sm}=\Sbar\,\mathbf{1}_\Nc$ (the case of perfect coupling discussed in Subsection~\ref{subsec:TDPC} corresponds to $\Sbar=0$).
More precisely, we choose
$
  r_b=-\big(r_b'\big)^\dagger=\mathbf{1}_\Nc\,\Sbar
$  
and 
$
  t_b=t_b'=\mathbf{1}_\Nc\,\sqrt{1-|\Sbar|^2}
$.
This picture makes easy to recover the Poisson kernel describing the statistical properties of the $\Sm$-matrix for non ideal contacts \cite{MelPerSel85,Bro95,MelKum04}
\begin{equation}
  \label{eq:PoissonKernel}
  \mathrm{D}\Sm P_{\Sm}(\Sm)\propto\mathrm{D}\Sm\,|\det(\identity_\Nc-\Sbar^*\Sm)|^{-2-\beta(\Nc-1)}\,.
\end{equation}
where $\mathrm{D}\Sm$ is the uniform measure over the circular ensemble (which corresponds with the Haar measure in the unitary case)~;
since the distribution of $\Sm_0$ is uniform, Eq.~\eqref{eq:PoissonKernel} simply corresponds with the Jacobian of the transformation \eqref{eq:MappingIdealNonideal}, which now takes the form
$   
\Sm =
  \big( \Sbar\,\mathbf{1}_\Nc + \Sm_0 \big)\, \big(  \mathbf{1}_\Nc + \Sbar^*\,\Sm_0 \big)^{-1}
$
 \cite{GraSavTex18}. 

Introducing the notation $\Gamma=\WSm_s^{-1}$ for the inverse of the symmetrised Wigner-Smith matrix $\WSm_{s}=-\I\Sm^{-1/2}\,\partial_\varepsilon\Sm\,\Sm^{-1/2}$, the distribution of eigenvalues \eqref{eq:BrouwerFrahmBeenakker1997} for perfect contacts (superscript ``$^{(0)}$'') can be related to the matrix distribution
\begin{equation}
  \label{eq:BrouwerFrahmBeenakker1997v2}
  P_\invQ^{(0)}(\invQ)\propto ( \det\invQ )^{\beta\Nc/2}\,\exp\{-(\beta/2)\tr{\Gamma}\}
  \:.
\end{equation}
A key result of Refs.~\cite{BroFraBee97,BroFraBee99} is the statistical independence of the scattering matrix $\Sm_0$ and the symmetrised Wigner-Smith matrix $\WSm_{0s}$.
Making use of 
\\[0.125cm]
(\textit{i}) the statistical independence of $\Sm_0$ and $\WSm_{0s}$,  
\\[0.125cm]
(\textit{ii}) the uniform distribution of $\Sm_0$,
\\[0.125cm]
(\textit{iii}) BFB's distribution \eqref{eq:BrouwerFrahmBeenakker1997v2}, and 
\\[0.125cm]
(\textit{iv})
assuming that the barrier coefficients are independent of the energy, 
\\[0.125cm]
we have obtained in Ref.~\cite{GraSavTex18} the joint distribution 
\begin{align}
  \label{eq:GrabschSavinTexierJMPDF}
  &
  \mathrm{D}\Sm\,\,\mathrm{D}\invQ\, P_{\Sm,\invQ}(\Sm,\invQ)
  =
  \mathrm{D}\Sm\,\,\mathrm{D}\invQ\,
  c_{\Nc,\beta}  \,
  \Theta(\invQ)\,
  \\\nonumber
  &
  \hspace{1cm}
  \times
  \left|\det\big(\identity_\Nc-\Sbar^*\Sm\big)\right|^{\beta\Nc}\,
  \big(\det\invQ\big)^{\beta\Nc/2} \,
  \\\nonumber
  &
  \hspace{1cm}
  \times
  \exp\left[
     -\beta\frac{\tr{(\identity_\Nc-\Sbar^*\Sm)(\identity_\Nc-\Sbar\Sm^\dagger)\invQ}}{2(1-|\Sbar|^2)}\,
  \right],
\end{align}
where $c_{\Nc,\beta}$ is a normalisation constant. $\mathrm{D}\invQ$ is the Lebesgue measure over the set of Hermitian matrices.  
$\Theta(\invQ)=1$ when all eigenvalues of $\invQ$ are positive and zero otherwise.
The representation \eqref{eq:GrabschSavinTexierJMPDF} may be regarded as an extension of the Poisson kernel~\eqref{eq:PoissonKernel}.
For $\Sbar=0$, we recover Eq.~\eqref{eq:BrouwerFrahmBeenakker1997v2}.
Note that a representation similar to \eqref{eq:GrabschSavinTexierJMPDF} was given in Ref.~\cite{MarSchBee16}, with extensions to the BdG symmetry classes describing scattering in an Andreev billiard.

The marginal distribution of the Wigner-Smith matrix can be obtained from \eqref{eq:GrabschSavinTexierJMPDF} by a matrix integration over $\Sm$.
We have prefered a more convenient form in terms of a matrix integral over Hermitian matrices $\Kmat$~\cite{GraSavTex18}
\begin{align}
  \label{eq:GrabschSavinTexierPDFinQ}
  P_\invQ(\invQ) 
  &= b_{\Nc,\beta}
    \Theta(\invQ)\,
    ( \det\invQ )^{\beta\Nc/2}
  \\\nonumber
  & \times
    \int\mathrm{D}\Kmat\,
    \frac{\det(\identity_\Nc+\Kmat^2)^{\beta\Nc/2}}
         {\det(\identity_\Nc+\fss^2\Kmat^2)^{1-\frac{\beta}{2}+\beta\Nc}}\,
  \\\nonumber
  & \hspace{1cm}
  \times
    \exp\left[
      -\frac{\beta}{2}\fss\,\tr{ \frac{\identity_\Nc+\Kmat^2}{\identity_\Nc+\fss^2\Kmat^2}\invQ }
    \right]
\end{align}
where $b_{\Nc,\beta}$ is a normalisation constant  and the coupling constant $\fss>0$ is related to the transmission probability $\coupl$ by
\begin{equation}
  \coupl \equiv 1-|\Sbar|^2 = \frac{4\fss}{(1+\fss)^2}
  \:.
\end{equation}
For perfect contacts ($\coupl=\fss=1$), we recover the Laguerre distribution  \eqref{eq:BrouwerFrahmBeenakker1997v2}, as it should.
It is interesting to point that the Hermitian matrix $\Kmat$ in the matrix integral in Eq.~\eqref{eq:GrabschSavinTexierPDFinQ} corresponds to the Wigner's reaction matrix characterizing the QD for perfect contacts. 
Hence \eqref{eq:GrabschSavinTexierPDFinQ} can be deduced by integrating \eqref{eq:GrabschSavinTexierJMPDF} over $\Sm$ and eventually performing the two changes of variable
$   
\Sm =
  \big( \Sbar\,\mathbf{1}_\Nc + \Sm_0 \big)\, \big(  \mathbf{1}_\Nc + \Sbar^*\,\Sm_0 \big)^{-1}
$
and
$\Sm_0=\big(\identity_\Nc-\I\,\Kmat\big)\big(\identity_\Nc+\I\,\Kmat\big)^{-1}$.
Note that the uniform distribution of $\Sm_0$ corresponds to the Cauchy distribution 
\begin{equation}
 \label{eq:Cauchy}
  P_\Kmat^{(0)} (\Kmat) \propto \big[\det(\identity_\Nc+\Kmat^2)\big]^{-1-\beta(\Nc-1)/2}
\:,
\end{equation}
defined over the set of Hermitian matrices.
We can indeed check that, removing the matrix integral in \eqref{eq:GrabschSavinTexierPDFinQ}, we recover $P_\invQ^{(0)}(\invQ)\:P_\Kmat^{(0)} (\Kmat)$ for $\fss=1$.

Remarks~:
\begin{enumerate}[(i)]
\item 
Another integral representation, equivalent to \eqref{eq:GrabschSavinTexierPDFinQ}, was derived in Ref.~\cite{GraSavTex18} in terms of a matrix integral over the unitary group, which turned out to be more convenient for numerical simulations.

\item 
The case of unequal couplings can also be analysed along the same lines~:
the diagonal matrix $\mean{\Sm}$ is not proportional to the identity matrix, and a set of $\Nc$ transmission coefficients $\coupl_1,\cdots,\coupl_\Nc$ (or $\fss_a$'s) should be introduced.
The form \eqref{eq:GrabschSavinTexierPDFinQ} was extended in Ref.~\cite{GraSavTex18}. 
\end{enumerate}

\subsubsection{Proper time delays}

The joint distribution for the proper time delays $\tau_a=1/\gamma_a$ can in principle be deduced from \eqref{eq:GrabschSavinTexierPDFinQ} by integrating over all degrees of freedom related to the eigenvectors of the matrix. 
This was done in Ref.~\cite{GraSavTex18} in the unitary case by using Harish-Chandra-Itzykson-Zuber integral~:
\begin{align}
  \label{eq:GrabschSavinTexierJPDFpt}
  &
  \mathcal{P}(\gamma_1,\cdots,\gamma_\Nc) \propto
  \Delta_\Nc(\gamma) \,\prod_n \heaviside(\gamma_n)\,\gamma_n^\Nc \,
  \nonumber\\
  &
\times
  \int_{\mathbb{R}^\Nc}\D k_1\cdots\D k_\Nc\,
  \frac{\Delta_\Nc(k)^2}{\Delta_\Nc(k^2)}
  \prod_n \frac{(1+k_n^2)^{\Nc}}{(1+\fss^2k_n^2)^{\Nc+1}}
  \nonumber\\
  &\hspace{1.5cm}
\times
  \det\left[
    \exp\left(
      -\fss\,\frac{1+k_i^2}{1+\fss^2k_i^2}\,\gamma_j
    \right)
  \right]
\end{align}
where $\Delta_\Nc(\gamma)=\prod_{i<j}(\gamma_i-\gamma_j)$ denotes the Vandermonde determinant
(here, I have slightly simplified the form given in \cite{GraSavTex18}). 
This form reduces to \eqref{eq:BrouwerFrahmBeenakker1997} for~$\fss\to1$, as we now check.
In the limit $\fss\to1^-$ we have $\fss\simeq1-2\Sbar$.  
The limit of the last determinant of the integral can be analysed by using the property
\begin{equation}
  \det\left( \EXP{\epsilon\, x_iy_j} \right)_{i,\,j=1}^N 
  \underset{\epsilon\to0}{\simeq}
  \frac{\epsilon^{N(N-1)/2}}{G(N+1)}\, \Delta_N(x) \,  \Delta_N(y) 
\end{equation}
where $G(N+1)=\prod_{n=1}^{N-1}n!$ is the Barnes-$G$ function, we obtain
\begin{align}
   &\det\left[
    \exp\left(
      -\fss\,\frac{1+k_i^2}{1+\fss^2k_i^2}\,\gamma_j
    \right)
  \right]
  \\\nonumber
  &\underset{\fss\to1}{\simeq}
     \frac{\big(4\Sbar\big)^{\Nc(\Nc-1)/2}}{G(\Nc+1)}
     \frac{\prod_n\EXP{-\gamma_n}}{\prod_n(1+k_n^2)^{\Nc-1}} \,\Delta_N(\gamma) \,  \Delta_N(k^2) 
\end{align}
Finally, we deduce that the limit $\fss\to1$ of \eqref{eq:GrabschSavinTexierJPDFpt} is 
\begin{align}
  \label{eq:106}
  &\mathcal{P}(\gamma_1,\cdots,\gamma_\Nc) \propto
  \Delta_\Nc(\gamma)^2 \,\prod_n \heaviside(\gamma_n)\,\gamma_n^\Nc \, \EXP{-\gamma_n}
   \nonumber
   \\
  &\times
  \int_{\mathbb{R}^\Nc}\D k_1\cdots\D k_\Nc\,
  \Delta_\Nc(k)^2 \prod_n(1+k_n^2)^{-\Nc}    
\end{align}
where the multiple integral is a constant corresponding to the normalisation of the Cauchy ensemble \eqref{eq:Cauchy}.  
Eq.~\eqref{eq:106} is precisely the BFB result \eqref{eq:BrouwerFrahmBeenakker1997} for $\beta=2$, as it should.

\begin{figure}[!ht]
\centering
\includegraphics[width=0.45\textwidth]{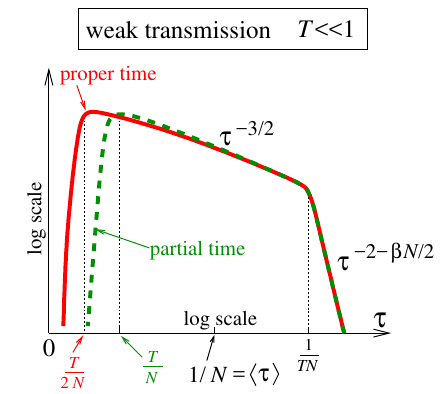}
\caption{\it Sketch of the distributions for perfect contacts~; compare with Fig.~\ref {fig:SketchDistribPerfect}.
  Figure from Ref.~\cite{GraSavTex18}.}
\label{fig:SketchDistribWeak}
\end{figure}

Although complicate, the form \eqref{eq:GrabschSavinTexierJPDFpt} was used in order to derive the distribution of the Wigner time delay~\cite{GraSavTex18}. 
Extracting the marginal distribution of the proper time or correlations seems however difficult.
We now review some results obtained earlier by other means.
The marginal distribution of the proper time delays was obtained in the unitary case by Sommers, Savin and Sokolov~\cite{SomSavSok01}~: the result was given under a rather complicate form involving a sum of multiple derivative of Bessel function. 
Here, we only give the limiting behaviours of the distribution for weak coupling \cite{SomSavSok01,GraSavTex18}~:
\begin{align}
  \label{eq:LimitsMarginalProperTimes}
  &\frac{\coupl}{4} \mprop{\Nc}{2}\left(\tau\simeq\frac{\coupl}{4}\,\rt\right)
  \\\nonumber
  &\underset{\coupl\to0}{\simeq}
  \left\{
    \begin{array}{ll}
    \displaystyle
    c_\Nc\,\rt^{-2\Nc-1/2}\,\EXP{-1/\rt}
    & \mbox{for } \rt\lesssim 1/\Nc
    \\[0.25cm]
    \displaystyle
    b_\Nc  \,\rt^{-3/2}
    & \mbox{for }  1/\Nc \lesssim \rt \lesssim \frac{1}{\Nc\coupl^2}
    \\[0.25cm]
    \displaystyle
    \frac{a_\Nc\coupl^3}{8} \left(\frac{4}{\coupl^2\rt}\right)^{\Nc+2}
    & \mbox{for } \rt\gtrsim \frac{1}{\Nc\coupl^2}
    \end{array}
  \right.
\end{align}
The two coefficients $a_\Nc$ and $b_\Nc$ are given below, Eqs.~(\ref{eq:CoeffBn},\ref{eq:CoeffAn}), and the last coefficient is  
\begin{align}
  \label{eq:CoeffCn}
   c_\Nc &= \frac{2^{2(\Nc-1)}}{\sqrt{\pi}\,\Nc(2\Nc-1)!}
  \:.
\end{align}
Thus $\mprop{\Nc}{\beta}(\tau)$ and $\mpart{\Nc}{\beta}(\tau)$ coincide in the two regimes above the scale $\tlow\sim\coupl/\Nc$ (Fig.~\ref{fig:SketchDistribWeak}).
The form of the distribution shows that the universal power law $\mprop{\Nc}{\beta}(\tau)\sim\tau^{-3/2}$ (which holds for any $\beta$) is cut off at the scale $ \tup \simeq 8\,\mathrm{e}/(\Nc\coupl)$. 
This cutoff controls the positive moments 
\begin{equation}
  \mean{\tau_a^k}  \sim b_\Nc\sqrt{\coupl}\,\tup^{k-1/2}
\end{equation} 
for $1\leq k<1+\beta\Nc/2$
(and $\smean{\tilde\tau_a^k}=\smean{\tau_a^k}=\infty$ for $k\geq1+\beta\Nc/2$).

\subsubsection{Partial time delays}

The marginal distribution of the partial time delays is known exactly for any symmetry class \cite{FyoSom97,FyoSavSom97,SavFyoSom01}.
A convenient integral representation was derived in Appendix~C of \cite{GraSavTex18} by extending the method of Ref.~\cite{GopMel98} to arbitrary number of channels.
Here, we only give the asymptotic behaviours
\begin{align}
\label{eq:LimitsMarginalPartialTimes}
&\frac{\coupl}{4} \mpart{\Nc}{\beta}\left(\tau\simeq\frac{\coupl}{4} \,\rt\right)
  \\\nonumber  
  &\underset{\coupl\to0}{\simeq}
  \left\{
    \begin{array}{ll}
    \displaystyle
    \tilde{c}_\Nc\,\rt^{-\beta\Nc/2-3/2}\,\EXP{-1/\rt}
    & \mbox{for }     \rt\lesssim 1/\Nc
    \\[0.25cm]
    \displaystyle
    b_\Nc \,\rt^{-3/2}
    & \mbox{for }    1/\Nc \lesssim\rt \lesssim
    \frac{1}{\Nc\coupl^2}
    \\[0.25cm]
    \displaystyle
    \frac{a_\Nc\coupl^3}{8} \left(\frac{4}{\coupl^2\rt}\right)^{2+\beta\Nc/2}
    & \mbox{for } \rt\gtrsim \frac{1}{\Nc\coupl^2}
    \end{array}
  \right.
\end{align}
involving the three coefficients 
\begin{align}
  \label{eq:CoeffCnTilde}
  \tilde{c}_\Nc &= \frac{1}{\sqrt{\pi}\,\Gamma(1+\beta\Nc/2)}
  \:,
  \\
  \label{eq:CoeffBn}
  b_\Nc &= \frac{1}{\pi}\, \frac{\Gamma(1/2+\beta\Nc/2)}{\Gamma(1+\beta\Nc/2)}
  \:,
  \\
  \label{eq:CoeffAn}
  a_\Nc &= \frac{2^{1+\beta\Nc}}{\sqrt\pi}\, \frac{\Gamma(1/2+\beta\Nc/2)}{\Gamma(1+\beta\Nc/2)^2}
   \:.
\end{align}
In Ref.~\cite{GraSavTex18}, we have also stressed that the weak coupling limit ($\coupl\to0$) of the marginal distribution takes the simple form
\begin{align}
  \label{eq:MarginalPartialTimeGoparMelloMethod}
  &\lim_{\coupl\to0}\frac{\coupl}{4}  \mpart{\Nc}{\beta}\left(\tau\simeq\frac{\coupl}{4} \,\rt\right)
  \\\nonumber
  &= \frac{1}{\sqrt{\pi}\,\Gamma(1+\beta\Nc/2)}
  \frac{\EXP{-1/\rt}}{\rt^{2+\beta\Nc/2}}\,
  U\left(  \frac{1}{2} , \frac{\beta\Nc+3}{2} , \frac{1}{\rt} \right)
  \:,
\end{align}
in terms of the Kummer function $U(a,c,z)$~\cite{AbrSte64}.
It is quite remarkable to obtain such a universal form in this limit.

The variance of the partial times are exactly known for $\beta=2$~\cite{FyoSom97}~:
\begin{equation}
  \label{eq:FyodorovSommers1997Eq167}
  \mathrm{var}(\tilde{\tau}_a)= \frac{2\Nc(\coupl^{-1}-1)+1}{\Nc^2(\Nc-1)}
   \underset{\coupl\ll1}{\simeq}
   \frac{2}{\coupl\,\Nc^2}
   \:.
\end{equation}
By combining \eqref{eq:FyodorovSommers1997Eq167} with the variance of the Wigner time delay given below, Eq.~\eqref{eq:FyodorovSommers1997Eq195}, we get the covariance~:
\begin{align}
  \label{eq:CovPartialWeakCoupling}
  &\mathrm{cov}(\tilde{\tau}_a,\tilde{\tau}_b)=  \frac{1}{\coupl^2\Nc(\Nc-1)^2}
  \\\nonumber
  &\hspace{0.5cm}\times
  \left[
     \frac{2\left[1-(1-\coupl)^{\Nc+1}\right]}{\Nc+1}
     -2\coupl(1-\coupl) - \frac{\coupl^2}{\Nc}
  \right]
  \\
  &   
  \label{eq:CovPartialWeakCoupling1}
   \underset{\coupl\ll1}{\simeq}
  -\frac{1}{\Nc^2}\times
  \left\{
  \begin{array}{ll}
    \displaystyle
    \frac{2}{\coupl \Nc} & \mbox{for } \Nc\coupl\gg1
    \\[0.25cm]
    \displaystyle
    1 & \mbox{for } \Nc\coupl\ll1
  \end{array}
  \right.
\end{align}
Contrary to the ideal contact case, where the correlations are always positive, the covariances change both in sign and scaling with $\Nc$ as transmission crosses over from perfect to weak coupling.
At this point it is important to stress that the \textit{weak transmission} $\coupl\ll1$ regime and \textit{weak coupling} $\Nc\coupl\ll1$ regime should not be confused.

\subsubsection{Wigner time delay}

The statistical properties of the Wigner time delay $\Wt=\tr{\invQ^{-1}}$ is in principle encoded in the matrix distribution \eqref{eq:GrabschSavinTexierPDFinQ}, however explicit calculations from the matrix integral seems a difficult task. We rather collect the known results obtained by other means.
The variance of the Wigner time delay for non ideal coupling was obtained in the unitary ($\beta=2$) case \cite{FyoSom96,FyoSom97}~:
\begin{equation}
  \label{eq:FyodorovSommers1997Eq195}
   \mathrm{var}(\Wt) = \frac{2\left[1-(1-\coupl)^{\Nc+1}\right]}{\coupl^2\Nc^2(\Nc^2-1)}
   \:.
\end{equation}
In the limit of weak transmission per channel $\coupl\ll1$, we get from \eqref{eq:FyodorovSommers1997Eq195} two different possible behaviours, which depend on the product $\Nc\coupl$ describing the degree of the resonance overlap (thus controlling the overall coupling to the leads)~:
\begin{equation}  \label{eq:VarianceWT}
  \frac{ \mathrm{var}(\Wt) }{ \mean{\Wt}^2 }
  \underset{\coupl\ll1}{\simeq}
  \left\{
  \begin{array}{ll}
    \displaystyle
    \frac{4}{\beta(\Nc\coupl)^2} \ll 1 & \mbox{for } \Nc\coupl\gg1
    \\[0.5cm]
    \displaystyle
    \frac{2}{\Nc\coupl} \gg 1 & \mbox{for } \Nc\coupl\ll1
  \end{array}
  \right. .
\end{equation}
Here we have reintroduced $\beta$ in order to match with the known $\beta=1$ result \cite{LehSavSokSom95}.

Starting from \eqref{eq:GrabschSavinTexierJPDFpt}, the distribution of the Wigner time delay was obtained in \cite{GraSavTex18} in the weak coupling regime $\Nc\coupl\ll1$, by a careful analysis of the characteristic function~:
\begin{align}
  &\mathscr{P}_{\Nc}^{(\beta)}(\tau)
  \\\nonumber
  &\sim
  \begin{cases}
    \tau^{-\frac{\beta\Nc^2}{2}-\frac32}\,
  \EXP{-\beta\Nc\coupl/(8\tau)}
    & \mbox{for } \tau\ll \coupl
  \\[0.125cm]
  \frac{1}{\coupl}(\coupl/\tau)^{3/2}
    & \mbox{for }
  \coupl \ll \tau \ll \frac{1}{\Nc^2\coupl}
  \\[0.125cm]
  \coupl^2\Nc^3
  \,
  \left( \coupl\Nc^2\,\tau\right)^{-2-\beta\Nc/2}
  & 
  \mbox{for }
  \tau \gg \frac{1}{\Nc^2\coupl}
  \end{cases}
\end{align}
These behaviours were verified by numerical calculations.
Some heuristic analysis was given in Ref.~\cite{GraSavTex18}, relating the distribution of the Wigner time delay and that of the resonance widths.
The power law $\mathscr{P}_{\Nc}^{(\beta)}(\tau)\sim\tau^{-3/2}$ was then related to typical resonance with, while the power law $\mathscr{P}_{\Nc}^{(\beta)}(\tau)\sim\tau^{-2-\beta\Nc/2}$ is due to atypically narrow resonances.

\subsection{Energy correlations for perfect coupling}
\label{sec:EnergyCorrelations}

The knowledge of the correlation function 
$\mean{\Wt(\varepsilon)\Wt(\varepsilon')}_c=
\mean{\Wt(\varepsilon)\Wt(\varepsilon')}-\mean{\Wt(\varepsilon)}\mean{\Wt(\varepsilon')}$ 
is of importance and has practical applications~: 
for example, this information was used by Polianski and B\"uttiker in order to study the effect of thermal fluctuations on the non-linear conductance of a quantum dot \cite{PolBut07a}. 
The correlation function was obtained by Lehmann \textit{et al.}~\cite{LehSavSokSom95} by a random matrix analysis in the orthogonal case ($\beta=1$), within the HA of random matrices (see also \cite{FyoSom97})~:
\begin{equation}
  \label{eq:LehmanSavinSokolovSommers1995}
 \frac{\mean{\Wt(\varepsilon)\Wt(\varepsilon')}_c}{\mean{\Wt}^2}
 \simeq \frac{4}{N^2} \, 
 \frac{1-(\omega\tau_d)^2}{\left[1+(\omega\tau_d)^2\right]^2}
 \:,
\end{equation} 
where $\omega=\varepsilon-\varepsilon'$.
For $\varepsilon=\varepsilon'$ we recover the cumulant \eqref{eq:MezzadriSimm2} at leading order $1/N^4$.
Eq.~\eqref{eq:LehmanSavinSokolovSommers1995} shows that correlations occur on the scale $1/\tau_d\sim N\Delta$ much larger than the mean resonance spacing.~\footnote{
  In a ballistic QD of size $L$, the Thouless energy, which sets the energy scale for correlations in a closed QD, is $\EThouless=v_F/L$, where $v_F$ is the Fermi energy.
  The open QD is characterised by the number of channels $N=k_Fw/\pi$, controlled by the width $w$ of the contacts. In this case, as we have seen, the relevant scale is a different Thouless energy~\cite{Bee97} $\EThouless^\mathrm{open}=1/\tau_d$. We have $\EThouless^\mathrm{open}/\EThouless\sim w/L\ll1$.
}
We have introduced
\begin{equation}
  \tau_d=\mean{\Wt}=\frac{2\pi}{N\Delta}=\Ht/N
  \:,
\end{equation}
the dwell time for an electron in the cavity 
(Eq.~\eqref{eq:LehmanSavinSokolovSommers1995} is valid for strongly overlapping resonances, $1/\tau_d\gg\Delta$, i.e. $N\gg1$).
The result \eqref{eq:LehmanSavinSokolovSommers1995} was later generalised~\cite{FyoSavSom97} in order to include parametric correlations and describe the crossover between orhtogonal and unitary cases.

Time delay correlations can also be determined within the stochastic approach~: the model described in Refs.~\cite{AleBroGla02,PolBro03,BroLamFle05} and used by Polianski and B\"uttiker \cite{PolBut06,PolBut07,PolBut07a} provides some information about the energy and magnetic field dependence of the $\Sm$-matrix correlator, which reads~\cite{PolBut07a}
\begin{equation}
  \label{eq:CorrelatorPolianskiBrouwer}
  \mean{\Sm_{ab}(\varepsilon,\mathcal{B})\Sm_{cd}^*(\varepsilon',\mathcal{B}')}
  = \delta_{ac}\delta_{bd}\,\mathscr{D}_{\varepsilon-\varepsilon'}
  +\delta_{ad}\delta_{bc}\,\mathscr{C}_{\varepsilon-\varepsilon'}
  \:.
\end{equation}
$\mathscr{D}_\omega$ and $\mathscr{C}_\omega$ are the (zero-dimensional) analogues of the Diffuson and the Cooperon appearing in the diagrammatic approach for weakly disordered metals~\cite{AkkMon07},~\footnote{
  This can be understood from the Fisher and Lee relation \cite{FisLee81} between scattering matrix and Green's function. 
}
given by~\footnote{
  Note that \eqref{eq:CorrelatorPolianskiBrouwer} only satisfies unitarity at leading order in the orthogonal case~: one gets  
  $\smean{\big(\Sm^\dagger\Sm)_{aa}-1}=\mathscr{C}_0=1/N$ instead of zero.
  This can be corrected by using the expression \cite{AleBroGla02}
  $
  \mean{\Sm_{ab}(\varepsilon)\Sm_{cd}^*(\varepsilon')}
  = \big[\delta_{ac}\delta_{bd}+(2/\beta-1)\delta_{ad}\delta_{bc}\big]\,\mathscr{D}_{\varepsilon-\varepsilon'}
  $
  with $\mathscr{D}_{\omega}=\big[\Nc+2/\beta-1-2\I\pi\omega/\Delta\big]^{-1}$.
}
\begin{equation}
  \label{eq:PolianskiBrouwer2003}
  \left.
  \begin{array}{c}
     \mathscr{D}_\omega \\ \mathscr{C}_\omega
  \end{array}
  \right\}
  = \frac{1}{N\tau_d}\,\frac{1}{1/\tau_{\mathscr{D},\mathscr{C}}-\I\omega}
  \:.
\end{equation}
The two characteristic times
\begin{equation}
  \frac{1}{\tau_{\mathscr{D},\mathscr{C}}}
  = \frac{1}{\tau_d} + \frac{1}{\tau_{(\mathcal{B}\mp\mathcal{B'})/2}}
\end{equation}
combines a contribution describing the escape rate $1/\tau_d$ from the cavity and dephasing due to the magnetic field. 
The magnetic dephasing rate is $1/\tau_\mathcal{B}=(v_F\ell/\mathrm{Surf})(\Phi/\phi_0)^2$ where $v_F$ is the Fermi velocity, $\Phi=\mathrm{Surf}\,\mathcal{B}$ the magnetic flux through the cavity and $\phi_0=h/e$ the quantum flux. $\ell$ is the size of the cavity in the ballistic case or the elastic mean free path in the weakly disordered (diffusive) case.

As a simple application of the formula we get
$\mean{\Wt(\varepsilon)}\simeq\tau_d(N^2\mathscr{D}_0^2+N\mathscr{C}_0^2)$,
where we have used 
$-\I\partial_\varepsilon\mathscr{D}_{\varepsilon-\varepsilon'}=\Ht\,\mathscr{D}_{\varepsilon-\varepsilon'}^2$.
We deduce 
$\mean{\Wt(\varepsilon)}/\tau_d\simeq1+\mathcal{O}(1/N)$,  
whose leading order term coincides with \eqref{eq:MeanWTD}. 

The analysis of the correlations requires additional information~:
simply applying Wick's theorem with the correlator \eqref{eq:CorrelatorPolianskiBrouwer}, we get
$
  \mean{\Wt(\varepsilon,\mathcal{B})\Wt(\varepsilon',\mathcal{B}')}_c^\mathrm{(Gaussian)}
  \simeq \Ht^2\,\big( 
      \left|\mathscr{D}_{\varepsilon-\varepsilon'}\right|^4
    + \left|\mathscr{C}_{\varepsilon-\varepsilon'}\right|^4
  \big)
$,
which leads, at zero magnetic field, to the incomplete expression
$
\mean{\Wt(\varepsilon,0)\Wt(\varepsilon',0)}_c^\mathrm{(Gaussian)}
/\mean{\Wt}^2
\simeq (2/N^2) \, \big[ 1+(\omega\tau_d)^2\big]^{-2}
$.
This calculation, known as the ``diagonal approximation'' in the context of semiclassical methods, disagrees with~\eqref{eq:LehmanSavinSokolovSommers1995}~: not only the energy dependence differs but also the value at $\omega=0$ is half the correct result [compare with the variance \eqref{eq:MezzadriSimm2}].
This emphasizes the importance of \textit{non Gaussian} contributions to the correlator, which are taken into account in the more precise expression of the four-point correlation function $\mean{S_{ab}S^*_{cd}S_{ij}S^*_{kl}}$ given in Appendix~B of Ref.~\cite{BroLamFle05} (see also \cite{BroBee96}).

This question has been much discussed in semiclassical approach~:
this discrepancy was underlined by Lewenkopf and Vallejos~\cite{LewVal04}.
Kuipers and Sieber~\cite{KuiSie08} identified the nature of the contributions (``trajectory quadruplets'') to be added to the diagonal approximation in order to recover the random matrix result \eqref{eq:LehmanSavinSokolovSommers1995} (doing so they have reconciled the two semiclassical approaches for a Wigner time delay analysis~: the periodic orbit expansion based on the relation with the density of states~\cite{Eck93b}, and the scattering approach involving trajectories entering and leaving the system).
The expression of the correlator, determined earlier within a semiclassical approach by Vallejos \textit{et al.}~\cite{ValOzoLew98} (for $\mathcal{B}=\mathcal{B}'$), is the sum of two contributions which can be identified as Diffuson and Cooperon contributions. Reintroducing the effect of the field mismatch, we get the structure obtained within the HA of random matrices in Ref.~\cite{FyoSavSom97,FyoSom97}~:
\begin{align}
  \label{eq:FyodorovSavinSommers97VallejosOzorioLewenkopf98}
   &\frac{\mean{\Wt(\varepsilon,\mathcal{B})\Wt(\varepsilon',\mathcal{B}')}_c}{\mean{\Wt}^2}
  \\\nonumber
  &
  \simeq \frac{2}{\Ht^2}\left\{
  \frac{1/\tau_{\mathscr{D}}^2 - \omega^2}{\left[1/\tau_{\mathscr{D}}^2 + \omega^2\right]^2}
  + \frac{1/\tau_{\mathscr{C}}^2 - \omega^2}{\left[1/\tau_{\mathscr{C}}^2 + \omega^2\right]^2}
  \right\}
  \:,
\end{align}
where we recall that $\Ht=2\pi/\Delta$.
This expression now agrees with \eqref{eq:LehmanSavinSokolovSommers1995} for $\tau_{\mathscr{D}}=\tau_{\mathscr{C}}=\tau_d=\Ht/N$.

Few remarks~:
\begin{enumerate}[(i)]

\item 
  The time delay may be represented in terms of as 
  $\Wt(\varepsilon)=(1/N)\sum_\alpha\Gamma_\alpha/\big[(\varepsilon-\varepsilon_\alpha)^2+\Gamma_\alpha^2/4\big]$, where the sum runs over the resonances~\cite{LehSavSokSom95,FyoSom97}.
  This establishes a connection between Wigner time delay fluctuations and Ericson fluctuations of the cross-section for strongly overlapping resonances $\Gamma\gg\Delta$ \cite{Eri60,Eck93b}.
  
\item
  Semiclassical methods were widely used in order to analyse Wigner time delay correlations~\cite{Eck93b,ValOzoLew98,LewVal04,KuiSie07,KuiSie08}~; correspondence between the random matrix and semiclassical approaches holds for strongly overlapping resonaces $1/\tau_d\gg1/\Ht\sim\Delta$ (i.e. $N\gg1$) \cite{ValOzoLew98}.
  
\item
  More recently, an improved semiclassical approach for the analysis of the Wigner time delay statistics was developped by Kuipers, Savin and Sieber \cite{KuiSavSie14}, who carried out a diagrammatic calculation of the moments of the Wigner time delay.
  A semiclassical derivation of the moments of the time delay was also proposed by Novaes~\cite{Nov15b}.
\end{enumerate}

\subsection{Other symmetry classes}

As we have already mentioned, the three Wigner-Dyson symmetry classes (denoted AI, A and AII in the Altland-Zirnbauer classification) were completed by three chiral classes (chiral orthogonal BDI, chiral unitary AIII and chiral symplectic CII) and four Bogoliubov-de Gennes classes (C, CI, D, DIII)~\cite{Zir96,AltZir97,EveMir08}.
The new symmetry classes have attracted a lot of attention during the last years in relation with topological insulators~\cite{HasKan10,QiZha11} and topological superconductors~\cite{Bee13,Bee15} (see also \cite{RyuSchFurLud10} were a classification of the various types of topological insulators was provided for the different symmetry classes and dimensions).

The analysis of the Wigner time delay distribution for different symmetry classes has been initiated by the work of Fyodorov and Ossipov \cite{FyoOss04} for the chiral-unitary class (AIII) in the $\Nc=1$ channel case~:
\begin{equation}
  \mathscr{P}_1^{(\mathrm{AIII})}(\tau)=\tau^{-2}\EXP{-1/\tau}
\end{equation}
in appropriate units.
This should be compared with the expression \eqref{eq:GoparMelloButtiker1996} for the unitary case (class A)~: $\mathscr{P}_1^{(2)}(\tau)=\tau^{-3}\EXP{-1/\tau}$.
We see that the exponent of the tail is increased from $2$ to $3$ when the chiral symmetry is broken,  as for the 1D disordered case reviewed above~:
compare \eqref{eq:TailDirac} in the presence of the chiral symmetry and \eqref{eq:ComtetTexier1997LimitLaw} when the chiral symmetry is broken.

More recently, the result \eqref{eq:BrouwerFrahmBeenakker1997} for the joint distribution of proper time delays was extended to the four BdG symmetry classes in Refs.~\cite{MarBroBee14,SchMarBee15}.
This question is reviewed in another article of this special issue~\cite{MarSchBee16}.
The generalisation to the three chiral symmetry classes remains an open problem.


\section{Generalised concepts~: injectance, emittance, etc.}
\label{sec:InjectanceEtc}

The concept of scattering matrix is central in the Landauer-B\"uttiker approach.
For this reason, the Krein-Friedel relation~\eqref{eq:TexierButtiker2003} plays a very important role as it allows to express the density of states (DoS) of the \textit{open} conductor in terms of scattering properties. 
In the following, we will simply write
\begin{equation}
  \label{eq:KreinFriedel}
  \DoS(\varepsilon)
  \simeq \frac{1}{2\I\pi} \tr{ \Sm^\dagger\derivp{\Sm}{\varepsilon} }
  =\frac{1}{2\pi\hbar}\tr{\WSm}
  \:,
\end{equation}
where we have neglected the term $\tr{\Sm-\Sm^\dagger}/\varepsilon$ in Eq.~\eqref{eq:TexierButtiker2003} (this is justified in the metallic regime).
Such relation allows to characterize the amount of charge injected in an open coherent conductor, a crucial tool which has been used by B\"uttiker in order to describe screening properties.
The out-of-equilibrium situation where the conductor is connected to several contacts (terminals) with different chemical potentials however requires to identify the contributions of each terminal to the DoS, which has led B\"uttiker to introduce several generalisations of \eqref{eq:KreinFriedel} as the partial DoS, the injectance and the emittance.
These concepts have been used to develop a theory of non-linear transport \cite{But93,ChrBut96a} and AC transport \cite{But93,ButPreTho93,ButThoPre93,ButThoPre94,PreThoBut96,ChrBut96} in coherent conductors (see the reviews \cite{ButChr97,But99,But00a,ButPol05}, or chapter~1 of \cite{Tex10hdr}).

\subsection{Partial DoS, injectance and emittance}

Let us consider a multi-terminal structure, whose scattering properties are characterised by a basis of stationary scattering states $\psi_{\varepsilon,\alpha}(x)$ where the index $\alpha$ labels the terminals. 
We assume that the terminals support each a \textit{single} conducting channel (i.e. contact wires are effectively one-dimensional), what simplify the analysis (the generalisation to contacts with several channels is straightforward).
Furthermore, this would allow to illustrate the discussion by explicit formulae by considering the case of metric graphs for which explicit construction of the scattering matrix is possible \cite{TexMon01,TexBut03}.

B\"uttiker introduced the concept of ``\textit{partial density of states}'' \cite{But93,ButThoPre94}
\begin{equation}
  \label{eq:PartialDoS}
  \DoS_{\alpha\beta}(\varepsilon) 
  \simeq \frac{1}{4\I\pi} 
  \left(
    \Sm_{\alpha\beta}^*\derivp{\Sm_{\alpha\beta}}{\varepsilon} 
    - \derivp{\Sm_{\alpha\beta}^*}{\varepsilon}\Sm_{\alpha\beta}
  \right)
\end{equation}
measuring the contribution to the DoS \eqref{eq:KreinFriedel} of particles incoming from terminal $\beta$ and outgoing at contact $\alpha$. 
Another quantity which appears in B\"uttiker's work is the ``\textit{injectance}'', obtained by summation of \eqref{eq:PartialDoS} over the first index
\begin{equation} 
  \label{eq:Injectance}
  \overline{\DoS}_{\alpha}(\varepsilon) 
  =\sum_\beta \DoS_{\beta\alpha}(\varepsilon) 
  \simeq \frac{1}{2\I\pi} 
  \left(
    \Sm^\dagger\derivp{\Sm}{\varepsilon} 
  \right)_{\alpha\alpha}
  =\frac{\WSm_{\alpha\alpha}}{2\pi} 
  \:.
\end{equation}
It provides the contribution to the DoS of the scattering states incoming from terminal $\alpha$.
Summation over the second index leads to the ``\textit{emittance}''
\begin{equation}
  \label{eq:Emittance}
  \underline{\DoS}_{\alpha}(\varepsilon) 
  =\sum_\beta \DoS_{\alpha\beta}(\varepsilon) 
  \simeq \frac{1}{2\I\pi} 
  \left(
    \derivp{\Sm}{\varepsilon} \Sm^\dagger
  \right)_{\alpha\alpha}
  =\frac{\widetilde\WSm_{\alpha\alpha}}{2\pi} 
\end{equation}
which corresponds to the contribution of particles outgoing at terminal~$\beta$. 
Injectance and emittance are related by magnetic field reversal,
\begin{equation}
  \label{eq:SymmetryInjectanceEmittance}
   \underline{\DoS}_{\alpha}(\varepsilon;\mathcal{B})
  =\overline{\DoS}_{\alpha}(\varepsilon;-\mathcal{B})
  \:, 
\end{equation}
which follows from $\Sm(-\mathcal{B})=\Sm(\mathcal{B})^\mathrm{T}$.
Obviously
\begin{equation}
  \DoS(\varepsilon) = \sum_\alpha\overline{\DoS}_{\alpha}(\varepsilon) 
  =\sum_\alpha\underline{\DoS}_{\alpha}(\varepsilon) 
  =\sum_{\alpha,\beta} \DoS_{\alpha\beta}(\varepsilon) 
  \:.
\end{equation}
Injectance and emittance allow for a preselection and a postselection, respectively, when identifying the contribution to the density inside the conductor from particles passing through it~\cite{GasChrBut96}.

Few remarks~:
\begin{enumerate}[(i)]
\item
  All formulae can be straightforwardly generalised to the case of multichannel contacts by adding some traces over channels.

\item 
  These concepts have been further generalised in order to deal with \textit{local} properties, leading to the concept of partial local DoS, inject\textit{ivity} and emi\textit{ssivity}~\cite{But93,GasChrBut96}.
  This can be understood from the relation
  \begin{equation}
  \label{eq:NonDiagElements}
  -\frac{1}{2\I\pi} \left(  \Sm^\dagger \derivf{\Sm}{\PotEnerg(x)} \right)_{\alpha\beta}
  = \psi^*_{\varepsilon,\alpha}(x)\psi_{\varepsilon,\beta}(x)
\end{equation}
where $\PotEnerg(x)$ is the potential (see Appendix of Ref.~\cite{But00} or Refs.~\cite{Tex02,TexBut03,TexDeg03}).
  This leads to introduce the inject\textit{ivity} 
  \begin{equation}
  \overline{\DoS}_{\alpha}(x;\varepsilon) 
  =-\frac{1}{2\I\pi} 
  \left(
    \Sm^\dagger\derivf{\Sm}{\PotEnerg(x)} 
  \right)_{\alpha\alpha}    
  = \left|\psi_{\varepsilon,\alpha}(x)\right|^2
  \end{equation}
  which measures the contribution of the scattering state to the local DoS
  $\DoS(x;\varepsilon)=\sum_\alpha\overline{\DoS}_{\alpha}(x;\varepsilon)$.
  Similarly, B\"uttiker introduced the concepts of partial \textit{local} DoS and emi\textit{ssivity}~\cite{But93,GasChrBut96}.
  
\item
  The Fisher and Lee relation \cite{FisLee81} 
\begin{equation}
  \Sm_{\alpha\beta}(\varepsilon)=-\delta_{\alpha\beta}
  + \I\sqrt{v_\alpha v_\beta}\,
  G^\mathrm{R}(\alpha,\beta;\varepsilon)
  \:,
\end{equation}
  where $v_\alpha$ is the group velocity in terminal $\alpha$ and where the retarded Green's function is measured at the two terminals,
  makes the functional derivation transparent~:
\begin{equation}
    \derivf{\Sm_{\alpha\beta}}{\PotEnerg(x)}
    =\I\sqrt{v_\alpha v_\beta}\,
  G^\mathrm{R}(\alpha,x;\varepsilon)G^\mathrm{R}(x,\beta;\varepsilon)
\:.
  \end{equation}
  Cf. also the discussion in \cite{TexBut03}.  
  
\item
  An illustration~:
  the knowledge of the injectivities allows to express the density of electrons in the conductor out-of-equilibrium 
  \begin{equation}
     \label{eq:DensityOOE}
     n(x) = \sum_\alpha\int\D\varepsilon\,f(\varepsilon-eV_\alpha)\,
     \overline{\DoS}_{\alpha}(x;\varepsilon) 
     \:,
  \end{equation}  
  where $f(\varepsilon)$ is the Fermi function and $V_\alpha$ the electric potential at contact $\alpha$.  
  
\item
  The introduction of the injectivities allows to avoid the approximation made in \eqref{eq:KreinFriedel} by neglecting the term $\tr{\Sm-\Sm^\dagger}/\varepsilon$ of Eq.~\eqref{eq:TexierButtiker2003}, and provides a better definition for the partial DoS, injectance and emittance.
  Introducing a uniform potential $\PotEnerg(x)=U$ \textit{inside} the conductor (but not in the contacts), we can write 
\begin{equation}
  -\frac{1}{2\I\pi} \left(  \Sm^\dagger \derivp{\Sm}{U} \right)_{\alpha\beta}
  = \int_\mathrm{QD}\D x\,\psi^*_{\varepsilon,\alpha}(x)\psi_{\varepsilon,\beta}(x)
  \:,
\end{equation}
where $\int_\mathrm{QD}\cdots$ denotes integration inside the conductor [the boundaries of integration are given by the place where the scattering states are matched with plane waves in order to define scattering amplitudes, like in Eq.~\eqref{eq:ScatteringState}].
In other terms~\cite{TexBut03,TexDeg03}
\begin{equation}
 - \Sm^\dagger \derivp{\Sm}{U}
 = \Sm^\dagger \derivp{\Sm}{\varepsilon} 
 + \frac{\Sm-\Sm^\dagger}{4\varepsilon}
 \:.
\end{equation}
  Therefore, instead of \eqref{eq:Injectance} which involves a high energy approximation, a more rigorous definition of the injectance should be 
  \begin{equation}
  \overline{\DoS}_{\alpha}(\varepsilon) 
  =-\frac{1}{2\I\pi}\left(\Sm^\dagger\derivp{\Sm}{ U}\right)_{\alpha\alpha}  
  \:.
  \end{equation}
  The remark also holds for partial DoS and emittances, Eqs.~(\ref{eq:PartialDoS},\ref{eq:Emittance}).
  The DoS takes the form 
  $\DoS(\varepsilon) =-(2\I\pi)^{-1}\partial_U\ln\det\Sm$.

\end{enumerate}

\subsection{A generalisation of the Feynman-Hellmann theorem for a continuous spectrum}

B\"uttiker's idea of relating the spectral properties to the scattering matrix is not limited to the DoS and the local DoS and can be generalised.
This may be viewed as an extension of the famous Feynman-Hellmann theorem which applies to bounded problems with discrete energy spectra.
Let us first recall this well-known theorem~:
consider a Hamiltonian $H$ characterised by the discrete spectrum $\big\{\varepsilon_n,\,\ket{\psi_n}\big\}$. 
The determination of the diagonal matrix elements of an observable $X=-\partial H/\partial f$, where $f$ is a conjugate force, does not require the knowledge of the eigenvectors but only of the eigenvalues~: $\bra{\psi_n}X\ket{\psi_n}=-\partial\varepsilon_n/\partial f$.
In Ref.~\cite{TexDeg03}, we proposed that an extension of this theorem to the case of Hamiltonians with continuous spectra, characterised by scattering stationary states $\big\{\ket{\psi_{\varepsilon,\alpha}}\big\}$, is  
\begin{equation}
  \label{eq:TexierDegiovanni2003}
  \bra{\psi_{\varepsilon,\alpha}}X\ket{\psi_{\varepsilon,\beta}}
  = \frac{1}{2\I\pi} 
  \left(
    \Sm^\dagger(\varepsilon) \derivp{\Sm(\varepsilon) }{f} 
  \right)_{\alpha\beta}
  \:.    
\end{equation}
This relation was proved in \cite{TexBut03,TexDeg03} for various observables in metric graphs~:~\footnote{
  Note that the expression~\eqref{eq:TexierDegiovanni2003} does not account for the contributions of BICs if present, as the DoS~\cite{TexDeg03}.
}
for example, choosing the local density $X\to\rho(x)=\ket{x}\bra{x}$ with $f\to-\PotEnerg(x)$, we recover \eqref{eq:NonDiagElements}.
In \cite{TexDeg03}, we considered also the case of the current density $I_a$ in a wire $(a)$ of the graph, which involves a derivation with respect to the magnetic flux $\phi_a$ along the wire
$\bra{\psi_{\varepsilon,\alpha}}I_a\ket{\psi_{\varepsilon,\beta}}=(2\I\pi)^{-1}\big(\Sm^\dagger\partial\Sm/\partial\phi_a\big)_{\alpha\beta}$.
A trace over indices leads to the well-known formula obtained by Akkermans, Auerbach, Avron and Shapiro \cite{AkkAueAvrSha91}~\footnote{See also the comment \cite{ComMorOuv95}}~:
$\mean{I}_\varepsilon=(2\I\pi)^{-1}\partial_\phi\ln\det\Sm$.
Note that similar considerations were used in order to analyse current fluctuations at equilibrium in Ref.~\cite{Tan01} (a formula for current correlations out-of-equilibrium was obtained in \cite{TexDeg03}).

\subsection{Statistical analysis in quantum dots}

\subsubsection{Mean values and covariances}

  In the case of chaotic quantum dots with several contacts with perfect couplings (Fig.~\ref{fig:qd}), Brouwer and B\"uttiker \cite{BroBut97} obtained the first two moments of the partial DoS \eqref{eq:PartialDoS} within the ASA of random matrices~:~\footnote{
  Brouwer and B\"uttiker considered the dimensionless AC conductances $G_{\alpha\beta}(\omega)=G_{\alpha\beta}(0)-\I\omega\,E_{\alpha\beta}+\mathcal{O}(\omega^2)$, which, in the unscreened limit, can be related to the partial DoS \eqref{eq:PartialDoS}~:
  $
  \lim_{C\to\infty}E_{\alpha\beta}=
  E_{\alpha\beta}^\mathrm{u}
  =2\pi\int\D\varepsilon\,(-\partial_\varepsilon f)\,\DoS_{\alpha\beta}(\varepsilon)$.
}
  \begin{align}
    \label{eq:MeanPDOS}
    \mean{\DoS_{\mu\nu}}
    =\frac{1}{\Delta}
    \bigg[
      \frac{N_\mu N_\nu}{N^2}
      -\frac{1}{N}\left(\frac{2}{\beta}-1\right)
      \left(
         \frac{N_\mu N_\nu}{N^2} - \frac{N_\mu}{N}\delta_{\mu\nu}
      \right)
     \bigg]
  \end{align}
  where $N_\mu$ is the number of open channels at contact $\mu$ and $N=\sum_\mu N_\mu$.
  We deduce $\smean{\overline{\DoS}_{\alpha}}=\smean{\underline{\DoS}_{\alpha}}=(N_\alpha/N)\,\Delta^{-1}$, as expected in an ergodic device.
   
  In the unitary case ($\beta=2$), the covariances are~\cite{BroBut97}
  \begin{align}
    \label{eq:CovariancePDOS}
    &\mathrm{Cov}(\DoS_{\mu\nu},\DoS_{\rho\sigma})
    = \frac{1}{(N\Delta)^2}
    \bigg[
      \frac{N_\mu N_\nu}{N^2}
        \left(
             \delta_{\mu\rho}\frac{N_\sigma}{N}
           + \delta_{\nu\sigma}\frac{N_\rho}{N}
        \right)   
      \nonumber\\
      &+
      \frac{3}{2}
      \left(
         \frac{N_\mu N_\rho}{N^2} - \frac{N_\mu}{N}\delta_{\mu\rho}
      \right)
      \left(
         \frac{N_\nu N_\sigma}{N^2} - \frac{N_\nu}{N}\delta_{\nu\sigma}
      \right)
     \bigg]
     \:.
  \end{align}
  The correlator for the orthogonal case ($\beta=1$) is obtained by adding similar terms obtained by permutation $\rho\leftrightarrow\sigma$ in order to fulfill the symmetry $\DoS_{\rho\sigma}=\DoS_{\sigma\rho}$ (for $\mathcal{B}=0$)~; the symplectic case receives additionally a factor $1/4$. 
  Eq.~\eqref{eq:CovariancePDOS} shows that mesoscopic fluctuations are of order 
  $\delta\DoS_{\alpha\beta}\sim \smean{\DoS_{\alpha\beta}}/N\sim1/(N\Delta)$.
  Note that \eqref{eq:CovariancePDOS} corresponds to the leading order terms at large $\Nc$ (the remark also holds for the remaing expressions given in the subsection).

  The correlations of injectances and emittances are also of interest.
  We deduce (for $\beta=1,\:2$)
  \begin{align}
    \label{eq:CovarianceInjectances}
    &\mathrm{Cov}(\overline{\DoS}_\mu,\overline{\DoS}_\nu)
    =\mathrm{Cov}(\underline{\DoS}_\mu,\underline{\DoS}_\nu)
    \nonumber\\
    =&\frac{1}{(N\Delta)^2}
    \left[
       \frac{N_\mu N_\nu}{N^2} + \delta_{\mu\nu}\frac{N_\mu}{N}
       +\left(\frac{2}{\beta}-1\right)
       \frac{2N_\mu N_\nu}{N^2}
    \right]
  \end{align}
  and 
  \begin{align}
    \label{eq:CovarianceInjectanceEmittance}
    &\mathrm{Cov}(\overline{\DoS}_\mu,\underline{\DoS}_\nu)
    =\frac{1}{(N\Delta)^2}
    \nonumber\\
    &\times
    \left[
      \frac{2N_\mu N_\nu}{N^2}
       +\left(\frac{2}{\beta}-1\right)
       \left(
         \frac{N_\mu N_\nu}{N^2} + \delta_{\mu\nu}\frac{N_\mu}{N}
       \right)
    \right]
    \:.
  \end{align}
  The interchange of the two contributions between \eqref{eq:CovarianceInjectances} and \eqref{eq:CovarianceInjectanceEmittance} is analogous to the exchange between Diffuson and Cooperon in weakly disordered metals \cite{TexMit18}.
  These expressions allow us to quantify the difference between injectance and emittance
  \begin{equation}
  \smean{(\overline{\DoS}_\alpha-\underline{\DoS}_\alpha)^2}
  =\left(2-\frac{2}{\beta}\right)\frac{2}{(N\Delta)^2}
  \frac{N_\alpha}{N}\left(1 - \frac{N_\alpha}{N}\right)
  \:.
  \end{equation}
  The difference vanishes in the orthogonal case ($\beta=1$), as a consequence of the symmetry \eqref{eq:SymmetryInjectanceEmittance}.
 This quantity has found some application in \cite{SanBut04}.


\subsubsection{Distribution of the injectance/emittance in the unitary case}

As mentioned at the end of Subsection~\ref{subsec:TDPC}, the full distribution of the injectance has been obtained in an Appendix of Ref.~\cite{GraMajTex17b} in the unitary case. 
This analysis was based on a result of Ref.~\cite{SavFyoSom01} for the distribution of a diagonal subblock $\WSm_K$ of size $K\times K$ of the $\Nc\times\Nc$ Wigner-Smith matrix $\WSm$ (we recall that $\WSm$ and $\WSm_s$ have the same statistical properties only in the unitary case).
Denote $\Gamma_K=\WSm_K^{-1}$, its distribution is~\cite{SavFyoSom01,GraMajTex17b}
$P(\Gamma_K)\propto\big(\det\Gamma_K\big)^\Nc\exp\big[-\tr{\Gamma_K}\big]$.
We now introduce the dimensionless variable
\begin{equation}
  s = \Delta\,\overline{\DoS}_\alpha= \frac{\Delta}{2\pi}\tr{\WSm_K} =  \frac{1}{\Ht}\sum_{a=1}^K \WSm_{aa} 
  \:.
\end{equation}
The distribution was obtained in the large $\Nc$ limit, with $\kappa=K/\Nc$ fixed \cite{GraMajTex17b}~:
\begin{align}
  \label{eq:DistributionInjectanceGMT2017}
  P^{(\beta=2)}_{\Nc,\kappa}(s) 
  \underset{\Nc\to\infty}{\sim}
  \begin{cases}
    s^{\Nc^2\kappa(\kappa+2)/2}\EXP{-\kappa\Nc^2/s}
    & \mbox{for } s\to0
    \\
    \exp\left\{ -\Nc^2\frac{(s-\kappa)^2}{2\kappa(\kappa+1)} \right\}
    & \mbox{for } s\sim\kappa
    \\
    s^{-\Nc-2}
    & \mbox{for } s\to\infty
  \end{cases}
\end{align}
The variance $\mathrm{Var}(s)\simeq\kappa(\kappa+1)/\Nc^2$ corresponds with \eqref{eq:CovarianceInjectances}.
In the limit $\kappa\to1$, the distribution \eqref{eq:DistributionInjectanceGMT2017} coincides with \eqref{eq:DistributionWTDTexierMajumdar2013}.

In the unitary case, $\WSm_s$, $\WSm=\Sm^{-1/2}\WSm_s\Sm^{1/2}$ and $\widetilde\WSm=\Sm^{1/2}\WSm_s\Sm^{-1/2}$ have the same distribution,~\footnote{This follows from the fact that eigenvectors of $\WSm$ and $\Sm$ are independent for $\beta=2$ \cite{BroFraBee99}.} thus \eqref{eq:DistributionInjectanceGMT2017} is also the distribution of the emittance.

\section{Non-linear transport in coherent conductors}
\label{sec:Nonlinear}

This section gives a brief presentation of the theory of non-linear transport in coherent conductors proposed by B\"uttiker and Christen~\cite{But93,ChrBut96a,ButChr97},
illustrating the interest of the quantities introduced in the previous section.
We consider the case of conductors of mesoscopic dimensions such that they can be considered in the ergodic regime (like quantum dots), which slightly simplifies the presentation, although the theory was developed in a more general context in Ref.~\cite{But93,ChrBut96a} (see also \cite{Tex10hdr}).

\begin{figure}[!ht]
\centering
\includegraphics[width=0.2\textwidth]{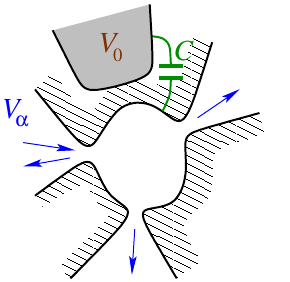}
\caption{\it A conductor of mesoscopic dimension connected to $M=3$ macroscopic reservoirs and capacitively coupled to a gate.
When the potential at contact $\alpha$ is raised, current is injected in the conductor from this contact (blue arrows).}
\label{fig:qd}
\end{figure}

\subsection{Non-linear conductances}

We consider a multiterminal mesoscopic structure (Fig.~\ref{fig:qd}). 
For small voltages $V_\alpha\to0$, current in terminal $\alpha$ can be written under the form of an expansion
\begin{equation}
  I_\alpha = \frac{2_se^2}{h}\sum_\beta g_{\alpha\beta} \,V_\beta
  + \frac{2_se^3}{h}\sum_\beta g_{\alpha\beta\gamma} \,V_\beta V_\gamma
  + \cdots
\end{equation}
where $2_s$ is the spin degeneracy.
$g_{\alpha\beta}$ is the dimensionless Landauer-B\"uttiker linear conductance \cite{But92}
\begin{equation}
  g_{\alpha\beta} = \int\D\varepsilon\,\left(-\derivp{f}{\varepsilon}\right)\,
  g_{\alpha\beta}(\varepsilon)
  \:.
\end{equation}
$f(\varepsilon)$ is the Fermi distribution and $g_{\alpha\beta}(\varepsilon)=N_\alpha\,\delta_{\alpha,\beta}-\mathrm{Tr}\big\{\Sm_{\alpha\beta}^\dagger(\varepsilon)\Sm_{\alpha\beta}(\varepsilon)\big\}$ the zero temperature dimensionless conductance, $N_\alpha$ being the number of conducting channels in contact $\alpha$.

The Landauer-B\"uttiker formula $I_\alpha=(2_se/h)\sum_\beta\int\D\varepsilon\,f(\varepsilon-eV_\beta)\,g_{\alpha\beta}(\varepsilon)$ already produces a contribution to the non-linear conductance
\begin{equation}
  \label{eq:NonInteractingNLC}
  g_{\alpha\beta\gamma}^0
  =\frac{1}{2}\delta_{\beta\gamma}
  \int\D\varepsilon\,\left(-\derivp{f}{\varepsilon}\right)\,
  g_{\alpha\beta}'(\varepsilon)
  \:.
\end{equation}
B\"uttiker emphasized the importance of screening effect which produces a second contribution that we now determine. The first step is to analyse the electrostatic inside the conductor.

\subsection{Characteristic potentials}

Let us recall the main ingredients involved in the description of screening in good metals~\cite{Zim72}, which B\"uttiker adapted to deal with the non-equilibrium situation.
These analysis is based on three ingredients.
(\textit{i}) 
When charges are introduced in the conductor, the total density of carriers $\delta n_\mathrm{tot}(x)=\delta n_\mathrm{ext}(x)+\delta n_\mathrm{ind}(x)$ is the sum of the density $\delta n_\mathrm{ext}(x)$ of charges injected in the conductor and the density $\delta n_\mathrm{ind}(x)$ induced by screening.
The concept of injectance introduced above allows us to write the number of injected charges when the conductor is out-of-equilibrium. An expansion of $\delta n_\mathrm{ext}(x)=n(x)-n_\mathrm{equil}(x)$ where $n(x)$ is given by \eqref{eq:DensityOOE} and $n_\mathrm{equil}(x)$ is the equilibrium density, gives
\begin{equation}
  \label{eq:ExternalCharge}
  \delta n_\mathrm{ext}
  =\int_\mathrm{QD}\D x\,\delta n_\mathrm{ext}(x) 
  \simeq \sum_{\alpha=1}^M  eV_\alpha \,
  \overline{\DoS}_{\alpha}(\varepsilon_F) 
  \:,
\end{equation}
where the summation runs over the $M$ contacts.
We assume zero temperature for simplicity~; the finite temperature formula involves an additional convolution with a Fermi function $\int\D\varepsilon\,\big(-\partial{f}/\partial{\varepsilon}\big)\cdots$.
(\textit{ii}) The induced density is related to the potential energy $\PotEnerg(x)$ by linear response theory
\begin{equation}
  \label{eq:LinearResponse}
  \delta n_\mathrm{ind}(x) = -\int\D x'\,\Pi(x,x')\,\PotEnerg(x')
\end{equation}
where $\Pi(x,x')$ is the (static) density-density correlation function of the non interacting electron gas (Lindhard function~\cite{BruFle04}).
Assuming ergodic properties, the potential energy can be considered uniform in the conductor $\PotEnerg(x)\simeq U$.
Thomas-Fermi approximation relates the response to the DoS~:
\begin{equation}
  \label{eq:ThomasFermi}
  \delta n_\mathrm{ind}
  =\int_\mathrm{QD}\D x\,\delta n_\mathrm{ind}(x) \simeq -\DoS(\varepsilon_F)\, U
  \:.
\end{equation}
(\textit{iii}) 
The potential energy and the density are related by the Poisson equation $\Delta \PotEnerg(x)=-4\pi e^2\,\delta n_\mathrm{tot}(x)$ which encodes the Coulomb interaction in the conductor.
The fact that \eqref{eq:LinearResponse} involves the potential $\PotEnerg(x)$ and not the electrostatic potential related to the external density $\delta n_\mathrm{ext}(x)$ makes the approach \textit{self-consistent}.
For a uniform potential, the Poisson equation is replaced by 
\begin{equation}
  \label{eq:Coulomb}
  C\, (U-eV_0) = e^2(\delta n_\mathrm{ext}+\delta n_\mathrm{ind})
\end{equation}
where $C$ is the capacitance of the conductor and $V_0$ the electrostatic potential of the gate, labelled with index $0$ (Fig.~\ref{fig:qd}).
Excess charge inside the conductor occurs when the system is driven out-of-equilibrium by the external potentials. At linear order, it is convenient to decompose the potential over contributions of the different external voltages
\begin{equation}
  \label{eq:CharactPot}
  U \simeq  \sum_{\alpha=0}^M  u_\alpha\, eV_\alpha
\end{equation}
where the ``\textit{characteristic potential}'' $u_a$ measures the response of the potential $U$ to a shift of the external voltage $V_\alpha$ \cite{But93,ChrBut96a} (one should not forget the response of the gate voltage controlled by $u_0$).
Injecting (\ref{eq:ExternalCharge},\ref{eq:ThomasFermi},\ref{eq:CharactPot}) in \eqref{eq:Coulomb} provides the characteristic potentials
\begin{equation}
  \label{eq:ExpressionCharacPot}
  u_\alpha 
  = \frac{C_\mu}{C}\,\frac{\overline{\DoS}_{\alpha}(\varepsilon_F) }{\DoS(\varepsilon_F) }
\end{equation}
for $\alpha\in\{1,\dots,M\}$ 
and 
\begin{equation}
  u_0 = 1 - \frac{C_\mu}{C} = \frac{C_\mu}{C_q}
  \:.
\end{equation}
We have introduced the ``\textit{mesoscopic capacitance}''
\begin{equation}
  \label{eq:CapaMeso}
  \frac{1}{C_\mu} = \frac{1}{C} + \frac{1}{C_q}
  \:,
\end{equation}
combining the ``\textit{geometric capacitance}'' $C$ and the ``\textit{quantum capacitance}''
\begin{equation}
  \label{eq:QuantumCapa1}
  C_q = e^2\,\DoS(\varepsilon_F)
  \:.
\end{equation}
We can check the sum rule 
\begin{equation}
  \sum_{\alpha=0}^M u_\alpha =1
  \:.
\end{equation}

The ratio $\gamma_\mathrm{int}=C_\mu/C$ in Eq.~\eqref{eq:ExpressionCharacPot} may be viewed as an interaction constant as it measures the efficiency of screening ($\gamma_\mathrm{int}\simeq1$ for efficient screening in a good metal and $\gamma_\mathrm{int}\ll1$ for weak screening).
The response of the gate $u_0=1-\gamma_\mathrm{int}$ is thus important when screening is weak, however the presence of the gate (or the surrounding medium) can be forgotten in good metals.

\subsection{Non-linear conductances}

We can now determine the non-linear conductances as follows.
The redefinition of the electrostatic potential inside the conductor leads to a modification of its scattering properties which depends on the potential, what contributes to the non-linear response~:
$g_{\alpha\beta}(\varepsilon_F)\to g_{\alpha\beta}(\varepsilon_F-U)\simeq g_{\alpha\beta}(\varepsilon_F)-g'_{\alpha\beta}(\varepsilon_F) \sum_\gamma u_\gamma\, eV_\gamma$.
After symmetrisation with respect to indices $\beta\leftrightarrow\gamma$ and reintroducing the Fermi function, we get the expression of the non-linear conductance~\cite{ChrBut96a,PolBut06}
\begin{align}
  \label{eq:NonLinearConductances}
  g_{\alpha\beta\gamma} &= \frac12\int\D\varepsilon\left(-\derivp{f}{\varepsilon}\right)\,
  \\\nonumber
  &\times
  \left[
    g_{\alpha\beta}'(\varepsilon)\,\delta_{\beta\gamma}
    -g_{\alpha\beta}'(\varepsilon)\, u_\gamma
    -g_{\alpha\gamma}'(\varepsilon)\, u_\beta
  \right]
  \:,
\end{align}
where the first term comes from the noninteracting theory, Eq.~\eqref{eq:NonInteractingNLC}.
For consistency, the $T=0$ expression~\eqref{eq:ExpressionCharacPot} should be replaced by 
\begin{equation}
 \label{eq:CharacteristicPotentials}
  u_\alpha = \frac{\int\D\varepsilon\,(-\partial_\varepsilon f)\,\overline{\DoS}_\alpha(\varepsilon)}
    {C/e^2+\int\D\varepsilon\,(-\partial_\varepsilon f)\,\DoS(\varepsilon)}
   \:.
\end{equation}
The second term of the denominator is the finite temperature quantum capacitance 
$C_q=e^2\int\D\varepsilon\,(-\partial_\varepsilon f)\,\DoS(\varepsilon)$.

Few remarks~:
\begin{enumerate}[(i)]
\item
  As emphasized above, all quantities involved in Eqs.~(\ref{eq:NonLinearConductances},\ref{eq:CharacteristicPotentials}), $g_{\alpha\beta}(\varepsilon)$, $\DoS(\varepsilon)$ and $\overline{\DoS}_\alpha(\varepsilon)$, are expressed in terms of the $\Sm$-matrix.
  The derivative of the conductance $g'_{\alpha\beta}(\varepsilon)$ can also be related to the \textit{sensitivies} introduced by Gasparian, Christen and B\"uttiker~\cite{GasChrBut96}~:
\begin{equation}
  \label{eq:Sensitivies}
    \eta_{\alpha\beta}(\varepsilon) 
  \simeq \frac{1}{4\pi} 
  \left(
    \Sm_{\alpha\beta}^*\derivp{\Sm_{\alpha\beta}}{\varepsilon} 
    + \derivp{\Sm_{\alpha\beta}^*}{\varepsilon}\Sm_{\alpha\beta}
  \right)
\end{equation}
  (note that Ref.~\cite{GasChrBut96} rather introduced a local version of the sensitivies).

\item 
  A statistical analysis of the non-linear conductance \eqref{eq:NonLinearConductances} in chaotic quantum dots  requires the statistical properties of the conductance's derivative provided in Ref.~\cite{BroLanFraButBee97} and those of the characteristic potentials, i.e. of the injectivities, obtained in \cite{BroBut97}, see Eqs.~(\ref{eq:MeanPDOS},\ref{eq:CovariancePDOS}).
\end{enumerate}

\subsection{Recent developments}

\begin{enumerate}[(i)]
\item 
As mentioned, the B\"uttiker's scattering formalism can be extended to discuss conductors which are not in the ergodic regime~\cite{But93,ChrBut96a,ButChr97}.
The equivalence between this formalism and the non equilibrium Green's function approach for non-linear transport was established by Hern\'andez and Lewenkopf~\cite{HerLew13} (see also \cite{TexMit18}).

\item 
An interesting aspect was identified by S\'anchez and B\"uttiker~\cite{SanBut04} and Spivak and Zyuzin~\cite{SpiZyu04}~:  the interacting part of the non-linear conductance 
$g_{\alpha\beta\gamma}^\mathrm{int} 
= -(1/2)\big[g_{\alpha\beta}'(\varepsilon_F)\, u_\gamma
   +g_{\alpha\gamma}'(\varepsilon_F)\, u_\beta\big]$
is not constrained to any specific symmetry under reversal of the magnetic field, contrary to the non-interacting part $g_{\alpha\beta\gamma}^0=(1/2)g_{\alpha\beta}'(\varepsilon_F)\delta_{\beta\gamma}$. This is due to the asymmetry of the characteristic potentials $u_\alpha$ (i.e. of the injectance $\overline{\DoS}_\alpha$).
The analysis of the asymmetry of the non-linear conductance under magnetic field reversal 
$\big[g_{111}(\mathcal{B})-g_{111}(-\mathcal{B})\big]/2=-(C_\mu/C)\,g'_{11}(\varepsilon_F)\,\big(\overline{\DoS}_1-\underline{\DoS}_1\big)/(2\DoS)$
was thus proposed as a new way for probing electronic interactions in mesoscopic structures (the parameter $\gamma_\mathrm{int}=C_\mu/C$ introduced above was measured in \cite{AngZakDebGueBouCavGenPol07} for a mesoscopic ring).
Spivak and Zyuzin borrowed arguments from diagrammatic techniques in order to analyse the low magnetic field regime~\cite{SpiZyu04}.
S\'anchez and B\"uttiker~\cite{SanBut04} proposed a random matrix approach in the unitary case (strong magnetic field), which was later improved by B\"uttiker and Polianski who extended the statistical analysis in order to describe the crossover from orthogonal to unitary cases and include thermal effects~\cite{PolBut06,PolBut07} (for more details, cf. the excellent review article~\cite{PolBut07a}).
Note also the study for a ring made of strictly 1D wires~\cite{HerLew09}.

\item
Several experimental groups have analysed the non-linear transport in mesoscopic structures and specifically the asymmetry in magnetic field \cite{LetSanGotIhnEnsDriGos06,MarTayFaiShoLin06,ZumMarHanGos06,AngZakDebGueBouCavGenPol07,Ang07}.

\item
The case of disordered wires in the diffusive regime was analysed in Ref.~\cite{TexMit18}.
\end{enumerate}

\section{AC transport}
\label{sec:ACtransport}

\begin{figure}[!ht]
\centering
\includegraphics[width=0.35\textwidth]{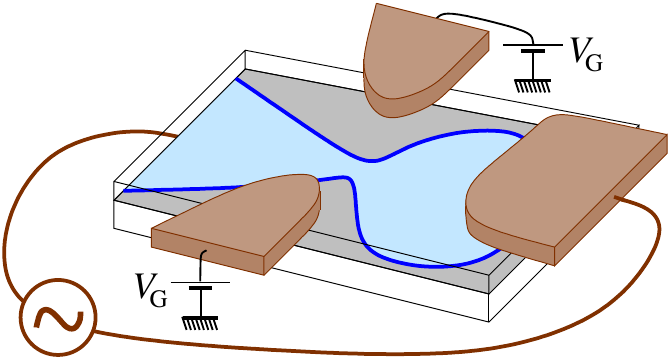}
\caption{\it The quantum RC circuit~: a coherent conductor patterned in a 2DEG (blue) is closed by a quantum point contact (QPC) controlled by side gates. 
A capacitive coupling to a top gate allow to analyse its AC response.
Figure from Ref.~\cite{GraTex15}.}
\label{fig:QRCcircuit}
\end{figure}

The search for fast control and manipulation of charge in coherent conductor has stimulated many developments (see the reviews \cite{GabFevBerPla12,BocFreParBerPlaWahRecJonMarGreFerDegFev14}).
B\"uttiker, Pr\^etre and Thomas have proposed a theory of time-dependent response in coherent conductors \cite{ButPreTho93,ButThoPre93,ButThoPre94} based on the scattering approach and a Hartree-Fock treatment (with Thomas-Fermi approximation) of electronic interactions (cf. chapter~1 of Ref.~\cite{Tex10hdr}).
In order to emphasize few ideas, we will restrict ourselves here to the case of the ``\textit{quantum RC circuit}'' which was studied experimentally in the integer quantum Hall regime during the PhD of Gabelli \cite{Gab06,Fev06,GabFevBerPlaCavEtiJinGla06} (Fig.~\ref{fig:QRCcircuit}). 
The RC circuit was later studied from the perspective of an emitter of electronic wave packets in the PhD of F\`eve \cite{Fev06} and as such is a building block for ``electron optics experiment'' (see the review \cite{BocFreParBerPlaWahRecJonMarGreFerDegFev14} and article \cite{HofDasFli16} of this special issue).
These latter developments have mostly considered the RC circuit in the integer quantum Hall regime, such that the current is carried by a single edge state along the boundary of the system, when one conducting channel is open at the constriction (quantum point contact, QPC).

Below, we do not consider the integer quantum Hall regime but we rather discuss the situation where the RC circuit of Fig.~\ref{fig:QRCcircuit} is submitted to a weak magnetic field, such that the electron dynamics inside the cavity is chaotic. This justifies a random matrix approach.
In this case, the opening of the constriction controls the number of channels~$N$.

\subsection{The AC response}
\label{subsec:ACresponse}

By using arguments similar to the ones described in Section~\ref{sec:Nonlinear}, B\"uttiker and coworkers have shown that the Coulomb interaction can be taken into account by adding the impedance of the non interacting electrons, $1/G_0(\omega)$, and the classical impedance of the capacitance
$Z(\omega)\equiv1/G(\omega)=1/G_0(\omega)+1/(-\I\omega C)$. The admittance is  \cite{ButPreTho93}
\begin{align}
  \label{eq:Admittance}
  G_0(\omega)=\frac{2_se^2}{h}
  \int\D\varepsilon\,
 & \tr{1- \Sm^\dagger(\varepsilon)\Sm(\varepsilon+\omega) }	
  \nonumber\\
  &\times\frac{f(\varepsilon)-f(\varepsilon+\omega)}{\omega}
  \:,
\end{align}
where $2_s$ denotes the spin degeneracy.
One can write the low frequency expansion of the impedance under the form given by the elementary laws of electrokinetics
\begin{equation}
  \label{eq:Impedance}
  Z(\omega) = \frac{1}{-\I\omega C_\mu} + R_q + \mathcal{O}(\omega)
  \:,
\end{equation}
where the ``\textit{mesoscopic capacitance}'' $C_\mu$ and the ``\textit{charge relaxation resistance}'' $R_q$ carry some information about the dynamics of electrons in a quantum coherent regime, as they control the relaxation time $\tau_{RC}=R_qC_\mu$.
The presence of the mesoscopic capacitance~\eqref{eq:CapaMeso} in the AC response comes from the $\omega\to0$ expansion of $G_0(\omega)$, 
whose first terms involve the Wigner-Smith matrix \eqref{eq:DefinitionWS}.  
Hence, this latter controls both the quantum capacitance \eqref{eq:QuantumCapa1}
\begin{equation}
  \label{eq:DefCq}
  C_q = \frac{2_se^2}{h}\tr{\WSm}
\end{equation}
and the charge relaxation resistance 
\begin{equation}
  \label{eq:DefRq}
  R_q = \frac{h}{2_s\times2e^2}\frac{\tr{\WSm^2}}{(\tr{\WSm})^2}
  \:.
\end{equation}  
In these equations, the Wigner-Smith matrix is taken at Fermi energy $\varepsilon_F$.

\subsection{The charge relaxation resistance}

In this section \textit{we forget about spin degeneracy factor $2_s$} and base the discussion on the expression 
\begin{equation}
  \label{eq:DefRq2}
  R_q = \frac{h}{2e^2}\frac{\tr{\WSm^2}}{(\tr{\WSm})^2}
  = \frac{h}{2e^2} \frac{\sum_a\tau_a^2}{\big(\sum_a\tau_a\big)^2}
  \:.
\end{equation}  
A first important observation is that the resistance belongs to the interval~:
\begin{equation}
  \frac{h}{2 N e^2} \leq R_q \leq \frac{h}{2 e^2}
  \:.
\end{equation} 
In particular the $N=1$ channel case leads to the universal value $R_q = h/(2e^2)$~\cite{ButThoPre93} (this is a property of a coherent conductor, cf. remarks closing the section).

\paragraph{Case $N=2$ ---}
Assuming a random matrix description of the quantum dot, we can deduce the distribution of $R_q$ starting from \eqref{eq:BrouwerFrahmBeenakker1997}. The calculation only involves a double integral~\cite{GraMajTex16a,Gra18}~:
  \begin{align}
     p_2(\rr)= A_\beta\,(2\rr-1)^{(\beta-1)/2}(1-\rr)^\beta
  \end{align}
  where $r_q=(\Nc e^2/h)R_q$ is the dimensionless charge relaxation resistance \textit{per channel}~; $A_\beta=2^{\beta+1}/B(\beta+1,(\beta+1)/2)$ is a normalisation constant.
We deduce the mean value
\begin{equation}
  \mean{R_q} = \frac{h}{3e^2}
\end{equation}
which is surprisingly independent on the symmetry class.

\subparagraph{Remark~:}
The distribution of $R_q$ for $N=2$ channels was analysed by Pedersen, van Langen and B\"uttiker in \cite{PedLanBut98}, however these authors introduced a slightly different averaging procedure~:
  following \cite{BroLanFraButBee97} the authors weighted the joint distribution of proper times by the DoS, corresponding to a ``canonical'' averaging.
They deduced 
$p_2^{(\mathrm{PvLB})}(\rr)=B_\beta\,(2\rr-1)^{(\beta-1)/2}(1-\rr)^{\beta-1}$
where the normalisation is $B_\beta=2^{\beta}/B(\beta,(\beta+1)/2)$. 
As a result the mean resistance depends on the symmetry index
$\mean{R_q}^{(\mathrm{PvLB})}=(3/8)\,h/e^2$ for $\beta=1$ and $\mean{R_q}^{(\mathrm{PvLB})}=(5/14)\,h/e^2$ for $\beta=2$.

\paragraph{Mean value ---}
B\"uttiker, Pr\^etre and Thomas \cite{ButThoPre93} already pointed out that the charge relaxation resistance should scale with the number of channels as $R_q\sim1/\Nc$, which corresponds to the addition of resistances in parallel, although the precise prefactor was not given.
The fact that the one channel case (spin polarised) leads to the universal \textit{half} quantum of resistance $R_q=h/(2e^2)$ has produced a certain confusion for the large $\Nc$ case, between $R_q=h/(2\Nc e^2)$ (incorrect) and $R_q=h/(\Nc e^2)$ (correct).

The first precise calculation of the mean value $\mean{R_q}$ in the $\Nc\gg1$ limit is contained in the study of the complex admittance $G(\omega)$ by Brouwer and B\"uttiker \cite{BroBut97} within the ASA of random matrices~: 
the low frequency expansion of $\mean{G(\omega)}$ (Eq.~12 of Ref.~\cite{BroBut97}) leads to
\begin{equation}
  \mean{R_q} = \frac{h}{\Nc e^2}\,\left[1+\left(\frac2\beta-1\right)\frac{1}{\Nc}+\mathcal{O}(\Nc^{-2})\right]
  \:,
\end{equation}
(without spin degeneracy $2_s$). 
The second term can be interpreted as a weak localisation correction.~\footnote{
  Although the geometries are different, we can compare this expansion with those of the DC resistance of a coherent quantum dot closed by two constrictions with $N_1$ and $N_2$ channels \cite{Bee97}~:
  $\mean{R_\mathrm{dc}}=R_\mathrm{class}\,\big[1+(2/\beta-1)N^{-1}+\cdots\big]$
  where $R_\mathrm{class}=(h/e^2)\,(1/N_1+1/N_2)$ and $N=N_1+N_2$.
}
A first review paper by B\"uttiker \cite{But99} incorrectly quoted this result by introducing a spurious factor of $1/2$ (see Eq.~17 of Ref.~\cite{But99}).
In a second review \cite{But00}, B\"uttiker gave the correct result [Eq.~9 of this reference]
\begin{equation}
  \label{eq:BeenakkerButtikerTexier1999}
  \mean{R_q} \simeq \frac{h}{e^2}\, \frac{1}{\Nc\mathcal{T}}
\end{equation}
for a non perfect contact with $\Nc\mathcal{T}\gg1$ (without spin degeneracy), where $\mathcal{T}$ is the transmission probability through the contact. B\"uttiker refered to a private communication with Carlo Beenakker and some unpublished work with myself~\cite{TexBut99unpub}, that I describe in \ref{app:ButtikerTexier}. 
This value coincides with the DC resistance of the constriction, what was expected in the large $N$ limit.



We can summarize this discussion in the following table giving the dimensionless charge relaxation resistance \textit{per channel} $r_q=(Ne^2/h)R_q$~:
\begin{center}
\begin{tabular}{lc}
  & $\smean{r_q}=(Ne^2/h)\smean{R_q}$ \\
  \hline
$N=1$   & $1/2$ 
\\  
$N=2$   & $2/3$
\\
$N\gg1$ &  $1+ (2/\beta-1)N^{-1}$
\\
\hline 
\end{tabular}
\end{center}

\paragraph{Fluctuations ---}

As the quantum capacitance is directly proportional to the Wigner time delay, its statistical properties were determined in \cite{TexMaj13} for a contact with many channels.
More recently, we have studied with Grabsch \cite{GraTex15,Gra18} the statistical properties of the charge relaxation resistance, which involves the analysis of the ratio of two ``linear statistics'' $\sum_a\tau_a^2/\big(\sum_a\tau_a\big)^2$.
We have obtained the mean values
$\mean{C_q}\simeq e^2/\Delta$ and $\mean{R_q}\simeq h/(\Nc e^2)$ (without spin degeneracy), in correspondence with the DC resistance of the QPC, as it should.
The variance of the capacitance is given by \cite{LehSavSokSom95,BroBut97,TexMaj13,MezSim13,Cun15}
\begin{equation}
  \label{eq:VarianceCq}
  \frac{ \mean{\delta C_q^2} }{ \mean{C_q}^2 } \simeq \frac{4}{\beta\Nc^2}
\end{equation}
(this coincides with the leading order term of  cumulant \eqref{eq:MezzadriSimm2} for $N\gg1$). 
The variance of the resistance
\begin{equation}
  \label{eq:VarianceRq}
   \frac{ \mean{\delta R_q^2} }{ \mean{R_q}^2 } \simeq \frac{8}{\beta\Nc^2}
\end{equation} 
and the correlations with the quantum capacitance
\begin{equation}
  \label{eq:CovarianceCqRq}
  \frac{ \mean{\delta C_q\delta R_q} }{ \sqrt{\smean{\delta C_q^2}\smean{\delta R_q^2}} }
  =+1/\sqrt{2}
\end{equation}
were found in Ref.~\cite{GraTex15}, where we have also analysed the large deviations (see also \cite{Gra18} for further developments).

\subsection{Final remarks}

\begin{enumerate}[(i)]
\item 
  The multiterminal version of the general formulae reviewed in the section was provided in Refs.~\cite{ButThoPre93,ButThoPre94}.

\item 
  The finite temperature formulae are obtained by introducing convolutions with Fermi functions 
  $\int\D\varepsilon\,\big(-\partial f/\partial\varepsilon\big)$~:
  the traces over channel indices in Eqs.~(\ref{eq:DefCq},\ref{eq:DefRq}) are then affected by  additional integrations over energy.
   \\
  In particular, using the representation 
  $C_q=(e^2/h)\int\D\varepsilon\,(-\partial_\varepsilon f)\,\tr{\WSm(\varepsilon)}$ with the correlator \eqref{eq:FyodorovSavinSommers97VallejosOzorioLewenkopf98},
  the model leads to the following expression 
  \begin{align}
   & \frac{\mean{\delta C_q(\mathcal{B}) \delta C_q(\mathcal{B}')}}{\mean{C_q}^2}
   \simeq
    \int\D\omega\,\delta_T(\omega)\,
    \\ \nonumber 
    &\times
    \frac{1}{\Ht^2}
    \derivp{^2}{\omega^2}
    \ln\left[
      \left(1+(\omega\tau_{\mathscr{D}})^2\right)\left(1+(\omega\tau_{\mathscr{C}})^2\right)
 \right]
  \end{align}
  where $\delta_T(\omega)$ is a (normalised) thermal function of width $T$ which arises from the convolution of the two Fermi function's derivatives.~\footnote{
      $\delta_T(\omega)=F\big(\omega/(2T)\big)/(2T)$ with $F(x)=(x\coth x-1)/\sinh^2x$.
  }
  For $T\tau_d\gg1$ this leads to the decay 
  $\mean{\delta C_q(\mathcal{B})^2}/\mean{C_q}^2\simeq\big[\kappa/(\beta\Nc^2)\big] \, (T\tau_d)^{-2}$ where $\kappa=-\int_0^\infty\D x\,F'(x)/x\simeq0.365$ (the function $F(x)$ is defined in the footnote). The dependence in the Dyson index $\beta$ can be replaced by the precise magnetic field dependence straightforwardly.
  \\
  Note that in the metallic limit ($e^2/C\gg\Delta$, i.e. $C\ll C_q$) the fluctuation of the mesoscopic capacitance is $\delta C_\mu\simeq (C/\mean{C_q})^2\,\delta C_q$.

\item
  A more precise discussion of the capacitive response of quantum dots has been provided in Ref.~\cite{AleBroGla02}, beyond mean field and also analysing the limit of a weakly transmitting constriction in the presence of Coulomb blockade.

\item 
  Nigg and B\"uttiker have considered in Ref.~\cite{NigBut08} the case of a single spin polarised channel. They showed that strong enough dephasing in the cavity can induce the transition from the universal value $R_q=h/(2e^2)$ to the DC resistance $\rdc =h/(e^2\mathcal{T})$ of the QPC, where $\mathcal{T}$ is the transmission probability throuhg the QPC.

\item
  The universal charge relaxation resistance $R_q=h/(2e^2)$ for a single spin non degenerate channel  \cite{ButThoPre93,ButThoPre94,But00} has stimulated several works.
  This value was later obtained within more accurate treatments of electronic interactions~:
  within a mean-field Hartree-Fock analysis~\cite{NigLopBut06,RinImrEnt08} improving the original Thomas-Fermi treatment, and beyond mean field~\cite{HamJonKatMar10,MorLeh10}, within Matveev formalism \cite{Mat95} of QPC (see also~\cite{EtzHorLed11}).
  These treatments have accounted for possible Coulomb blockade effects.  
\\
  The treatment of interaction in a multichannel contact was extended more recently in \cite{DutSchMorLeh13}.

\end{enumerate}


\section{Conclusion}

Despite the huge diversity of B\"uttiker's work, this article has tried to show that a possible Ariane's thread is furnished by the concept of time delay and related quantities.
B\"uttiker's interest in this matter goes back to his early fundamental study of times in quantum mechanics~; this question has led to many possible definitions, from scattering phase shifts (Wigner time delay, transmission group delays,...), clock approaches, etc.
When randomness is introduced, these quantities are characterised by non-trivial distributions.
I have reviewed in Section~\ref{sec:DisorderedSystems} the statistical analysis for the most simple of these fundamental times in the 1D situation in the presence of disorder. 
In particular I have emphasized the role of symmetry by considering the effect of a chiral symmetry.
In this case, a new observation was the relation between the presence of a chiral zero mode and the existence of a limit law for the Wigner time delay when the size of the disordered region goes to infinity.

One of B\"uttiker's major motivation for Wigner-Smith time delay matrix' analysis comes from its central role in the scattering approach for non-linear (Section~\ref{sec:Nonlinear}) and AC (Section~\ref{sec:ACtransport}) coherent electronic transport.
This has led to introduce several new concepts (Section~\ref{sec:InjectanceEtc}) which have been analysed within the frame of random matrix theory~:
B\"uttiker, with Gopar and Mello, obtained one of the first important result in the Wigner time delay statistical analysis \cite{GopMelBut96}, motivated by the statistical analysis of the mesoscopic capacitance of a chaotic quantum dot.
The first moments of the AC response were obtained with Brouwer \cite{BroBut97} within the random matrix theory  and the non-linear response with S\'anchez and Polianski \cite{SanBut04,PolBut06}.
In the present article I have provided a review of the known results on time delay statistics within the random matrix approach~: 
joint distribution of proper times, moments of Wigner time delay and its correlation function and statistical properties of the partial DoS.


\section*{Acknowledgments}

The material reviewed in this article is based on work done with several collaborators who are deeply thanked~:
Markus B\"uttiker, Alain Comtet, Pascal Degiovanni, Aur\'elien Grabsch, Satya Majumdar, Gilles Montambaux and Dmitry Savin.
I thank Fabio Cunden for useful discussions.
I am indebted to Dmitry Savin for numerous insightful comments. 
I am grateful to Alain Comtet for several remarks on the manuscript.
Finally I thank the referee for helpful remarks.

\begin{appendix}


\section{Calculation of $\mean{R_q}$ for $N\gg1$ -- Mapping to the study of absorption in the cavity}
\label{app:ButtikerTexier}

This section reproduces some unpublished contribution due to myself and B\"uttiker \cite{TexBut99unpub} (quoted in  Ref.~\cite{But00} but remained unpublished so far). 

\begin{figure}[!ht]
\centering
\includegraphics[width=0.25\textwidth]{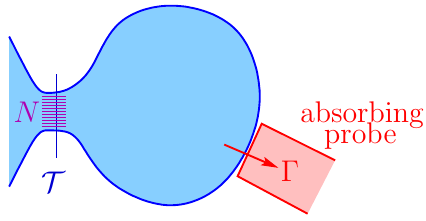}
\caption{\it A conductor closed by a constriction with $N$ channels (quantum point contact, QPC) and with an absorbing probe.}
\label{fig:condabsorb}
\end{figure}

Our aim is to estimate the charge relaxation resistance for a one contact conductor in the good contact limit $\Nc\mathcal{T}\gg1$, where $\Nc$ is the number of conducting channels and $\mathcal{T}$ the transmission probability through the QPC.
Our starting point is the expression \eqref{eq:DefRq2} (without spin degeneracy)~\cite{ButThoPre93}. 
As we mentioned in the text, the trace in the denominator is a measure of the density of states (DoS) of the cavity $\DoS\simeq\str{\WSm}/h$,  therefore $\smean{\str{\WSm}}=2\pi/\Delta$ where $\Delta$ is the mean level spacing and the main question left is to estimate $\str{\WSm^2}$. 
We remark that it appears in the expansion of the dimensionless AC conductance~\eqref{eq:Admittance}
\begin{align}
  \label{eq:ExpansionSdaggerS}
  G_0(\omega) &=\Nc - \str{\Sm^\dagger(\varepsilon-\frac12\omega)\Sm(\varepsilon+\frac12\omega)}
  \nonumber\\
  &= - \I\omega\,\str{\WSm}
  +\frac{\omega^2}{2}\,\str{\WSm^2}+\cdots
\end{align}
B\"uttiker suggested to use analytic properties of the scattering matrix, $\omega\to\I\Gamma$, i.e. to map the problem of AC transport onto the study of DC transport for a cavity in the presence of absorption, whose dimensionless conductance reads
\begin{align}
  \label{eq:AbsorbingContact}
  \widetilde{G}&=G_0(\I\Gamma) 
    \\\nonumber
  &= \Nc 
  - \tr{\left[\Sm\left(\varepsilon+\frac\I2\Gamma\right)\right]^\dagger\Sm\left(\varepsilon+\frac\I2\Gamma\right)}
  \\\nonumber
  &= \Gamma\,\str{\WSm}
  -\frac{\Gamma^2}{2}\,\str{\WSm^2}+\cdots
  \:.
\end{align}
This corresponds to a situation where absorption takes place uniformly in space with absorbing rate $\Gamma$.
In the ergodic regime where RMT holds, this also describes an absorbing contact (Fig.~\ref{fig:condabsorb}).~\footnote{
  The relation between the dynamical problem (real $\omega$) and the static problem with absorption (complex $\omega$) was used in another context in Refs.~\cite{KlySai92,RamKum00,Bee01,BeeBro01,Fyo03}. 
}

\begin{figure}[!ht]
\centering
\includegraphics[width=0.2\textwidth]{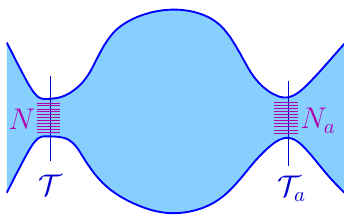}
\caption{\it A two terminal conductor.}
\label{fig:cond2term}
\end{figure}

The idea is then to estimate $\widetilde{G}$ by noticing that a conductor with one contact and an absorbing probe is equivalent to a two terminal conductor (Fig.~\ref{fig:cond2term}). In this latter case, the conductance is dominated by the resistance of the constrictions, hence
$1/\widetilde{G}=1/(\Nc\mathcal{T})+1/(N_a\mathcal{T}_a)$, 
where $N_a$ and $\mathcal{T}_a$ characterise the second contact.
We deduce
\begin{align}
  \label{eq:TwoTerminal}
  \Nc-\widetilde{G}=
  \Nc \, \frac{ \Nc\mathcal{T}+N_a\mathcal{T}_a(1-\mathcal{T})}{\Nc\mathcal{T}+N_a\mathcal{T}_a}  
  \:.
\end{align}

Expansion \eqref{eq:AbsorbingContact} in powers of $\Gamma$ and expansion of \eqref{eq:TwoTerminal} in powers of $N_a\mathcal{T}_a$ can be identified by setting $N_a\mathcal{T}_a=c\,(\Gamma/\Delta)$, where $c$ is a dimensionless constant. Hence
\begin{align}
  &
  \smean{\str{\big[\Sm(\varepsilon+\I\frac\Gamma2)\big]^\dagger\Sm(\varepsilon+\I\frac\Gamma2)}}
  =\Nc \frac{\Nc\mathcal{T}+(1-\mathcal{T})\,c\,\frac{\Gamma}{\Delta}}{\Nc\mathcal{T}+c\,\frac{\Gamma}{\Delta}}
  \nonumber\\
  &=
  \Nc - c\,\frac{\Gamma}{\Delta} + \frac{1}{\Nc\mathcal{T}} \left(c\,\frac{\Gamma}{\Delta} \right)^2+\mathcal{O}(\Gamma^3)
  \:.
\end{align}
Comparison with \eqref{eq:AbsorbingContact} and the fact that $\smean{\str{\WSm}}=2\pi/\Delta$ shows that $c=2\pi$. We deduce 
\begin{equation}
  \smean{\str{\WSm^2}} = \smean{\sum_i\tau_i^2} \simeq \frac{8\pi^2}{\Nc\mathcal{T}\Delta^2}
  \:,
\end{equation}
leading to \eqref{eq:BeenakkerButtikerTexier1999}.

\section{Partial \textit{versus} proper times}
\label{Appendix:PartialVsProper}

Let us make few remarks on the relation between partial time delays and proper time delays.
We consider a scattering situation with $\Nc$ channels.
The $\Sm$-matrix may be represented under the form
\begin{equation}
  \Sm=\mathcal{U}\,\EXP{\I\Theta}\,\mathcal{U}^\dagger
\end{equation}
where $\Theta=\mathrm{diag}(2\eta_1,\cdots,2\eta_N)$ gathers the $\Nc$ phase shifts.
We deduce the following representation for the Wigner-Smith matrix~:  
\begin{equation}
  \label{eq:ProperVsPartial}
  \WSm=
  \mathcal{U}\,(\partial_\varepsilon\Theta)\,\mathcal{U}^\dagger
  +\I\,\big[
    \Sm^\dagger (\mathcal{U}\partial_\varepsilon\mathcal{U}^\dagger) \Sm
    - \mathcal{U}\partial_\varepsilon\mathcal{U}^\dagger
  \big]
\end{equation}
The first term involves the partial time delays~:
$\partial_\varepsilon\Theta=\mathrm{diag}(\tilde{\tau}_1,\cdots,\tilde{\tau}_N)$, which are intrinsic characteristics of the scattering matrix (independent on the basis).
The second term controlled by $\mathcal{U}$, i.e. by the choice of basis, is the origin of the difference between partial and proper times.

We now say a little bit more on the unitary matrix $\mathcal{U}$ in the $N=2$ channel case. Going back to the notations introduced in \S~\ref{subsec:CharacteristicTimes}, we can rewrite the relation between partial waves~\eqref{eq:PartialWave} (eigenstates of the $\Sm$-matrix) and the left/right scattering states in the matricial form
\begin{equation}
  \begin{pmatrix}  
    \phi_{\varepsilon,1}(x) \\ \phi_{\varepsilon,2}(x)
  \end{pmatrix}
  = 
  \begin{pmatrix}
    A_{L,1} & A_{R,1} \\ A_{L,2} & A_{R,2} 
  \end{pmatrix}
  \begin{pmatrix}  
    \psi_{\varepsilon,L}(x) \\ \psi_{\varepsilon,R}(x)
  \end{pmatrix}
\end{equation} 
which involves the transpose of the matrix which diagonalises the $\Sm$-matrix
\begin{equation}
  \mathcal{U}(\varepsilon)
  =  \begin{pmatrix}
    A_{L,1} & A_{L,2}  \\ A_{R,1} & A_{R,2}
  \end{pmatrix}
\end{equation}
therefore, when eigenstates are properly normalised (associated to a measure $\D\varepsilon$), we can write
$\scalar{\phi_{\varepsilon,a}}{\psi_{\varepsilon',b}}=\mathcal{U}_{ba}(\varepsilon)\,\delta(\varepsilon-\varepsilon')$
where $a\in\{1,\,2\}$ and $b\in\{L,\,R\}$.

\subsection{Symmetric case}

We first consider the symmetric case $r=r'$ and $t=t'$ [in 1D this corresponds to a symmetric potential $V(-x)=V(x)$].
The scattering states incoming from left and right are simply related by 
$\psi_{\varepsilon,R}(x)=\psi_{\varepsilon,L}(-x)$ and the two partial waves are the symmetric/antisymmetric eigenstates $\phi_{\varepsilon,a}(-x)=(-1)^{a+1}\phi_{\varepsilon,a}(x)$, hence the unitary matrix is independent on the energy in this case
\begin{equation}
  \mathcal{U}
  =  \frac{1}{\sqrt{2}}
  \begin{pmatrix}
    1 & 1  \\ 1 & -1
  \end{pmatrix}
\end{equation}
and the partial and proper times coincide
\begin{equation}
  \label{eq:SymPotPartialEqualProper}
  \tau_1=\tilde\tau_1
  \quad\mbox{and}\quad
  \tau_2=\tilde\tau_2
  \:.
\end{equation}

\begin{figure}[!ht]
\centering
\includegraphics[width=0.4\textwidth]{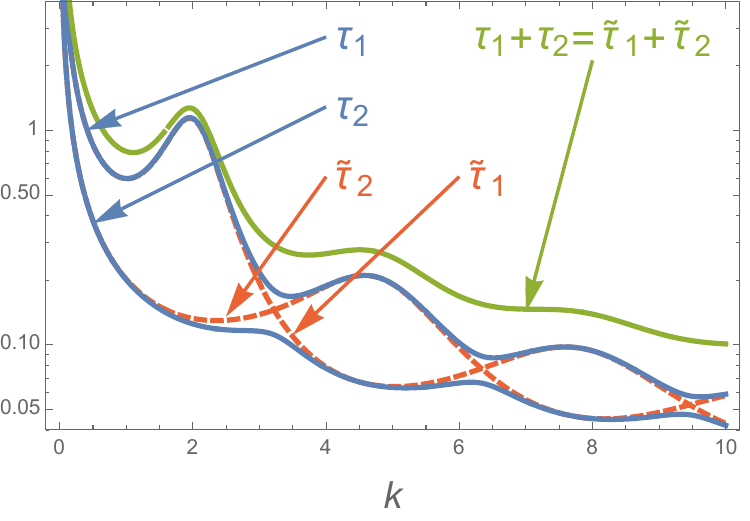}
\caption{\it 
  Proper time delays (continous blue line) and partial time delays (dashed red line) as a function of $k=\sqrt{\varepsilon}$ in a particular example of 1D Hamiltonian with potential $V(x)=\lambda_1\,\delta(x)+\lambda_2\,\delta(x-a)$ with $\lambda_1=2$ and $\lambda_2=4$ ($a=1$).}
\label{fig:PartialProper}
\end{figure}

\subsection{Asymmetric case}

In general the matrix $\mathcal{U}$ carries some energy dependence.
In order to illustrate the difference between the two sets of characteristic times, we consider a specific 1D example~:
we compute these times for the Hamiltonian 
$H=-\partial_x^2+\lambda_1\,\delta(x)+\lambda_2\,\delta(x-a)$.
We consider the basis of left and right scattering states in which the $\Sm$-matrix takes the form~\eqref{eq:2by2Smatrix}.
The three coefficients $r$, $t$ and $r'$ can be computed easily thanks to a transfer matrix approach. 
Diagonalisation of the $\Sm$-matrix provides the phase shifts and hence the partial time delays $\tilde\tau_1$ and $\tilde\tau_2$.
We also compute the Wigner-Smith matrix in the basis of left/right scattering states (cf.~\S~\ref{subsec:CharacteristicTimes}) and diagonalise it, we get the proper time delays $\tau_1$ and $\tau_2$.
We check that for the symmetric potential ($\lambda_1=\lambda_2$) the two sets exactly coincide, Eq.~\eqref{eq:SymPotPartialEqualProper}.
In the asymmetric case ($\lambda_1\neq\lambda_2$) the four times are plotted in Fig.~\ref{fig:PartialProper} as a function of $\sqrt{\varepsilon}$.
We see that proper and partial times are very close (they get closer as the energy grows). At the crossing points where the two partial times are equal we can however observe that partial times and proper times differ. 
This follows from the fact that the second term of Eq.~\eqref{eq:ProperVsPartial} can be understood as a perturbation which produces anticrossing of the eigenvalues of~$\WSm$.

\end{appendix}



\addcontentsline{toc}{section}{\protect\bibname}
\begin{thebibliography}{300}

\bibitem{AbrAndLicRam79}
E.~Abrahams, P.~W. Anderson, D.~C. Licciardello and T.~V. Ramakrishnan, Scaling
  theory of localization: absence of quantum diffusion in two dimensions, Phys.
  Rev. Lett. {\bf 42}(10), 673 (1979).

\bibitem{AbrSte64}
M.~Abramowitz and I.~A. Stegun (editors), {\em Handbook of Mathematical
  functions\/}, Dover, New York (1964).

\bibitem{AkkAueAvrSha91}
E.~Akkermans, A.~Auerbach, J.~E. Avron and B.~Shapiro, Relation between
  persistent currents and the scattering matrix, Phys. Rev. Lett. {\bf 66}(1),
  76--79 (1991).

\bibitem{AkkMon07}
E.~Akkermans and G.~Montambaux, {\em Mesoscopic physics of electrons and
  photons\/}, Cambridge University Press, Cambridge, UK (2007).

\bibitem{AleBroGla02}
I.~L. Aleiner, P.~W. Brouwer and L.~I. Glazman, Quantum effects in Coulomb
  blockade, Phys. Rep. {\bf 358}(5-6), 309--440 (2002).

\bibitem{AltZir97}
A.~Altland and M.~R. Zirnbauer, Nonstandard symmetry classes in mesoscopic
  normal-superconducting hybrid structures, Phys. Rev.~B {\bf 55}(2),
  1142--1161 (1997).

\bibitem{AndThoAbrFis80}
P.~Anderson, D.~J. Thouless, E.~Abrahams and D.~S. Fisher, New method for a
  scaling theory of localization, Phys. Rev. B {\bf 22}(8), 3519--3526 (1980).

\bibitem{Ang07}
L.~Angers, {\em Rectification et supraconductivit\'e de proximit\'e dans des
  anneaux m\'esoscopiques\/}, Ph.D. thesis, Universit\'e Paris-Sud (2007),
  \url{http://tel.archives-ouvertes.fr/tel-00156703}.

\bibitem{AngZakDebGueBouCavGenPol07}
L.~Angers, E.~Zakka-Bajjani, R.~Deblock, S.~Gu\'{e}ron, H.~Bouchiat,
  A.~Cavanna, U.~Gennser and M.~Polianski, Magnetic-field asymmetry of
  mesoscopic $dc$ rectification in Aharonov-Bohm rings, Phys. Rev.~B {\bf 75},
  115309 (2007).

\bibitem{AntPasSly81}
T.~N. Antsygina, L.~A. Pastur and V.~A. Slyusarev, Localization of states and
  kinetic properties of one-dimensional disordered systems, Sov. J. Low Temp.
  Phys. {\bf 7}(1), 1--21 (1981).

\bibitem{AviBan85}
Y.~Avishai and Y.~B. Band, One-dimensional density of states and the phase of
  the transmission amplitude, Phys. Rev.~B {\bf 32}, 2674--2676 (1985).

\bibitem{Azb83}
M.~{\relax{Ya}}. Azbel, Resonance tunneling and localization spectroscopy,
  Solid State Commun. {\bf 45}(7), 527--530 (1983).

\bibitem{Bee97}
C.~W.~J. Beenakker, Random-matrix theory of quantum transport, Rev. Mod. Phys.
  {\bf 69}(3), 731--808 (1997).

\bibitem{Bee01}
C.~W.~J. Beenakker, Dynamics of localization in a waveguide, in {\em Photonic
  Crystals and Light Localization in the 21st Century\/}, edited by
  C.~Soukoulis, NATO Science Series C563, pp. 489--508, Kluwer, Dordrecht
  (2001).

\bibitem{Bee13}
C.~W.~J. Beenakker, Search for Majorana Fermions in Superconductors, Annu. Rev.
  Cond. Mat. Phys. {\bf 4}, 113--136 (2013).

\bibitem{Bee15}
C.~W.~J. Beenakker, Random-matrix theory of Majorana fermions and topological
  superconductors, Rev. Mod. Phys. {\bf 87}, 1037--1066 (2015).

\bibitem{BeeBro01}
C.~W.~J. Beenakker and P.~W. Brouwer, Distribution of the reflection
  eigenvalues of a weakly absorbing chaotic cavity, Physica E {\bf 9}, 463--466
  (2001).

\bibitem{BenJay02}
C.~Benjamin and A.~M. Jayannavar, Wave attenuation to clock sojourn time, Solid
  State Commun. {\bf 121}, 591--595 (2002).

\bibitem{BerLec02}
D.~Bernard and A.~LeClair, A classification of 2D random Dirac fermions,
  J.~Phys.~A: Math. Gen. {\bf 35}, 2555--2567 (2002).

\bibitem{BetUhl37}
E.~Beth and G.~E. Uhlenbeck, The quantum theory of the non-ideal gas II.
  Behaviour at low temperatures, Physica {\bf 4}(10), 915--924 (1937).

\bibitem{BieTex08}
T.~Bienaim\'e and C.~Texier, Localization for one-dimensional random potentials
  with large fluctuations, J.~Phys.~A: Math. Theor. {\bf 41}, 475001 (2008).

\bibitem{BirKre62}
M.~S. Birman and M.~G. Krein, On the theory of wave operators, Dokl. Akad. Nauk
  SSSR {\bf 144}, 475--478 (1962), [Soviet Math. Dokl. \textbf{3}, 740--743
  (1962)].

\bibitem{BluSmi88}
R.~{Bl\"umel} and U.~Smilansky, Classical irregular scattering and its
  quantum-mechanical implications, Phys. Rev. Lett. {\bf 60}(6), 477--480
  (1988).

\bibitem{BocFreParBerPlaWahRecJonMarGreFerDegFev14}
E.~Bocquillon, V.~Freulon, F.~D. Parmentier, J.-M. Berroir, B.~{Pla\c{c}ais},
  C.~Wahl, J.~Rech, T.~Jonckheere, T.~Martin, C.~Grenier, D.~Ferraro,
  P.~Degiovanni and G.~{F{\`e}ve}, Electron quantum optics in ballistic chiral
  conductors, Ann. Phys. (Berlin) {\bf 526}(1-2), 1--30 (2014).

\bibitem{BolOsb76}
D.~Boll{\'e} and T.~A. Osborn, Concepts of multiparticle time delay, Phys.
  Rev.~D {\bf 13}, 299--311 (1976).

\bibitem{BolLamFalPriEps99}
C.~J. Bolton-Heaton, C.~J. Lambert, V.~I. {Fal'ko}, V.~Prigodin and A.~J.
  Epstein, Distribution of time constants for tunneling through a
  one-dimensional disordered chain, Phys. Rev. B {\bf 60}(15), 10569--10572
  (1999).

\bibitem{BouComGeoLeD90}
J.-P. Bouchaud, A.~Comtet, A.~Georges and P.~{Le~Doussal}, Classical diffusion
  of a particle in a one-dimensional random force field, Ann. Phys. (N.Y.) {\bf
  201}, 285--341 (1990).

\bibitem{Bro95}
P.~W. Brouwer, Generalized circular ensemble of scattering matrices for a
  chaotic cavity with nonideal leads, Phys. Rev. B {\bf 51}, 16878--16884
  (1995).

\bibitem{BroBee96}
P.~W. Brouwer and C.~W.~J. Beenakker, Diagrammatic method of integration over
  the unitary group, with applications to quantum transport in mesoscopic
  systems, J. Math. Phys. {\bf 37}(10), 4904--4934 (1996).

\bibitem{BroBut97}
P.~W. Brouwer and M.~B{\"u}ttiker, Charge-relaxation and dwell time in the
  fluctuating admittance of a chaotic cavity, Europhys. Lett. {\bf 37}(7),
  441--446 (1997).

\bibitem{BroFraBee97}
P.~W. Brouwer, K.~M. Frahm and C.~W. Beenakker, Quantum mechanical time-delay
  matrix in chaotic scattering, Phys. Rev. Lett. {\bf 78}(25), 4737 (1997).

\bibitem{BroFraBee99}
P.~W. Brouwer, K.~M. Frahm and C.~W. Beenakker, Distribution of the quantum
  mechanical time-delay matrix for a chaotic cavity, Waves Random Media {\bf
  9}, 91--104 (1999).

\bibitem{BroLamFle05}
P.~W. Brouwer, A.~Lamacraft and K.~Flensberg, Nonequilibrium theory of Coulomb
  blockade in open quantum dots, Phys. Rev. B {\bf 72}, 075316 (2005).

\bibitem{BroLanFraButBee97}
P.~W. Brouwer, S.~A. van Langen, K.~M. Frahm, M.~B\"uttiker and C.~W.~J.
  Beenakker, Distribution of Parametric Conductance Derivatives of a Quantum
  Dot, Phys. Rev. Lett. {\bf 79}, 913--916 (1997).

\bibitem{BruFle04}
H.~Bruus and K.~Flensberg, {\em Many-body quantum theory in condensed matter
  physics\/}, Oxford University Press, Oxford (2004).

\bibitem{Bus62}
V.~S. Buslaev, Trace formulas for the three-dimensional Schr\"odinger operator,
  Dokl. Akad. Nauk SSSR {\bf 143} (1962).

\bibitem{But83}
M.~B{\"u}ttiker, Larmor precession and the traversal time for tunneling, Phys.
  Rev.~B {\bf 27}(10), 6178--6188 (1983).

\bibitem{But90}
M.~B{\"u}ttiker, Traversal, reflection and dwell time for quantum tunneling, in
  {\em Electronic properties of multilayers and low-dimensional semiconductors
  structures\/}, edited by J.~M. {Chamberlain et al.}, p. 297, Plenum Press,
  New York (1990).

\bibitem{But92}
M.~B{\"u}ttiker, Scattering theory of current and intensity noise correlations
  in conductors and wave guides, Phys. Rev.~B {\bf 46}(19), 12485--12507
  (1992).

\bibitem{But93}
M.~B{\"u}ttiker, Capacitance, admittance, and rectification properties of small
  conductors, J.~Phys.~Cond. Matter {\bf 5}(50), 9361--9378 (1993).

\bibitem{But99}
M.~B{\"u}ttiker, Charge relaxation resistances and charge fluctuations in
  mesoscopic conductors, J. Korean Phys. Soc. {\bf 34}, S121--130 (1999),
  (cond-mat/99\,02\,054).

\bibitem{But00}
M.~B{\"u}ttiker, Charge fluctuations and dephasing in Coulomb coupled
  conductors, in {\em Quantum mesoscopic phenomena and mesoscopic devices\/},
  edited by I.~O. Kulik and R.~Ellialtioglu, volume 559, p. 211, Kluwer
  Academic Publishers, Dordrecht (2000), (cond-mat/99\,11\,188).

\bibitem{But00a}
M.~B{\"u}ttiker, Time-dependent transport in mesoscopic structures, J. Low
  Temp. Phys. {\bf 11}(5/6), 519--542 (2000).

\bibitem{But02a}
M.~B{\"u}ttiker, The local Larmor clock, Partial densities of states, and
  mesoscopic physics, in {\em Time in quantum mechanics\/}, edited by J.~G.
  Muga, R.~S. Mayato and I.~L. Egusquiza, Lecture Notes in Physics, p. 256,
  Springer (2002), preprint quant-ph/0103164.

\bibitem{ButChr97}
M.~B{\"u}ttiker and T.~Christen, Admittance and nonlinear transport in quantum
  wires, points contacts and resonant tunneling barriers, in {\em Mesoscopic
  Electron Transport\/}, edited by L.~L. Sohn, L.~P. Kouwenhoven and
  G.~Sch{\"o}n, pp. 259--289, Kluwer Academic Publishers, Dordrecht (1997).

\bibitem{ButLan82}
M.~B{\"u}ttiker and R.~Landauer, Traversal time for tunneling, Phys. Rev. Lett.
  {\bf 49}, 1739 (1982).

\bibitem{ButLan85}
M.~B{\"u}ttiker and R.~Landauer, Traversal time for tunneling, Phys. Scr. {\bf
  32}, 429--434 (1985).

\bibitem{ButPol05}
M.~B\"{u}ttiker and M.~L. Polianski, Charge fluctuation in open chaotic
  cavities, J.~Phys.~A: Math. Theor. {\bf 38}, 10559--10585 (2005).

\bibitem{ButPreTho93}
M.~B{\"u}ttiker, A.~Pr{\^e}tre and H.~Thomas, Dynamic conductance and the
  scattering matrix of small conductors, Phys. Rev. Lett. {\bf 70}(26), 4114
  (1993).

\bibitem{ButThoPre93}
M.~B{\"u}ttiker, H.~Thomas and A.~Pr{\^e}tre, Mesoscopic capacitors, Phys.
  Lett.~A {\bf 180}, 364--369 (1993).

\bibitem{ButThoPre94}
M.~B{\"u}ttiker, H.~Thomas and A.~Pr{\^e}tre, Current partition in multiprobe
  conductors in the presence of slowly oscillating potentials, Z. Phys. B {\bf
  94}, 133--137 (1994).

\bibitem{ChaGen01}
A.~A. Chabanov and A.~Z. Genack, Statistics of Dynamics of Localized Waves,
  Phys. Rev. Lett. {\bf 87}, 233903 (2001).

\bibitem{ChrBut96a}
T.~Christen and M.~B{\"u}ttiker, Gauge invariant nonlinear electric transport
  in mesoscopic conductors, Europhys. Lett. {\bf 35}(7), 523--528 (1996).

\bibitem{ChrBut96}
T.~Christen and M.~B{\"u}ttiker, Low-frequency admittance of quantized Hall
  conductors, Phys. Rev.~B {\bf 53}, 2064--2072 (1996).

\bibitem{CohRotSha88}
A.~Cohen, Y.~Roth and B.~Shapiro, Universal distributions and scaling in
  disordered systems, Phys. Rev. B {\bf 38}(17), 12125--12132 (1988).

\bibitem{ComDesTex05}
A.~Comtet, J.~Desbois and C.~Texier, Functionals of the Brownian motion,
  localization and metric graphs, J.~Phys.~A: Math. Gen. {\bf 38}, R341--R383
  (2005).

\bibitem{ComMon96}
A.~Comtet and C.~Monthus, Diffusion in one-dimensional random medium and
  hyperbolic Brownian motion, J.~Phys.~A: Math. Gen. {\bf 29}(7), 1331--1345
  (1996).

\bibitem{ComMonYor98}
A.~Comtet, C.~Monthus and M.~Yor, Exponential functionals of Brownian motion
  and disordered systems, J. Appl. Probab. {\bf 35}, 255 (1998).

\bibitem{ComMorOuv95}
A.~Comtet, A.~Moroz and S.~Ouvry, Persistent current of free electrons in the
  plane, Phys. Rev. Lett. {\bf 74}(5), 828 (1995).

\bibitem{ComTex97}
A.~Comtet and C.~Texier, On the distribution of the Wigner time delay in
  one-dimensional disordered systems, J.~Phys.~A: Math. Gen. {\bf 30},
  8017--8025 (1997).

\bibitem{ComTex98}
A.~Comtet and C.~Texier, One-dimensional disordered supersymmetric quantum
  mechanics: a brief survey, in {\em Supersymmetry and Integrable Models\/},
  edited by H.~Aratyn, T.~D. Imbo, W.-Y. Keung and U.~Sukhatme, Lecture Notes
  in Physics, Vol. {\bf502} (available as arXiv:cond-mat/97\,07\,313), pp.
  313--328. Springer (1998).

\bibitem{Cun15}
F.~D. Cunden, Statistical distribution of the Wigner-Smith time-delay matrix
  moments for chaotic cavities, Phys. Rev. E {\bf 91}, 060102 (2015).

\bibitem{CunMezSimViv16}
F.~D. Cunden, F.~Mezzadri, N.~Simm and P.~Vivo, Correlators for the
  Wigner-Smith time-delay matrix of chaotic cavities, J. Phys. A: Math. Theor.
  {\bf 49}, 18LT01 (2016).

\bibitem{CunMezSimViv16b}
F.~D. Cunden, F.~Mezzadri, N.~Simm and P.~Vivo, Large-$N$ expansion for the
  time-delay matrix of ballistic chaotic cavities, J. Math. Phys. {\bf 57}(11),
  111901 (2016).

\bibitem{CunViv14}
F.~D. Cunden and P.~Vivo, Universal covariance formula for linear statistics on
  random matrices, Phys. Rev. Lett. {\bf 113}, 070202 (2014).

\bibitem{DasMaBer69}
R.~Dashen, S.-K. Ma and H.~J. Bernstein, $S$-matrix formulation of statistical
  mechanics, Phys. Rev. {\bf 187}(1), 345--370 (1969).

\bibitem{CarNus02}
C.~A.~A. {de Carvalho} and H.~M. Nussenzveig, Time delay, Phys. Rep. {\bf 364},
  83--174 (2002).

\bibitem{DeyLisAlt00}
L.~I. Deych, A.~A. Lisyansky and B.~L. Altshuler, Single Parameter Scaling in
  One-Dimensional Localization Revisited, Phys. Rev. Lett. {\bf 84}(12), 2678
  (2000).

\bibitem{DeyLisAlt01}
L.~I. Deych, A.~A. Lisyansky and B.~L. Altshuler, Single Parameter Scaling in
  1-D Anderson localization. Exact analytical solution, Phys. Rev.~B {\bf 64},
  224202 (2001).

\bibitem{DutSchMorLeh13}
P.~Dutt, T.~L. Schmidt, C.~Mora and K.~Le~Hur, Strongly correlated dynamics in
  multichannel quantum RC circuits, Phys. Rev. B {\bf 87}, 155134 (2013).

\bibitem{Eck93b}
B.~Eckhardt, Correlations in quantum time delay, Chaos {\bf 3}(4), 613--617
  (1993).

\bibitem{Eis48}
L.~Eisenbud, {\em The formal properties of nuclear collisions\/}, Ph.D. thesis,
  Princeton (1948).

\bibitem{Eri60}
T.~Ericson, Fluctuations of Nuclear Cross Sections in the "Continuum" Region,
  Phys. Rev. Lett. {\bf 5}, 430--431 (1960).

\bibitem{EtzHorLed11}
Y.~Etzioni, B.~Horovitz and P.~Le~Doussal, Rings and boxes in dissipative
  environments, Phys. Rev. Lett. {\bf 106}, 166803 (2011).

\bibitem{EveMir08}
F.~Evers and A.~D. Mirlin, Anderson transitions, Rev. Mod. Phys. {\bf 80}(4),
  1355--1417 (2008).

\bibitem{FarTsa94}
W.~G. Faris and W.~J. Tsay, Time delay in random scattering, SIAM J. Appl.
  Math. {\bf 54}(2), 443--455 (1994).

\bibitem{Fev06}
G.~{F\`eve}, {\em Quantification du courant alternatif~: la bo{\^\i}te
  quantique comme source d'\'electrons uniques subnanoseconde\/}, Ph.D. thesis,
  Universit\'e Paris 6 (2006),
  \url{http://tel.archives-ouvertes.fr/tel-00119589}.

\bibitem{FisLee81}
D.~S. Fisher and P.~A. Lee, Relation between conductivity and transmission
  matrix, Phys. Rev.~B {\bf 23}(12), 6851--6854 (1981).

\bibitem{For12}
P.~J. Forrester, Large deviation eigenvalue density for the soft edge Laguerre
  and Jacobi $\beta$-ensembles, J. Phys. A: Math. Theor. {\bf 45}, 145201
  (2012).

\bibitem{Fri52}
J.~Friedel, The distribution of electrons round impurities in monovalent
  metals, Phil. Mag. {\bf 43}, 153--189 (1952).

\bibitem{Fri58}
J.~Friedel, Metallic alloys, Nuovo Cimento Suppl. {\bf 7}, 287--311 (1958).

\bibitem{FrieMel85}
W.~A. Friedman and P.~A. Mello, Information theory and statistical nuclear
  reactions II. Many-channel case and Hauser-Feshbach formula, Ann. Phys. {\bf
  161}(2), 276--302 (1985).

\bibitem{FriFroSchSul73}
U.~Frisch, C.~Froeschle, J.-P. Scheidecker and P.-L. Sulem, Stochastic
  resonance in one-dimensional random media, Phys. Rev.~A {\bf 8}(3),
  1416--1421 (1973).

\bibitem{Fyo03}
Y.~Fyodorov, Induced vs. spontaneous breakdown of $S$-matrix unitarity:
  Probability of no return in quantum chaotic and disordered systems, JETP
  Letters {\bf 78}, 250--254 (2003).

\bibitem{FyoOss04}
Y.~V. Fyodorov and A.~Ossipov, Distribution of the Local Density of States,
  Reflection Coefficient, and Wigner Delay Time in Absorbing Ergodic Systems at
  the Point of Chiral Symmetry, Phys. Rev. Lett. {\bf 92}, 084103 (2004).

\bibitem{FyoSavSom97}
Y.~V. Fyodorov, D.~V. Savin and H.-J. Sommers, Parametric correlations of phase
  shifts and statistics of time delays in quantum chaotic scattering: Crossover
  between unitary and orthogonal symmetries, Phys. Rev. E {\bf 55},
  R4857--R4860 (1997).

\bibitem{FyoSom96}
Y.~V. Fyodorov and H.-J. Sommers, Parametric correlations of scattering phase
  shifts and fluctuations of delay times in few-channel chaotic scattering,
  Phys. Rev. Lett. {\bf 76}(25), 4709 (1996).

\bibitem{FyoSom97}
Y.~V. Fyodorov and H.-J. Sommers, Statistics of resonance poles, phase shifts
  and time delays in quantum chaotic scattering: Random matrix approach for
  systems with broken time-reversal invariance, J. Math. Phys. {\bf 38}(4),
  1918--1981 (1997).

\bibitem{Gab06}
J.~Gabelli, {\em Mise en \'evidence de la coh\'erence quantique dans des
  conducteurs en r\'egime dynamique\/}, Ph.D. thesis, Universit\'e Paris 6
  (2006), \url{http://tel.archives-ouvertes.fr/tel-00011619}.

\bibitem{GabFevBerPla12}
J.~Gabelli, G.~{F\`eve}, J.-M. Berroir and B.~{Pla\c{c}ais}, A coherent RC
  circuit, Rep. Prog. Phys. {\bf 75}, 126504 (2012).

\bibitem{GabFevBerPlaCavEtiJinGla06}
J.~Gabelli, G.~{F\`eve}, J.-M. Berroir, B.~{Pla\c{c}ais}, A.~Cavanna,
  B.~Etienne, Y.~Jin and C.~Glattli, Violation of Kirchhoff's laws for a
  coherent RC circuit, Science {\bf 313}(5786), 499--502 (2006).

\bibitem{GasChrBut96}
V.~Gasparian, T.~Christen and M.~B{\"u}ttiker, Partial densities of states,
  scattering matrices and Green's functions, Phys. Rev. A {\bf 54}(5),
  4022--4031 (1996).

\bibitem{GenSebStoTig99}
A.~Z. Genack, P.~Sebbah, M.~Stoytchev and B.~A. {van Tiggelen}, Statistics of
  wave dynamics in random media, Phys. Rev. Lett. {\bf 82}(4), 715 (1999).

\bibitem{GerVas59}
M.~E. Gertsenshtein and V.~B. {Vasil'ev}, Waveguides with random
  inhomogeneities and Brownian motion in the Lobachevsky plane, Theory Prob.
  and Appl. {\bf 4}(4), 391--398 (1959).

\bibitem{GopMel98}
V.~A. Gopar and P.~A. Mello, The problem of quantum chaotic scattering with
  direct processes reduced to the one without, Europhys. Lett. {\bf 42}(2),
  131--136 (1998).

\bibitem{GopMelBut96}
V.~A. Gopar, P.~A. Mello and M.~B{\"u}ttiker, Mesoscopic capacitors: a
  statistical analysis, Phys. Rev. Lett. {\bf 77}(14), 3005 (1996).

\bibitem{Gra18}
A.~Grabsch, {\em Random matrices in statistical physics: quantum scattering and
  disordered systems\/}, Ph.D. thesis, Universit\'e Paris Saclay (2018),
  \url{https://tel.archives-ouvertes.fr/tel-01849097v1}.

\bibitem{GraMajTex16a}
A.~Grabsch, S.~Majumdar and C.~Texier, Time delays and coherent AC transport in
  chaotic cavities -- Random matrices and Coulomb gas, in preparation  (2016).

\bibitem{GraMajTex17}
A.~Grabsch, S.~N. Majumdar and C.~Texier, Truncated linear statistics
  associated with the top eigenvalues of random matrices, J. Stat. Phys. {\bf
  167}, 234--259 (2017).

\bibitem{GraMajTex17b}
A.~Grabsch, S.~N. Majumdar and C.~Texier, Truncated linear statistics
  associated with the eigenvalues of random matrices II. Partial sums over
  proper time delays for chaotic quantum dots, J. Stat. Phys. {\bf 167},
  1452--1488 (2017).

\bibitem{GraSavTex18}
A.~Grabsch, D.~V. Savin and C.~Texier, Wigner-Smith time-delay matrix in
  chaotic cavities with non-ideal contacts, J. Phys. A: Math. Theor. {\bf 51},
  404001 (2018), special issue ``\textit{Random Matrices: the first 90
  years}''.

\bibitem{GraTex15}
A.~Grabsch and C.~Texier, Capacitance and charge relaxation resistance of
  chaotic cavities -- Joint distribution of two linear statistics in the
  Laguerre ensemble of random matrices, Europhys. Lett. {\bf 109}, 50004
  (2015).

\bibitem{GraTex16b}
A.~Grabsch and C.~Texier, Distribution of spectral linear statistics on random
  matrices beyond the large deviation function -- Wigner time delay in
  multichannel disordered wires, J. Phys. A: Math. Theor. {\bf 49}, 465002
  (2016).

\bibitem{GraTex16}
A.~Grabsch and C.~Texier, Topological phase transitions in the 1D multichannel
  Dirac equation with random mass and a random matrix model, Europhys. Lett.
  {\bf 116}, 17004 (2016).

\bibitem{gragra}
I.~S. Gradshteyn and I.~M. Ryzhik, {\em Table of integrals, series and
  products\/}, Academic Press, fifth edition (1994).

\bibitem{GurSha12}
E.~Gurevich and B.~Shapiro, Statistics of resonances in one-dimensional
  disordered systems, Lithuanian Journal of Physics {\bf 52}, 115 (2012).

\bibitem{HamJonKatMar10}
Y.~Hamamoto, T.~Jonckheere, T.~Kato and T.~Martin, Dynamic response of a
  mesoscopic capacitor in the presence of strong electron interactions, Phys.
  Rev. B {\bf 81}, 153305 (2010).

\bibitem{HasKan10}
M.~Z. Hasan and C.~L. Kane, Colloquium: Topological insulators, Rev. Mod. Phys.
  {\bf 82}(4), 3045--3067 (2010).

\bibitem{HauSto89}
E.~H. Hauge and J.~A. St{\o}vneng, Tunneling times: a critical review, Rev.
  Mod. Phys. {\bf 61}(4), 917--936 (1989).

\bibitem{Hei90}
J.~Heinrichs, Invariant embedding treatment of phase randomization and
  electrical noise at disordered surfaces, J.~Phys.~Cond. Matter {\bf 2},
  1559--1568 (1990).

\bibitem{HerLew09}
A.~R. Hern\'andez and C.~H. Lewenkopf, Nonlinear Conductance in a Ballistic
  Aharonov-Bohm Ring, Phys. Rev. Lett. {\bf 103}, 166801 (2009).

\bibitem{HerLew13}
A.~R. Hern\'andez and C.~H. Lewenkopf, Nonlinear electronic transport in
  nanoscopic devices: nonequilibrium Green’s functions versus scattering
  approach, Eur. Phys. J. B {\bf 86}(4), 131 (2013).

\bibitem{HofDasFli16}
P.~P. Hofer, D.~Dasenbrook and C.~Flindt, Electron waiting times for the
  mesoscopic capacitor, Physica E {\bf 82}, 3--11 (2016), Frontiers in quantum
  electronic transport - In memory of Markus Büttiker.

\bibitem{JayVijKum89}
A.~M. Jayannavar, G.~V. Vijayagovindan and N.~Kumar, Energy dispersive
  backscattering of electrons from surface resonances of a disordered medium
  and $1/f$ noise, Z. Phys. B -- Condens. Matter {\bf 75}, 77--79 (1989).

\bibitem{JosJay98}
S.~K. Joshi and A.~M. Jayannavar, Distribution of Wigner delay time from single
  channel disordered systems, Solid State Commun. {\bf 106}(6), 363 (1998).

\bibitem{KlySai92}
V.~I. Klyatskin and A.~I. Saichev, Statistical and dynamic localization of
  plane waves in randomly layered media, Sov. Phys. Usp. {\bf 35}(3), 231--247
  (1992), [Usp. Fiz. Nauk {\bf162}, 161 (1992)].

\bibitem{Kot05}
T.~Kottos, Statistics of resonances and delay times in random media: beyond
  random matrix theory, J.~Phys.~A: Math. Theor. {\bf 38}, 10761--10786 (2005).

\bibitem{KotWei02}
T.~Kottos and M.~Weiss, Statistics of resonances and delay times: A criterion
  for metal-insulator transitions, Phys. Rev. Lett. {\bf 89}, 056401 (2002).

\bibitem{Kre53}
M.~G. Krein, Trace formulas in perturbation theory, Matem. Sbornik {\bf 33},
  597 (1953).

\bibitem{KuiSavSie14}
J.~Kuipers, D.~V. Savin and M.~Sieber, Efficient semiclassical approach for
  time delays, New J. Phys. {\bf 16}, 123018 (2014).

\bibitem{KuiSie07}
J.~Kuipers and M.~Sieber, Semiclassical expansion of parametric correlation
  functions of the quantum time delay, Nonlinearity {\bf 20}(4), 909--926
  (2007).

\bibitem{KuiSie08}
J.~Kuipers and M.~Sieber, Semiclassical relation between open trajectories and
  periodic orbits for the Wigner time delay, Phys. Rev. E {\bf 77}, 046219
  (2008).

\bibitem{KunSha08}
H.~Kunz and B.~Shapiro, Statistics of resonances in a semi-infinite disordered
  chain, Phys. Rev. B {\bf 77}, 054203 (2008).

\bibitem{LanLif66e}
L.~D. Landau and E.~Lifchitz, {\em Physique statistique\/}, Mir (1966), tome 5.

\bibitem{LanMar94}
R.~Landauer and T.~Martin, Barrier interaction time in tunneling, Rev. Mod.
  Phys. {\bf 66}(1), 217--228 (1994).

\bibitem{LeaAer89}
C.~R. Leavens and G.~C. Aers, Dwell time and phase times for transmission and
  reflection, Phys. Rev.~B {\bf 39}(2), 1202--1206 (1989).

\bibitem{LehSavSokSom95}
N.~Lehmann, D.~V. Savin, V.~V. Sokolov and H.-J. Sommers, Time delay
  correlations in chaotic scattering: random matrix approach, Physica D {\bf
  86}, 572--585 (1995).

\bibitem{LetSanGotIhnEnsDriGos06}
R.~Leturcq, D.~S\'{a}nchez, G.~G\"{o}tz, T.~Ihn, K.~Ensslin, D.~C. Driscoll and
  A.~C. Gossard, Magnetic Field Symmetry and Phase Rigidity of the Nonlinear
  Conductance in a Ring, Phys. Rev. Lett. {\bf 96}, 126801 (2006).

\bibitem{LevBut00}
A.~{Levy Yeyati} and M.~B\"uttiker, Scattering phases in quantum dots: An
  analysis based on lattice models, Phys. Rev. B {\bf 62}(11), 7307--7315
  (2000).

\bibitem{LewVal04}
C.~H. Lewenkopf and R.~O. Vallejos, Open orbits and the semiclassical dwell
  time, J. Phys. A: Math. Gen. {\bf 37}(1), 131--136 (2004).

\bibitem{Llo69}
P.~Lloyd, Exactly solvable model of electronic states in a three-dimensional
  disordered Hamiltonian: non-existence of localized states, J.~Phys.~C: Solid
  St. Phys. {\bf 2}(10), 1717--1725 (1969).

\bibitem{Lyu77}
V.~L. Lyuboshitz, On collision duration in the presence of strong overlapping
  resonance levels, Phys. Lett. {\bf 72B}(1), 41--44 (1977).

\bibitem{Lyu83}
V.~L. Lyuboshits, The probability distribution of the delay time of a wave
  packet in strong overlap of resonance levels, Sov. J. Nucl. Phys. {\bf 37},
  174--176 (1983), [Yad. Fiz. \textbf{37}, 292--297 (1983)].

\bibitem{Ma85}
S.-K. Ma, {\em Statistical mechanics\/}, World Scientific, Singapore (1985).

\bibitem{MarBroBee14}
M.~Marciani, P.~W. Brouwer and C.~W.~J. Beenakker, Time-delay matrix, midgap
  spectral peak, and thermopower of an Andreev billiard, Phys. Rev. B {\bf 90},
  045403 (2014).

\bibitem{MarSchBee16}
M.~Marciani, H.~Schomerus and C.~W.~J. Beenakker, Effect of a tunnel barrier on
  the scattering from a Majorana bound state in an Andreev billiard, Physica E
  {\bf 77}, 54--64 (2016), Frontiers in quantum electronic transport - In
  memory of Markus Büttiker.

\bibitem{MarTayFaiShoLin06}
C.~A. Marlow, R.~P. Taylor, M.~Fairbanks, I.~Shorubalko and H.~Linke,
  Experimental Investigation of the Breakdown of the Onsager-Casimir Relations,
  Phys. Rev. Lett. {\bf 96}, 116801 (2006).

\bibitem{MarMarGar14}
A.~M. Mart{\'\i}nez-Arg{\"u}ello, M.~Mart{\'\i}nez-Mares and J.~C. Garc{\'\i}a,
  Joint moments of proper delay times, J. Math. Phys. {\bf 55}(8), 081901
  (2014).

\bibitem{Mat95}
K.~A. Matveev, Coulomb blockade at almost perfect transmission, Phys. Rev. B
  {\bf 51}, 1743--1751 (1995).

\bibitem{Meh04}
M.~L. Mehta, {\em Random matrices\/}, Elsevier, Academic, New York, third
  edition (2004).

\bibitem{MelBar99}
P.~A. Mello and H.~U. Baranger, Interference phenomena in electronic transport
  through chaotic cavities: An information-theoretic approach, Waves Random
  Media {\bf 9}, 105--162 (1999).

\bibitem{MelKum04}
P.~A. Mello and N.~Kumar, {\em Quantum transport in mesoscopic systems --
  Complexity and statistical fluctuations\/}, Oxford University Press (2004).

\bibitem{MelPerSel85}
P.~A. Mello, P.~Pereyra and T.~H. Seligman, Information theory and statistical
  nuclear reactions. I. General theory and applications to few-channel
  problems, Ann. Phys. {\bf 161}(2), 254--275 (1985).

\bibitem{MezSim11}
F.~Mezzadri and N.~J. Simm, Moments of the transmission eigenvalues, proper
  delay times, and random matrix theory. I, J. Math. Phys. {\bf 52}, 103511
  (2011).

\bibitem{MezSim12}
F.~Mezzadri and N.~J. Simm, Moments of the transmission eigenvalues, proper
  delay times, and random matrix theory. II, J. Math. Phys. {\bf 53}, 053504
  (2012).

\bibitem{MezSim13}
F.~Mezzadri and N.~J. Simm, $\tau$-function theory of quantum chaotic transport
  with $\beta=1,\,2,\,4$, Commun. Math. Phys. {\bf 324}, 465--513 (2013).

\bibitem{MonCom94}
C.~Monthus and A.~Comtet, On the flux distribution in a one-dimensional
  disordered system, J.~Phys.~I (France) {\bf 4}(6), 635--653 (1994).

\bibitem{MorLeh10}
C.~Mora and K.~{Le Hur}, Universal resistances of the quantum
  resistance-capacitance circuit, Nat. Phys. {\bf 6}, 697--701 (2010).

\bibitem{MucJalPic97}
E.~R. Mucciolo, R.~A. Jalabert and J.-L. Pichard, Parametric statistics of the
  scattering matrix: from metallic to insulating quasi-unidimensional
  disordered systems, J.~Phys.~I (France) {\bf 7}, 1267 (1997).

\bibitem{NigBut08}
S.~E. Nigg and M.~B\"uttiker, Quantum to classical transition of the charge
  relaxation resistance of a mesoscopic capacitor, Phys. Rev. B {\bf 77},
  085312 (2008).

\bibitem{NigLopBut06}
S.~E. Nigg, R.~L\'{o}pez and M.~B\"{u}ttiker, Mesoscopic charge relaxation,
  Phys. Rev. Lett. {\bf 97}, 206804 (2006).

\bibitem{Nov15a}
M.~Novaes, Statistics of time delay and scattering correlation functions in
  chaotic systems. I. Random matrix theory, J. Math. Phys. {\bf 56}, 062110
  (2015).

\bibitem{Nov15b}
M.~Novaes, Statistics of time delay and scattering correlation functions in
  chaotic systems. II. Semiclassical approximation, J. Math. Phys. {\bf 56},
  062109 (2015).

\bibitem{OshMogMor93}
G.~Oshanin, A.~Mogutov and M.~Moreau, Steady flux in a continuous-space Sinai
  chain, J.~Stat. Phys. {\bf 73}(1/2), 379--388 (1993).

\bibitem{Oss18}
A.~Ossipov, Scattering approach to Anderson localisation, Phys. Rev. Lett. {\bf
  121}, 076601 (2018).

\bibitem{OssFyo05}
A.~Ossipov and Y.~V. Fyodorov, Statistics of delay times in mesoscopic systems
  as a manifestation of eigenfunction fluctuations, Phys. Rev.~B {\bf 71},
  125133 (2005).

\bibitem{OssKotGei00}
A.~Ossipov, T.~Kottos and T.~Geisel, Statistical properties of phases and delay
  times of the one-dimensional Anderson model with one open channel, Phys. Rev.
  B {\bf 61}, 11411--11415 (2000).

\bibitem{Pas92}
H.~M. Pastawski, Classical and quantum transport from generalized
  Landauer-B\"uttiker equations. II. Time-dependent resonant tunneling, Phys.
  Rev. B {\bf 46}, 4053--4070 (1992).

\bibitem{PedLanBut98}
M.~H. Pedersen, S.~A. {van Langen} and M.~B\"uttiker, Charge fluctuations in
  quantum point contacts and chaotic cavities in the presence of transport,
  Phys. Rev. B {\bf 57}(3), 1838--1846 (1998).

\bibitem{PenKirCas86}
J.~B. Pendry, P.~D. Kirkman and E.~Castano, Electrons at disordered surfaces
  and $1/f$ noise, Phys. Rev. Lett. {\bf 57}(23), 2983 (1986).

\bibitem{PolBro03}
M.~L. Polianski and P.~W. Brouwer, Scattering matrix ensemble for
  time-dependent transport through a chaotic quantum dot, J.~Phys.~A: Math.
  Gen. {\bf 36}, 3215--3236 (2003).

\bibitem{PolBut06}
M.~L. Polianski and M.~B\"{u}ttiker, Mesoscopic Fluctuations of Nonlinear
  Conductance of Chaotic Quantum Dots, Phys. Rev. Lett. {\bf 96}, 156804
  (2006).

\bibitem{PolBut07a}
M.~L. Polianski and M.~{B\"uttiker}, Magnetic-field symmetries of mesoscopic
  non-linear conductance, Physica E {\bf 40}(1), 67--75 (2007).

\bibitem{PolBut07}
M.~L. Polianski and M.~B\"{u}ttiker, Rectification and nonlinear transport in
  chaotic dots and rings, Phys. Rev.~B {\bf 76}, 205308 (2007).

\bibitem{PraKum94}
P.~Pradhan and N.~Kumar, Localization of light in coherently amplifying random
  media, Phys. Rev. B {\bf 50}, 9644--9647 (1994).

\bibitem{PreThoBut96}
A.~Pr\^etre, H.~Thomas and M.~B\"uttiker, Dynamic admittance of mesoscopic
  conductors: Discrete-potential model, Phys. Rev. B {\bf 54}(11), 8130--8143
  (1996).

\bibitem{QiZha11}
X.-L. Qi and S.-C. Zhang, Topological insulators and superconductors, Rev. Mod.
  Phys. {\bf 83}, 1057--1110 (2011).

\bibitem{RamKum00}
S.~A. Ramakrishna and N.~Kumar, Imaginary potential as a counter of delay time
  for wave reflection from a one-dimensional random potential, Phys. Rev.~B
  {\bf 61}(5), 3163--3165 (2000).

\bibitem{RamKum01}
S.~A. Ramakrishna and N.~Kumar, Distribution of the delay time and the dwell
  time for wave reflection from a long random potential, Eur. Phys. J. B {\bf
  23}(4), 509--513 (2001).

\bibitem{RamTex14}
K.~Ramola and C.~Texier, Fluctuations of random matrix products and 1D Dirac
  equation with random mass, J. Stat. Phys. {\bf 157}(3), 497--514 (2014).

\bibitem{RinImrEnt08}
Z.~Ringel, Y.~Imry and O.~Entin-Wohlman, Delayed currents and interaction
  effects in mesoscopic capacitors, Phys. Rev. B {\bf 78}, 165304 (2008).

\bibitem{RyuSchFurLud10}
S.~Ryu, A.~P. Schnyder, A.~Furusaki and A.~W.~W. Ludwig, Topological insulators
  and superconductors: tenfold way and dimensional hierarchy, New J. Phys. {\bf
  12}, 065010 (2010).

\bibitem{SanBut04}
D.~S\'{a}nchez and M.~B\"{u}ttiker, Magnetic-Field Asymmetry of Nonlinear
  Mesoscopic Transport, Phys. Rev. Lett. {\bf 93}, 106802 (2004).

\bibitem{Savin2016}
D.~V. Savin, private communication  (2016).

\bibitem{SavFyoSom01}
D.~V. Savin, Y.~V. Fyodorov and H.-J. Sommers, Reducing nonideal to ideal
  coupling in random matrix description of chaotic scattering: Application to
  the time-delay problem, Phys. Rev. E {\bf 63}, 035202 (2001).

\bibitem{SchMarBee15}
H.~Schomerus, M.~Marciani and C.~W.~J. Beenakker, Effect of Chiral Symmetry on
  Chaotic Scattering from Majorana Zero Modes, Phys. Rev. Lett. {\bf 114},
  166803 (2015).

\bibitem{SchTit03}
H.~Schomerus and M.~Titov, Band-center anomaly of the conductance distribution
  in one-dimensional Anderson localization, Phys. Rev.~B {\bf 67}, 100201
  (2003).

\bibitem{SebLegGen99}
P.~Sebbah, O.~Legrand and A.~Z. Genack, Fluctuations in photon local delay time
  and their relation to phase spectra in random media, Phys. Rev.~E {\bf
  59}(2), 2406--2411 (1999).

\bibitem{Smi17}
U.~Smilansky, Delay-time distribution in the scattering of time-narrow wave
  packets (I), J. Phys. A: Math. Theor. {\bf 50}, 215301 (2017).

\bibitem{Smi60}
F.~T. Smith, Lifetime matrix in collision theory, Phys. Rev. {\bf 118}(1),
  349--356 (1960).

\bibitem{SomSavSok01}
H.-J. Sommers, D.~V. Savin and V.~V. Sokolov, Distribution of Proper Delay
  Times in Quantum Chaotic Scattering: A Crossover from Ideal to Weak Coupling,
  Phys. Rev. Lett. {\bf 87}, 094101 (2001).

\bibitem{SouSuz02}
S.~Souma and A.~Suzuki, Local density of states and scattering matrix in
  quasi-one-dimensional systems, Phys. Rev.~B {\bf 65}, 115307 (2002).

\bibitem{SpiZyu04}
B.~Spivak and A.~Zyuzin, Signature of the electron-electron interaction in the
  magnetic-field dependence of nonlinear $I$-$V$ characteristics in mesoscopic
  systems, Phys. Rev. Lett. {\bf 93}, 226801 (2004).

\bibitem{SteCheFabGog99}
M.~Steiner, Y.~Chen, M.~Fabrizio and A.~O. Gogolin, Statistical properties of
  localization-delocalization transition in one dimension, Phys. Rev.~B {\bf
  59}(23), 14848--14851 (1999).

\bibitem{StiHer75}
F.~H. Stillinger and D.~R. Herrick, Bound states in the continuum, Phys. Rev. A
  {\bf 11}, 446--454 (1975).

\bibitem{Tan01}
T.~Taniguchi, Charge current density from the scattering matrix, Phys. Lett.~A
  {\bf 279}, 81--86 (2001).

\bibitem{TanBut99}
T.~Taniguchi and M.~B{\"u}ttiker, Friedel phases and phases of transmission
  amplitudes in quantum scattering systems, Phys. Rev.~B {\bf 60}, 13814--13823
  (1999).

\bibitem{Tex99}
C.~Texier, {\em Quelques aspects du transport quantique dans les syst\`emes
  d\'esordonn\'es de basse dimension\/}, Ph.D. thesis, Universit\'e Paris 6
  (1999), \url{http://lptms.u-psud.fr/christophe_texier/} or ,
  \url{http://tel.archives-ouvertes.fr/tel-01088853}.

\bibitem{Tex02}
C.~Texier, Scattering theory on graphs (2): the Friedel sum rule, J.~Phys.~A:
  Math. Gen. {\bf 35}, 3389--3407 (2002).

\bibitem{Tex10hdr}
C.~Texier, {\em D\'esordre, localisation et interaction -- Transport quantique
  dans les r\'eseaux m\'etalliques\/} (Habilitation \`a Diriger des Recherches,
  Universit\'e Paris-Sud, 2010),
  \url{http://tel.archives-ouvertes.fr/tel-01091550}.

\bibitem{Tex15book}
C.~Texier, {\em M\'ecanique quantique\/}, Dunod, Paris, second edition (2015).

\bibitem{TexBut99unpub}
C.~Texier and M.~B{\"u}ttiker  (1999), unpublished.

\bibitem{TexBut03}
C.~Texier and M.~B{\"u}ttiker, Local Friedel sum rule in graphs, Phys. Rev.~B
  {\bf 67}, 245410 (2003).

\bibitem{TexCom99}
C.~Texier and A.~Comtet, Universality of the Wigner time delay distribution for
  one-dimensional random potentials, Phys. Rev. Lett. {\bf 82}(21), 4220--4223
  (1999).

\bibitem{TexDeg03}
C.~Texier and P.~Degiovanni, Charge and current distribution in graphs,
  J.~Phys.~A: Math. Gen. {\bf 36}, 12425--12452 (2003).

\bibitem{TexHag10}
C.~Texier and C.~Hagendorf, The effect of boundaries on the spectrum of a
  one-dimensional random mass Dirac Hamiltonian, J.~Phys.~A: Math. Theor. {\bf
  43}, 025002 (2010).

\bibitem{TexMaj13}
C.~Texier and S.~N. Majumdar, Wigner time-delay distribution in chaotic
  cavities and freezing transition, Phys. Rev. Lett. {\bf 110}, 250602 (2013),
  Erratum: {\it ibid} {\bf112}, 139902 (2014).

\bibitem{TexMit18}
C.~Texier and J.~Mitscherling, Nonlinear conductance in weakly disordered
  mesoscopic wires: Interaction and magnetic field asymmetry, Phys. Rev. B {\bf
  97}, 075306 (2018).

\bibitem{TexMon01}
C.~Texier and G.~Montambaux, Scattering theory on graphs, J.~Phys.~A: Math.
  Gen. {\bf 34}, 10307--10326 (2001).

\bibitem{TitBroFurMud01}
M.~Titov, P.~W. Brouwer, A.~Furusaki and C.~Mudry, Fokker-Planck equations and
  density of states in disordered quantum wires, Phys. Rev. B {\bf 63}, 235318
  (2001).

\bibitem{TitFyo00}
M.~Titov and Y.~V. Fyodorov, Time-delay correlations and resonances in
  one-dimensional disordered systems, Phys. Rev.~B {\bf 61}(4), R2444 (2000).

\bibitem{TitSch03}
M.~Titov and H.~Schomerus, Anomalous Wave Function Statistics on a
  One-Dimensional Lattice with Power-Law Disorder, Phys. Rev. Lett. {\bf 91},
  176601 (2003).

\bibitem{TsaOsb75}
T.~Y. Tsang and T.~A. Osborn, The spectral property of time delay, Nucl.
  Phys.~A {\bf 247}, 43--50 (1975).

\bibitem{UhlBet36}
G.~E. Uhlenbeck and E.~Beth, The quantum theory of the non-ideal gas I.
  Deviations from the classical theory, Physica {\bf 3}(8), 729--745 (1936).

\bibitem{ValOzoLew98}
R.~O. Vallejos, A.~M. {Ozorio de Almeida} and C.~H. Lewenkopf, Quantum time
  delay in chaotic scattering: a semiclassical approach, J. Phys. A: Math. Gen.
  {\bf 31}(21), 4885--4897 (1998).

\bibitem{TigSebStoGen99}
B.~A. van Tiggelen, P.~Sebbah, M.~Stoytchev and A.~Z. Genack, Delay-time
  statistics for diffuse waves, Phys. Rev. E {\bf 59}, 7166--7172 (1999).

\bibitem{Viv14privcom}
P.~Vivo, private communication  (2014).

\bibitem{NeuWig29}
J.~{von~Neumann} and E.~Wigner, \"Uber merkw\"urdige diskrete Eigenwerte, Phys.
  Z. {\bf 30}, 465 (1929).

\bibitem{SebZycZak96}
P.~\v{S}eba, K.~\.{Z}yczkowski and J.~Zakrzewski, Statistical properties of
  random scattering matrices, Phys. Rev. E {\bf 54}, 2438--2446 (1996).

\bibitem{Wig55}
E.~P. Wigner, Lower limit for the energy derivative of the scattering phase
  shift, Phys. Rev. {\bf 98}(1), 145--147 (1955).

\bibitem{XuWan11}
F.~Xu and J.~Wang, Statistics of Wigner delay time in Anderson disordered
  systems, Phys. Rev. B {\bf 84}, 024205 (2011).

\bibitem{Yor00}
M.~Yor, {\em Exponential functionals of Brownian motion and related
  processes\/}, Springer (2000).

\bibitem{Zim72}
J.~M. Ziman, {\em Principles of the theory of solids\/}, Cambridge University
  Press (1972).

\bibitem{Zir96}
M.~R. Zirnbauer, Riemannian symmetric superspaces and their origin in
  random-matrix theory, J. Math. Phys. {\bf 37}(10), 4986--5018 (1996).

\bibitem{ZumMarHanGos06}
D.~M. Zumb\"{u}hl, C.~M. Marcus, M.~P. Hanson and A.~C. Gossard, Asymmetry of
  Nonlinear Transport and Electron Interactions in Quantum Dots, Phys. Rev.
  Lett. {\bf 96}, 206802 (2006).

\end{thebibliography}
%

\end{document}